\documentclass[preprint2]{aastex}

\usepackage{natbib}
\usepackage{url}
\usepackage[section]{placeins}
\usepackage{graphics,graphicx}
\usepackage[caption=false]{subfig}
\usepackage{pdflscape}
\usepackage{amsmath, amsthm, amssymb}

\newcommand{\feii}{[Fe \textsc{ii}]}

\newcommand{\brg}{Br$\gamma$}
\newcommand{\brd}{Br$\delta$}
\newcommand{\molhy}{H$_2$}
\newcommand{\paa}{Pa$\alpha$}
\newcommand{\pab}{Pa$\beta$}

\def\ltsima{$\; \buildrel < \over \sim \;$}
\def\simlt{\lower.5ex\hbox{\ltsima}}
\def\gtsima{$\; \buildrel > \over \sim \;$}
\def\simgt{\lower.5ex\hbox{\gtsima}}

\shorttitle{Nuclear Disks in Nearby (U)LIRGs}
\shortauthors{Medling et al.}

\begin{document}


\title{Stellar and Gaseous Nuclear Disks Observed in Nearby (U)LIRGs} 


\author{Anne M. Medling\altaffilmark{1,2,8}, 
Vivian U\altaffilmark{3,4}, 
Javiera Guedes\altaffilmark{5},
Claire E. Max\altaffilmark{2},  
Lucio Mayer\altaffilmark{6},
Lee Armus\altaffilmark{7},  
Bradford Holden\altaffilmark{2},
Rok Ro\v{s}kar\altaffilmark{6},
David Sanders\altaffilmark{4}
}

\altaffiltext{1}{Research School of Astronomy \& Astrophysics, Mount Stromlo Observatory, Australian National University, Cotter Road, Weston Creek, ACT 2611, Australia; anne.medling@anu.edu.au}

\altaffiltext{2}{Department of Astronomy \& Astrophysics, University of California, Santa Cruz, CA 95064, USA;  max@ucolick.org}

\altaffiltext{3}{Department of Physics and
   Astronomy, University of California, Riverside, 900 University
   Avenue, Riverside, CA 92521, USA; vivianu@ucr.edu}

\altaffiltext{4}{Institute for Astronomy, University of Hawaii,
   2680 Woodlawn Dr., Honolulu, HI 96822, USA}

\altaffiltext{5}{Institute for Astronomy, ETH Z\"urich, Wolfgang-Pauli-Strasse 27, 8093 Z\"urich, Switzerland}

\altaffiltext{6}{Institute for Theoretical Physics, University of Z\"urich, Winterthurerstrasse 190, 8057, Z\"urich, Switzerland}

\altaffiltext{7}{Spitzer Science Center, California Institute of
    Technology, 1200 E. California Blvd., Pasadena, CA 91125, USA}

\altaffiltext{8}{NSF Graduate Research Fellow}




\begin{abstract}

We present near-infrared integral field spectroscopy of the central kiloparsec of 17 nearby luminous and ultra-luminous infrared galaxies undergoing major mergers.  These observations were taken with OSIRIS assisted by the Keck I and II Adaptive Optics systems, providing spatial resolutions of a few tens of parsecs.  The resulting kinematic maps reveal gas disks in at least 16 out of 19 nuclei and stellar disks in 11 out of 11 nuclei observed in these galaxy merger systems.  In our late-stages mergers, these disks are young (stellar ages $<30$ Myr) and likely formed as gas disks which became unstable to star formation during the merger.  On average, these disks have effective radii of a few hundred parsecs, masses between $10^{8}$ and $10^{10} M_{\sun}$, and $v/\sigma$ between 1 and 5.  These disks are similar to those created in high-resolution hydrodynamical simulations of gas-rich galaxy mergers, and favor short coalescence times for binary black holes.  The few galaxies in our sample in earlier stages of mergers have disks which are larger ($r_{eff}\sim200-1800$ pc) and likely are remnants of the galactic disks that have not yet been completely disrupted by the merger.    

\end{abstract}


\keywords{Galaxies: kinematics and dynamics -- galaxies: nuclei -- galaxies: interactions}


\section{Introduction}

High spatial resolution hydrodynamic simulations of gas-rich galaxy mergers have predicted the formation of nuclear disks on scales of tens to hundreds of parsecs \citep[e.g.][]{Mayer08}.  These disks form as gravitational torques funnel the gas towards the nuclei of the merging galaxies, feeding gas down to the smallest possible scales resolved in the simulations ($\sim1$ pc).  The presence of this gas and the resulting dissipation can provide a mechanism to speed up the formation and final coalescence of binary black holes \citep{Kazantzidis05,Mayer07,Dotti06,Dotti07,Dotti08,Lodato09,Cuadra09,Chapon13}.  When the black hole binary is in the early stages, dynamical friction caused by the gas disk will remove angular momentum from the binary, causing its separation to decrease.  In later stages, when the binary dominates the local gravitational potential, it can create a density enhancement in the surrounding gas which causes a gravitational torque on the binary, removing further angular momentum.  Simulations by \citet{Escala05} further show that increasing the black hole mass (or, equivalently, the density of the gas disk) by a factor of ten will increase the dynamical friction such that the coalescence time is halved.  During final coalescence of the black holes, asymmetric emission of gravitational waves can give the merged black hole a kick, potentially ejecting it from the system in extreme cases.  Gas disks can both limit the magnitude of the kick through accretion-driven spin alignment of the black holes \citep{Bogdanovic07, Dotti10} and contribute to the retention of recoiling black holes near the center of the merger remnant by dynamical friction \citep{Guedes11}.  

To date, nuclear disks of varying scales (a few to a few hundred parsecs) have been found in a handful of nearby isolated galaxies, both spirals \citep[e.g.][]{Zasov99,Pizzella02,Dumas07,Haan09,GarciaBurillo12,Hicks13} and ellipticals \citep[e.g.][]{Scorza98, Krajnovic04,Kormendy05}.  These nuclear disks are identified using one or both of high-resolution imaging and spectroscopy to identify disky isophotes and Keplerian-like rotation curves.  Kinematically decoupled cores and nuclear disks on these scales have also been discovered in a number of early-type galaxies in the SAURON survey \citep{deZeeuw02,McDermid04,Emsellem04,McDermid06NAR,Krajnovic08}.  \citet{McDermid06} suggest that the most compact of these decoupled cores are linked with recent star formation.  Their formation is likely to have resulted from gas inflow caused by a galaxy merger \citep{Kormendy84,Bender88,Balcells90,Hernquist91} or interaction \citep{Bender88,Hau94}.  This hypothesis has been supported by the case of NGC~5953, whose kinematically decoupled core may have been caused by the current inflow of cold gas from its neighbor, NGC~5954 \citep{Falcon06,Falcon07}.
Recently, \citet{Hicks13} have shown that active galactic nuclei (AGN) are likely to host nuclear disks on scales of a few hundred parsecs, and that these disks are absent in similar quiescent galaxies; this suggests that disks may be an important mechanism for transporting fuel inwards to power AGN. 
Studies of merging galaxies have revealed nuclear disks, both with ALMA data \citep{Imanishi13} and from the radial distribution of supernovae \citep{Herrero11}.

In order to understand the origins of kinematically decoupled galaxy cores, and to determine when and how much gas is available to affect black hole merger rates, it is important to search for and characterize nuclear disks in nearby merging galaxies, or galaxies which have undergone recent mergers.  To accomplish this, we have targeted local luminous and ultraluminous infrared galaxies ((U)LIRGs), the majority of which have been identified as gas-rich major mergers in the local universe \citep[e.g.][]{IshidaPhD, Sanders88, Sanders96}.  These systems are bright in the infrared because of the large quantities of dust heated by either or both of an AGN or starburst \citep{Sanders96}.  

There is considerable evidence that infrared luminosity is linked with merging activity: morphological analysis of LIRGs reveals an increase of strongly interacting systems with higher infrared luminosity \citep[e.g.][]{IshidaPhD}; studies of pairs of Sloan Digital Sky Survey galaxies find an increasing fraction of LIRGs as pair separation decreases \citep{Ellison13}; and observations of (U)LIRGs at high redshift find mergers to be a major factor, though slightly less so than at low redshift \citep{Melbourne05, Dasyra08, Kartaltepe2010, Kartaltepe2012}.  Locally, the merger fraction can pass 90\% at the highest luminosities \citep{Sanders88,Melnick90,Clements96, Veilleux02, IshidaPhD,Haan11}.  With higher infrared luminosity, the mid-infrared slope and silicate depths also increase, suggesting nuclei are more obscured and more compact \citep{Stierwalt13}.  Additionally, as infrared luminosity increases, AGN fraction also increases \citep{Veilleux95, Iwasawa11, Ellison13, Koss13}.  AGN contribute significantly to the infrared luminosity in 18\% of (U)LIRGs, making up more than half of the system's $L_{IR}$ in 10\% of (U)LIRGs \citep{Petric11}.  Note, however, that though merger activity has been linked to triggering star formation and AGN activity, the converse is not necessarily true: star formation and AGN do not require mergers, and the most common low-luminosity AGNs are likely unassociated with mergers \citep{Cisternas11_merger,Kocevski12}.

An extensive body of multiwavelength observations of (U)LIRGs exists as the Great Observatory All-sky LIRG Survey \citep[GOALS;][]{Armus09}.  This survey consists of the brightest infrared galaxies at redshifts $z<0.1$ and provides imaging and spectroscopic observations with the Hubble Space Telescope, the Galaxy Evolution Explorer (GALEX), the Chandra X-Ray Observatory, the Spitzer Space Telescope, and the Herschel Space Observatory.  
The existing Hubble ACS images of these merging galaxies have sufficient spatial resolution to detect the disky isophotes commonly associated with nuclear disks \citep{Kim13}.  However, because (U)LIRGs exhibit high levels of extinction due to dust \citep{Scoville00}, the nuclear regions are too obscured in the optical bands to detect these disks.  High spatial resolution observations in the near-infrared have been difficult to obtain from space due to the limiting size of space telescope primary mirrors, though recently $H$-band imaging has been obtained using the Hubble Space Telescope \citep{Haan11,Haan13}.  The diffraction limit of resolution is set by the ratio of the wavelength to the diameter of the telescope, $\frac{\lambda}{D}$, and is $0\farcs15$ for Hubble in the $H$-band.  However, clumpy dust extinction is still a problem in $H$-band in these galaxies; for studying nuclear disks, one would prefer even longer wavelengths and higher spatial resolutions.  
This paper focuses on the central kiloparsec or two of each system; to understand the broader context of each (U)LIRG, we refer the reader to already-published wide-field imaging in the near-infrared \citep{Haan11} and the optical \citep{Kim13}.

In this paper, we present high spatial resolution integral field spectroscopy of the nuclear regions of 17 (U)LIRGs undergoing major mergers.  In \S\ref{obs} we describe our observations and data processing.  The resulting spectra are analyzed using a procedure described in \S\ref{methods} to produce flux and velocity maps, which are presented in \S\ref{results}.  In \S\ref{discussion}, we discuss the implications of our observed nuclear disks and compare our results to disks found in high-resolution merger simulations.  Our conclusions are summarized in \S\ref{conclusions}.  Figures for the complete sample are included in Appendix~\ref{extrafig}; details of GALFIT fits are included in Appendix~\ref{galfitdetails}.  We  adopt $H_0 = 70$\,km\,s$^{-1}$\,Mpc$^{-1}$,
$\Omega_{\rm m}$ = 0.28, and $\Omega_\Lambda$ = 0.72
throughout the paper.  Physical scales were calculated using Ned Wright's Cosmology Calculator\footnote{Available at \url{http://www.astro.ucla.edu/~wright/CosmoCalc.html}.} \citep{Wright06}.


\section{Data}
\label{obs}

\subsection{Sample Selection}
We selected our sample of gas-rich merging galaxies from the GOALS sample, which spans all nuclear spectral types and merger stages.  Our targets are primarily from later stages of mergers, in which the two nuclei are within $\sim5\arcsec$ (2-7 kpc at these redshifts), in order to target the time period when simulations predict nuclear disks are most prevalent.  Several galaxies at earlier stages of merging (IRAS F00359+1523, VV340a, and NGC~6090), originally observed for a different program, were included for comparison; this makes our sample of 17 galaxies (21 nuclei) comprised of 3 early- and 14 late-stage galaxy mergers (3 early and 18 late merging nuclei).  We further limited our sample to those GOALS galaxies with available Hubble Space Telescope ACS $B$- and $I$-band imaging \citep[][Evans et al. in prep]{Kim13}; the high spatial resolution of ACS was required to ensure accurate pointings, as the field of view of OSIRIS is quite small.  This sample represents about half of the late-stage mergers in the ACS-GOALS sample which are observable from the latitude of Keck Observatory and which have suitable guide stars for adaptive optics correction (see next section for guide star requirements).

\subsection{Observations}
We observed the nuclei of 17 (U)LIRGs undergoing major mergers with the OH-Suppressing InfraRed Imaging Spectrograph \citep[OSIRIS,][]{Larkin06}, on the W.~M. Keck II telescope using LGS AO between April 2007 and May 2012 and on the W.~M. Keck I telescope in May 2013.  OSIRIS is a near-infrared integral field spectrograph with a lenslet array capable of producing up to 3000 spectra at once.  The spectral resolution ranges from about 3400 in the largest pixel scale to 3800 in the three finer pixel scales.  This resolution is sufficient to resolve spectral regions between the OH emission lines from Earth's atmosphere, and enables improved sky subtraction.  

OSIRIS sits behind the Keck Observatory AO System \citep{Wiz00,vanDam04,Wiz06,vanDam06}, which measures wavefront distortions due to atmospheric turbulence and corrects for them using a deformable mirror.  This technique enables spatial resolutions in the near-infrared matching those of the Hubble Space Telescope at visible wavelengths ($\sim0\farcs05$).  OSIRIS observations may be taken at 0\farcs02, 0\farcs035, 0\farcs05 or 0\farcs1 per pixel.  In order to measure atmospheric distortions, AO systems require either a natural guide star (NGS) or a laser guide star (LGS) plus a tip-tilt star.  NGS AO at Keck Observatory requires a guide star brighter than $\sim$13th magnitude in the $R$-band and within $\sim30$\arcsec of the target.  For all of our systems except IRAS F01364-1042, we used the Keck LGS AO system, since the required tip-tilt star can be up to an arcminute away and as faint as $\sim$18th magnitude in the $R$-band.  In this case, the Keck LGS AO system uses a pulsed laser tuned to the 589 nm Sodium D$_2$ transition, exciting atoms in the sodium layer of the atmosphere (at $\sim$95 km) and causing spontaneous emission.  Thus, the laser creates a spot in the upper atmosphere which allows the AO system to monitor turbulence below the sodium layer via a Shack-Hartmann wavefront sensor and correct for it with a deformable mirror.  This laser guide star enables high-order corrections to the wavefront, but corrections for image motion are corrected using the tip-tilt star.

Our OSIRIS observations usually consisted of ten minute exposures taken in sets of three, using an  object-sky-object dither pattern.  Total exposure times on each target and their relevant sky frames are listed in Table~\ref{tbl:observingparams}.  In each case we also observed the tip-tilt star to estimate the point-spread function (PSF).  We selected filters and plate scales based on the redshift of each target so that our lines of interest fall within the observed waveband; these are also listed in Table~\ref{tbl:observingparams}.  We employed a combination of broadband observations (Kbb and Kcb: 1.965 $\mu$m - 2.381 $\mu$m) and narrowband observations (Hn4: 1.652 $\mu$m - 1.737 $\mu$m; Kn5: 2.292 $\mu$m - 2.408 $\mu$m).  

\subsection{Data Reduction}
We reduced our 2009-2012 OSIRIS observations using the OSIRIS Data Reduction Pipeline version 2.3\footnote{Available at \url{http://irlab.astro.ucla.edu/osiris/pipeline.html}.}, which includes modules to subtract sky frames, adjust channel levels, remove crosstalk, identify glitches, clean cosmic rays, extract a spectrum for each spatial pixel, assemble the spectra into a data cube, correct for atmospheric dispersion, perform telluric corrections, and mosaic frames together.  We implemented the updated OSIRIS wavelength solution for observations taken after October 2009, available now in the pipeline version 3.0.  When the atmosphere is particularly unstable, a standard sky subtraction may leave residuals around the atmospheric OH lines.  In such cases, we utilized the Scaled Sky Subtraction module based on the technique outlined in \citet{Davies07}, which scales the thermal continuum and OH line groups separately to provide optimal sky subtraction; we modified the module to include a smoother subtraction of the thermal continuum.  This is not possible for observations that lack strong OH lines in their spectral coverage (e.g. Kn5) and is no different from regular sky subtraction if OH lines do not vary.  For the two galaxies observed in 2013 we used the full OSIRIS Data Reduction Pipeline version 3.0 for updated calibrations of the reductions described above.

We imaged the tip-tilt star immediately before observing each galaxy in order to provide an estimate for the PSF.  When conditions allowed, we also imaged the tip-tilt star again at the end and/or half-way through a galaxy's observations.  Though additional time on the tip-tilt star provides higher signal-to-noise, we did not find the PSF to vary significantly between the start and finish of a set of observations (approximately 2-3 hours), except in cases of changing weather conditions.  A Moffat profile was fit to each tip-tilt star and then broadened according to the distance between the galaxy and the tip-tilt star.  At Mauna Kea, the isokinetic angle is $\sim75$\arcsec \cite[][]{vanDam06}; this represents the on-sky distance from the tip-tilt star at which the Strehl will be degraded by $1/e$.  Each galaxy's tip-tilt star was therefore broadened by a Gaussian broad enough to decrease the tip-tilt star's peak flux to match the predicted degradation based on the distance to the target.  Because most tip-tilt stars are significantly closer than 75\arcsec, this correction was usually small, about 10-30\% in peak flux.

\onecolumn
\begin{landscape}
\tiny
\begin{center}
 \begin{deluxetable}{lcccccccc}

    \centering
    \tabletypesize{\scriptsize}
    \tablewidth{0pt}
    \tablecolumns{9}
    \tablecaption{Details of Observations}
    \tablehead{   
      \colhead{Galaxy Name} &
      \colhead{Redshift} &
      \colhead{UT Date(s)} &
      \colhead{Filter} &
      \colhead{Plate Scale} &
      \colhead{Exp Time on} &
      \colhead{Exp Time on} & 
      \colhead{Tip-tilt Star} &
      \colhead{Tip-tilt Star} \\
      \colhead{} &
      \colhead{} &
      \colhead{YYMMDD} &
      \colhead{} &
      \colhead{(\arcsec / pixel)} &
      \colhead{Target (minutes)} &
      \colhead{Sky (minutes)} & 
      \colhead{$R$ Magnitude\tablenotemark{a}} &
      \colhead{Separation (\arcsec)} \\
      	}
    \startdata
    CGCG436-030 & 0.0315 & 120102 & Kcb & 0.1 & 30 & 10 & 11.2 & 33.1 \\
    IRAS F01364-1042 & 0.0493 & 101113, 101114, 120102 & Kcb&0.1&130 & 80 & 10.3 &30.0 \\
    IIIZw035 & 0.0281 & 111210 & Kbb & 0.035 & 100 & 50 & 12.6 &44.3 \\
    IRAS F03359+1523 & 0.0368 & 101114 & Kcb & 0.1 & 60 & 30 & 17.4 & 30.0 \\
    MCG+08-11-002 & 0.0198 & 110110, 120102 & Kcb & 0.1 & 60 & 30 & 16.4 &17.6\\
    NGC~2623 & 0.0199 & 110110, 110203 & Kcb & 0.1& 50 & 25 & 16.9 & 55.4 \\
    UGC5101 & 0.0393 & 101114 & Kcb & 0.1 & 30 & 20 & 14.7 &35.4\\
    UGC5101 & 0.0393 & 100304, 100305 & Kn5 & 0.035 & 80 & 50 & 14.7&35.4\\
    Mrk231/UGC8058 & 0.0436 & 110523 & Kbb & 0.035 & 36 & 18 & 10. & 0.0\tablenotemark{b}\\
    UGC8387 & 0.0239 & 130518 & Kcb & 0.1 & 30 & 10 & 14.2 & 16.2 \\
    Mrk273/UGC8696\tablenotemark{c} & 0.0383 & 120522 & Kcb & 0.1 & 50 & 30 & 16.1& 33.1\\
    VV340a & 0.0348 & 130518 & Kcb & 0.1 & 40 & 20 & 13.2 & 39.6 \\
    IRAS F15250+3608 & 0.0566 & 110523 & Kbb & 0.05 & 80 & 40 & 15.6&53 \\
    NGC~6090 & 0.0307 & 100305 & Kcb & 0.1 & 30 & 10 &15.7 &36.5\\
    NGC~6090 & 0.0307 & 110524 & Kn5 & 0.035 & 130 & 70 & 15.7 & 36.5\\
    NGC~6240N & 0.0247 &  090617 & Kn5 & 0.035 & 210 & 75 & 12.0 &35.8 \\
    NGC~6240S\tablenotemark{d} & 0.0247 &  070421 & Kn5 & 0.035 & 20 & 10 &12.0 &35.8 \\
    IRAS F17207-0014 & 0.0435 & 110523 & Kcb & 0.1 &  60 & 30 & 14.1 & 25.6\\
    IRAS F17207-0014 & 0.0435 & 110523, 110524 &  Hn4 & 0.035 & 40 & 20 &14.1 &25.6 \\
    IRAS 20351+2521 & 0.0343 & 110522 & Kcb & 0.1 & 30 & 20 & 12.2& 18.1\\
    IRAS 20351+2521 & 0.0343 & 110522 & Hn4 & 0.035 & 60 & 30 & 12.2& 18.1 \\
    IRAS F22491-1808 & 0.0784 & 101114 & Kcb & 0.1 & 20 & 10 & 16.4 &43.0 \\
    \enddata
    \tablenotetext{a}{Taken from the USNO B1 Catalog}
    \tablenotetext{b}{Mrk231 observations used the central quasar as the tip-tilt star; therefore the offset to the galaxy is zero.}
    \tablenotetext{c}{Originally presented in \citet{mrk273}}
    \tablenotetext{d}{Originally presented in \citet{medling11}}
    \label{tbl:observingparams}
  \end{deluxetable}

\end{center}
\end{landscape}
\twocolumn

\subsection{Tracers Analyzed for Each Galaxy}
For each galaxy, observations include a subset of the following emission and absorption features: \brg~2.16 $\mu$m,  \molhy~2.12 $\mu$m, \paa~1.875 $\mu$m, and the CO (2-0) and (3-1) bandheads at 2.293 $\mu$m and 2.323 $\mu$m.  The tracers observed for each galaxy are listed in Table~\ref{tbl:tracers}.  Stellar kinematics are most commonly unavailable, as the CO bandheads at 2.29 $\mu$m are redshifted out of the $K$-band.  Some galaxies in our sample have stellar kinematics measured from the CO bandheads in the $H$-band, though not all $H$-band observations have high enough signal-to-noise for this.  Several other lines are too faint or noisy for analysis (e.g. Mrk231); these are marked in Table~\ref{tbl:tracers} as well.  Lastly, we note that the data for NGC~6240 were observed originally for a stellar kinematics project only \citep{medling11}, and therefore gas kinematics are not available.  However, we direct the reader to \cite{Hauke}, in which an IFU analysis of NGC~6240 is presented; they find that the gas in NGC~6240 is not morphologically or kinematically consistent with the stars.

Our observations of UGC8387 are not well-centered on the core of the galaxy, and for this reason will be left out of the analysis in this paper.  Though we see portions of the core, they fall on the edge of the frame and we cannot be certain that the center and major axis are visible.  Still, the maps of flux and velocity are included in this paper for completeness to aid future observations of this object.

From this set of data, we are able to measure stellar kinematics of 10 galaxies and 11 nuclei and gas kinematics in 16 galaxies and 19 nuclei; combined we can investigate stellar or gas kinematics in 17 galaxies and 21 nuclei.  These numbers include UGC 8387, though it is left out of the more sophisticated analyses.  The galaxies with double nuclei are Mrk 273, NGC~6240, IRAS F17207-0014, and IRAS F22491-1808. 

\begin{center}
 \begin{deluxetable}{lccccc}
\tiny

    \centering
    \tabletypesize{\scriptsize}
    \tablewidth{0pt}
    \tablecolumns{5}
    \tablecaption{Tracers Observed}
    \tablehead{   
      \colhead{Galaxy Name} &
      \colhead{CO bandheads\tablenotemark{a}} &
      \colhead{\brg} &
      \colhead{\paa\tablenotemark{b}} &
      \colhead{\molhy~(2.12 $\mu$m)} &
      	}
    \startdata
    CGCG436-030 & Y & Y & N & Y \\
    IRAS F01364-1042 & N & Y & Y & Y \\ 
    IIIZw035 & Y & Y & N & Y \\
    IRAS F03359+1523 & N & Y & N & Y \\
    MCG+08-11-002 & Y & Y & N & Y \\
    NGC~2623 & Y & Y & N & Y \\
    UGC5101 & Y & Y & N & Y \\
    Mrk231/UGC8058 & Y/N\tablenotemark{c} & Y/N & N & Y \\
    UGC8387\tablenotemark{d} & Y & Y & N & Y \\
    Mrk273/UGC8696 & Y/N & Y & N & Y \\
    VV340a & Y & Y & N & Y \\
    IRAS F15250+3608 & N & Y & Y & Y \\
    NGC~6090 & Y & Y & N & Y \\
    NGC~6240\tablenotemark{e} & Y &  N & N & N \\
    IRAS F17207-0014 & Y/N & Y & N & Y \\
    IRAS 20351+2521 & Y & Y & N & Y \\
    IRAS F22491-1808 & N & Y & Y & N \\
    \enddata
    \tablenotetext{a}{The CO bandheads are used to trace stellar kinematics, and are redshifted out of the $K$-band at $z\sim0.34$.  CO bandheads are also present across the $H$-band, but as they are less deep, higher signal-to-noise is required to fit them.}
    \tablenotetext{b}{When \paa~is observed, we use it instead of \brg~even though both are available, since they trace the same atomic hydrogen gas and in all cases the \paa~line has the higher signal-to-noise ratio.}
    \tablenotetext{c}{Tracers marked with `Y/N' indicate that this tracer's wavelength falls within the observed band but that the signal-to-noise ratio is too low to provide meaningful analysis in this paper.}
    \tablenotetext{d}{As mentioned in the text, the observations of UGC8387 are off-center and therefore analysis requiring knowledge of the major axis is not possible.  It is left out of the analysis sections of this paper, but its data are included for completeness and for future reference.}
    \tablenotetext{e}{NGC~6240 was observed originally for \citet{medling11} with a narrow-band filter, rendering studies of the gas kinematics impossible.  However, similar observations were presented in \citet{Hauke}, showing that the \brg~and \molhy~gas do not match the stellar morphology or kinematics.}
    \label{tbl:tracers}
  \end{deluxetable}

\end{center}


\section{Line-Fitting Methods}
\label{methods}

The OSIRIS data cubes capture both gas emission lines and stellar absorption features, which are analyzed separately in the following two sections.  In each case, the data were binned using optimal Voronoi tessellations according to the code described in \citet{voronoi}\footnote{IDL routines available at \url{www-astro.physics.ox.ac.uk/~mxc/idl/}.}.  This technique retains the high spatial resolution in regions with sufficient signal-to-noise, while binning lower signal-to-noise ratio regions enough to obtain meaningful measurements.

The signal-to-noise ratio in each pixel was calculated using one of two methods.  For emission lines, a single Gaussian function was fit to the data; the continuum-subtracted area under this curve is the signal.  The corresponding noise is the root-mean-squared residual of the nearby continuum.  For absorption lines, the signal-to-noise ratio was calculated as described in \citet{medling11}, using flux in the continuum for signal and calculating the noise theoretically using published values of readnoise and dark current and Poisson noise for the galaxy and sky fluxes.  The theoretical noise estimates match well with other empirical noise measurements (e.g. strength of continuum vs rms of continuum, flux in emission line vs. rms of continuum, and flux of line vs. rms of fit).  

We then used the signal-to-noise ratio calculations to spatially bin the data to obtain regions of a sufficient minimum signal-to-noise ratio.  To measure an emission line, the pixels were Voronoi binned to provide a signal-to-noise ratio of at least 3 per bin.  For stellar absorption line fitting, the broad nature of molecular absorption features require a higher signal-to-noise ratio for reliable measurements.  Therefore, for stellar kinematics, the pixels were Voronoi binned to a signal-to-noise ratio threshold of 20 in $K$-band and, since the 1.6$\mu$m CO bandheads are intrinsically fainter, 40 in $H$-band.

\subsection{Probing the Gas}

Emission lines were fit using the methods outlined in \citet{mrk273}.  First the cubes were continuum-subtracted, by fitting a power law continuum to each pixel's spectrum.  Then the pixels were binned according to the optimal Voronoi tessellations.  The binned, continuum-subtracted spectra were then used as input for the appropriate line-fitting codes.  When a species had only one line falling within a band (e.g. \paa), that line was fit with a single Gaussian function.  The Brackett series of lines (\brg~and \brd) are frequently both observable in $K$-band; these two lines were fit simultaneously, requiring that the velocity and velocity dispersion be consistent between the two lines.  Similarly, the OSIRIS $K$ broadband hosts five molecular hydrogen lines which all have the same line profiles.  We simultaneously fit the five lines, allowing the fluxes in each line to vary but constraining the velocities and velocity dispersions to be the same.

No broad lines ($\sigma > 300$ km s$^{-1}$) are seen in our data, indicating that any AGNs present with broad line regions are obscured even in the $K$ band.  However, the narrow emission lines we do see may still be photoionized in part by an AGN.  The reader should consider this before attributing all \brg~emission to star formation.

\subsection{Probing the Stars}
Stellar kinematics were investigated for 10 systems (11 nuclei) using the CO molecular bandheads in the $H$- or $K$-bands.  In those cases, stellar kinematics were fit using the Penalized Pixel Fitting routine \citep[pPXF;][]{ppxf}, which produces a line-of-sight velocity distribution using a maximum penalized likelihood approach and a library of stellar templates.  In the same manner as in \citet{medling11}, the routine was adapted to low signal-to-noise spectra by setting the BIAS keyword to 1000; this biases the code against fitting higher order moments ($h_{3}$ and $h_{4}$) and instead only fits the velocity and velocity dispersion of the stars in each bin, effectively shutting off $h_{3}$ and $h_{4}$.  If we had chosen to increase the signal-to-noise ratio to 50 or higher in order to obtain reliable higher-order moments, the binning would sacrifice too much resolution to produce meaningful maps. 

Due to the recent star formation in these galaxies, the CO absorption bandheads are prominent stellar features in the $H$- and $K$-bands.  Therefore, to further improve the signal-to-noise of the fits, the spectral regions around the CO bandheads were weighted more heavily.  For $K$-band spectra, we used a selection of late-type giant and supergiant spectral templates from the GNIRS \citep{GNIRS} library.  Late-type giant and supergiant stars have been shown to be a good fit for starburst galaxies \citep[e.g.][]{Hauke, medling11} and host deep CO bandheads, excellent for fitting kinematics.  For $H$-band spectra, we used unpublished spectral templates of K and M giants and supergiants provided by Nate McCrady (personal communication), and posted by Nicholas McConnell on the OSIRIS wiki\footnote{See the Spectrograph Standard Stars section of the OSIRIS wiki at: \url{irlab.astro.ucla.edu/osiriswiki/}}.


\section{Results}
\label{results}

\subsection{Modeling the Flux Distribution with GALFIT}
\label{galfitsection}
To quantify the characteristic sizes of the flux profiles, we modeled the continuum and emission line flux maps with GALFIT \citep{Peng02, Peng10} using multiple components; at least one S\'ersic profile was fit in each case.  Because these galaxies frequently show substructure in the nuclear regions, additional components (Gaussians, PSFs, and S\'ersics) were added to the fit until the disk's properties stabilized.  This enabled us to fit the nuclear disks without confusion by spiral arms, tidal streams, AGNs, on- and off-nuclear star clusters.  Note that the continuum flux maps include continuum due to both star formation and AGN; to obtain an appropriate stellar disk model with GALFIT, we include a central component.

For galaxies in which two S\'ersics were fit, the nuclear disk was always the S\'ersic component with the smaller radius and smaller S\'ersic index.  For each galaxy (except Mrk 231, due to flux confusion from the central quasar), the continuum image was fit first.  Because emission line maps are noisier, certain disk parameters were kept fixed for the gas disks: coordinates of the center, the position angle, and the axis ratio of the disk.  In some cases, the center and position angle of the continuum were constrained to the kinematic center from stellar velocity maps (see \S\ref{kinematics}).  The kinematic center is midway between the positive and negative velocity peaks and usually lines up with the photometric center and isophote contours as expected.  In cases where dust or bright star clusters affect the photometric center, the velocity field is more reliable indicator of the center than the galfit modeling.  The adopted center is marked with a red X in all maps.

Six galaxies in our sample show extended \molhy~emission revealed by GALFIT modeling, usually along the minor axis of the disk.  Because \molhy~traces warm gas which may be shock-heated, these patches may be indicative of turbulent polar outflows.  They will be studied in detail in a future paper; for now they are simply fit with Gaussians and removed.  Evidence of these components can easily be noted by comparing the apparent position angle of \molhy~emission to that of the continuum and \brg~emission; in many cases the major axis of the \molhy~emission is offset by 90 degrees from that of the continuum or \brg~emission.  These components are also noted for individual galaxies in their respective figure captions.

\begin{figure}[ht]
\centering
\includegraphics[scale=0.8]{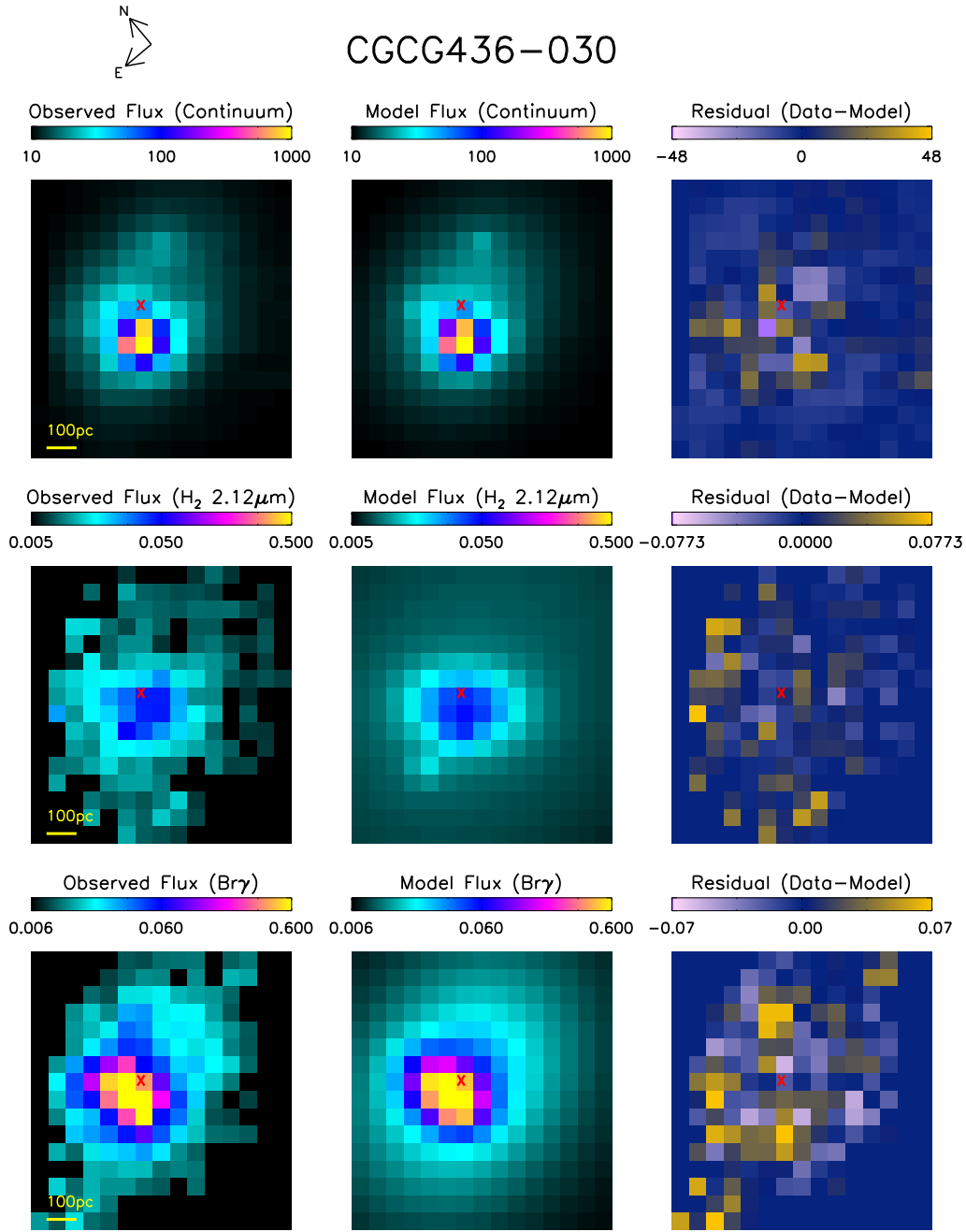}
\caption{Flux map (left panel), GALFIT model (center panel) and residual (right panel) for CGCG436-030.  The top row is the continuum flux, the middle row is \molhy~flux, and the bottom row is \brg~flux.  The flux maps (left and center panels) are shown on a log scale, while the residual map is shown on a linear scale, in units of counts per second.}
\label{galfit}
\end{figure}

In Table~\ref{tbl:galfitresults}, we include the GALFIT parameter results for each nucleus.  Note that the errors listed are the formal errors reported by GALFIT, and do not include the systematic errors associated with GALFIT fits (e.g. input parameters, number and selection of components fit).  Figure~\ref{galfit} shows the flux map, GALFIT model and residual for the first galaxy, CGCG436-030.  Analogous figures for the remainder of the sample are included in Appendix~\ref{extrafig}.  The flux profiles show an average S\'ersic index of $1.1\pm0.1$ for stars, $1.0\pm0.2$ for \molhy, and $0.8\pm0.1$ for atomic hydrogen line emission, consistent with $n=1$ exponential disk profiles.  The average effective radius (for a S\'ersic, $r_{eff}$ is defined such that half of the light is emitted interior to it) is $252\pm13$ pc for stars, $305\pm21$ pc for \molhy, and $307\pm34$ pc for atomic hydrogen line emission.  We exclude from these and subsequent averages the nuclei which are not confirmed by rotation (see next section) or higher resolution imaging \citep[as in][]{mrk273}.  We also exclude NGC~6090 and IRAS03359+1523, and the large disk of VV340a, which appear to be at such an early stage of merger that the disk measured is more likely a remnant disk from the progenitor rather than a nuclear disk formed during the merger (see \S\ref{formation}).

\onecolumn
\twocolumn

\begin{figure}[ht]
\centering
\includegraphics[scale=0.53]{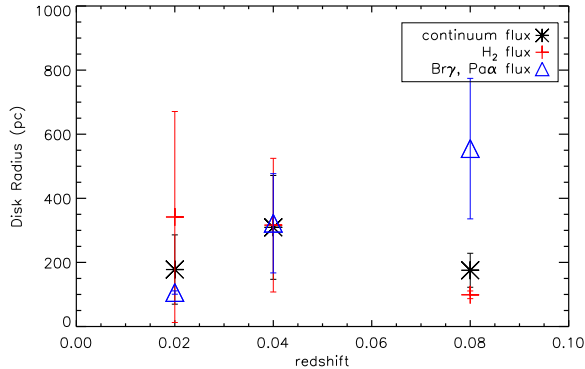}
\caption[Nuclear Disk Radius vs. Redshift]{Effective radii of nuclear disks, as measured from GALFIT parameters.  Galaxies are binned by 0.02 in redshift.  The lack of trend confirms that our disk size measurements are not merely a function of resolution.  Points represent the disk sizes measured from the continuum flux map (black asterisks), the \molhy~flux map (red crosses), and the \brg~or \paa~flux map (blue triangles).  The error bars show the standard deviation of the disk sizes within each bin.}
\label{disksize_z}
\end{figure}

We also include a plot of disk size for different redshift bins (Figure~\ref{disksize_z}) in order to determine if our modeling is limited by resolution; it does not appear to affect our results.

 \begin{deluxetable}{lcccccc}
    \centering
    \tabletypesize{\scriptsize}
    \tablewidth{0pt}
    \tablecolumns{6}
    \tablecaption{Disk Parameters}
    \tablehead{   
      \colhead{Galaxy Name} &
      \colhead{Tracer} &
      \colhead{Effective Radius} &
      \colhead{S\'ersic index} &
      \colhead{Axis Ratio\tablenotemark{a} } &
      \colhead{Position Angle} & \\
      \colhead{} & 
      \colhead{} & 
      \colhead{(pc)} &
      \colhead{n} &
      \colhead{(b/a)} & 
      \colhead{($^\circ$ E of N)} & 
      	}
    \startdata 
    CGCG436-030 & continuum & $293\pm6$ & $0.6\pm0.05$&$0.76\pm0.01$& -71 \\
       & \molhy & $775\pm200$&$0.42\pm0.17$&$0.76$&-71\\
       & \brg      & $450\pm30$ & $0.87\pm0.1$ & $0.76$&-71\\ 
    IRAS F01364-1042 & continuum & $211\pm21$ & $0.4\pm0.2$ & $0.49\pm0.03$&75 \\
       & \molhy & $173\pm20$&$1.2\pm0.5$&$0.49$&75 \\
       & \paa     & $182\pm15$ & $0.75\pm0.25$&$0.49$&75 \\
    IIIZw035 & continuum & $182\pm6$& $0.67\pm0.04$&$0.51\pm0.01$&35\\
       & \molhy & $132\pm5$&$0.45\pm0.04$&$0.51$& 35\\ 
       & \brg      & $103\pm5$&$0.5\pm0.1$&$0.51$& 35\\
    IRAS F03359+1523 & continuum & $1765\pm2$&$2.9\pm0.5$&$0.24\pm0.02$&72\\
       & \molhy & $725\pm25$&$0.4\pm0.1$&$0.24$&72 \\
       & \brg & $259\pm24$&$0.7\pm0.2$&$0.24$&72\\
    MCG+08-11-002 & continuum & $167\pm20$&$0.9\pm0.1$&$0.66\pm0.02$&65\\
       & \molhy & $720\pm85$&$0.4\pm0.1$&$0.4$&65\\ 
       & \brg      & $105\pm30$&$0.8\pm0.3$&$0.4$&65\\ 
    NGC~2623 & continuum & $142\pm3 $& $2.0\pm0.1$ & $0.81\pm0.01$ &-110 \\
       & \molhy & $172\pm18 $& $1.25\pm0.15$ & $0.81\pm0.01$ &-110 \\
       & \brg      & $112\pm7 $& $0.94\pm0.10$ & $0.81\pm0.01$ &-110 \\
    UGC5101 & continuum & $518\pm5$& $0.6\pm0.01$&$0.44\pm0.01$&79\\
       & \molhy & $505 \pm 16 $&$0.6\pm0.1$&$0.44$&79\\
       & \brg & $459 \pm 15$ &$0.3\pm0.1$&$0.44$&79\\
    Mrk231 
       & \molhy & $115\pm33$&$4.2\pm1.8$&$1.0\pm0.1$&100\\
    Mrk273 N\tablenotemark{b}& continuum & $562\pm3$ &$ 2.1\pm0.5$& $0.75\pm0.1$& 60 \\
       & \molhy & $205\pm3$ &$0.64\pm0.02$ & $0.75$& 60 \\
       & \brg      & $220\pm5$ & $0.43\pm0.03$& $0.75$& 60 \\
    Mrk273 SW\tablenotemark{b} & continuum & $55\pm2$&$1.6\pm0.5$&$0.44\pm0.03$&0\\
   VV340a & continuum & $238\pm23$ & $1.1\pm0.1$ & $0.33\pm0.02$& 180\\
       & \molhy & $211\pm20$ & $0.6\pm0.2$ & $0.33$ & 180\\
       & \brg & $1078\pm19$ & $0.1\pm0.1$ & $0.33$ & 180 \\
    IRAS F15250+3608 NW\tablenotemark{c} & continuum & $246\pm41$&$1.9\pm0.3$&$0.78\pm0.01$&136\\
       & \molhy & $220\pm10$&$2.0\pm0.9$&$0.78$&136\\ 
       & \paa     & $135\pm30$&$2.2\pm0.7$&$0.78$&136\\ 
    NGC~6090 & continuum & $780\pm50$&$1.5\pm0.1$&$0.74\pm0.01$&174\\
    NGC6240N & continuum & $350\pm140$&$1.9\pm0.4$&$0.61\pm0.02$&61\\
    NGC6240S & continuum & $50\pm1$&$0.4\pm0.1$&$0.50\pm0.01$&-15\\
    IRAS F17207-0014 E& continuum &$410\pm15$&$0.8\pm0.05$&$0.4\pm0.01$&-52\\
       & \molhy & $225\pm10$&$0.2\pm0.1$&$0.41$&-52\\
       & \brg & $485\pm50$&$1.1\pm0.2$&$0.041$&-52\\
    IRAS F17207-0014 W& continuum & $200\pm15$&$0.9\pm0.05$&$0.85\pm0.05$&40\\
       & \molhy & $330\pm30$&$1.2\pm0.2$&$0.85$&40\\
       & \brg & $96\pm7$&$0.75\pm0.24$&$0.85$&40\\
    IRAS 20351+2521 & continuum & $296\pm40$&$1.4\pm0.2$&$0.81\pm0.1$&15\\
       & \molhy & $295\pm130$&$0.85\pm0.4$&$0.81$&15\\
       & \brg & $360\pm155$&$1.1\pm0.4$&$0.81$&15\\
    IRAS F22491-1808 E & continuum & $213\pm137$&$1.9\pm1.6$&$0.9\pm0.1$&120\\
       & \molhy  & $100 \pm 10$&$1.0\pm0.9$&$0.85$&120\\
       & \paa & $710\pm350$&$1.2\pm0.07$&$0.85$&120\\
    IRAS F22491-1808 W & continuum & $138\pm10$&$1.0\pm0.2$&$0.6\pm0.1$&205 \\
       & \paa & $400\pm120$&$1.2\pm0.8$&$0.6$&205\\
    \enddata
    \tablecomments{(a) For galaxies with continuum and line emission maps, the axis ratio was fit using the continuum, then fixed to this best-fit value for additional line fits. (b) Originally presented in \citet{mrk273}. (c) Note that kinematic analysis of this galaxy does not show evidence of a disk (see \S\ref{kinematics}), but GALFIT parameters were included here for completeness.  This does not mean that the GALFIT fits are incorrect; there is a structure in the flux maps with these morphological parameters.  However, having a disky morphology does not confirm the presence of a disk; for confirmation we require kinematic evidence of rotation.}
    \label{tbl:galfitresults}
  \end{deluxetable}

Many of the local (U)LIRGs have been observed with the SubMillimeter Array (SMA) to have rotating gas disks \citep[radii 0.3-3.1 kpc;][]{Wilson08,Iono09}, but given the discrepancy in spatial resolution between their data and ours, the submillimeter measurements are likely upper limits on the sizes of the structures we measure.  These authors have not modeled inclinations, but a qualitative comparison of the position angles of their disks match ours in all overlapping cases except NGC~6240.  However, we only have stellar kinematics for NGC~6240, and it has been documented \citep{Hauke} that the stellar and gas kinematics for this merger are not broadly consistent.
Higher resolution data taken with the SMA's very extended configuration 
(\citealp[e.g. Arp 220 in][]{Sakamoto08}; \citealp[NGC~6240 in][]{U11ASPC})
find resolved disky structure of size $\sim200$ parsecs in the spatial distribution of CO(3-2), comparable to the disk sizes presented here.  Additional high-resolution data taken with ALMA \citep[e.g.][]{Imanishi13} for a large sample of galaxy mergers will be useful for tracing the evolution of these gaseous nuclear disks at various merger stages.

Several of the galaxies in our sample have been observed by the Jansky Very Large Array at the $Ka$- and $C$-bands (Barcos et al., in preparation).  These preliminary results show radio disks that are co-located and consistent in size with the near-infrared nuclear disks reported here, with radii of a few hundred parsecs.  A more detailed comparison between the structure of the radio and infrared disks will be presented in Barcos et al. (in preparation). 

Similar disks have also been seen in a handful of isolated Seyfert galaxies.  In particular, \citet{Hicks13} looked at 5 Seyfert and 5 quiescent field galaxies matched in many parameters (e.g. mass, inclination, Hubble type, angular size) and found rotational structure out to $\sim250$ parsecs present in all 5 Seyferts and none of the quiescent galaxies; this suggests that disks may be an important fueling mechanism for the central black holes.  However, the mechanism through which those disks are formed is not well-constrained; because the Seyfert galaxies have not undergone recent major mergers, the nuclear disks were likely formed through some other means like a minor merger, direct gas accretion, or bar instabilities.  

\subsection{Disk Kinematics}
\label{kinematics}

We find that in 19 of 21 nuclei (from 17 merger systems, 4 of which show a second nucleus in our field of view confirmed by either rotation or an AGN), there is strong rotation seen in stars, molecular hydrogen and/or atomic hydrogen, evidenced by steep velocity gradients across the major axis.  Velocity maps and a one-dimensional velocity profile across the major axis are shown for CGCG436-030 in Figure~\ref{vels}, with additional galaxies presented in Appendix~\ref{extrafig}.  

\onecolumn
\begin{figure}[ht]
\centering
\subfloat{\includegraphics[scale=1.5]{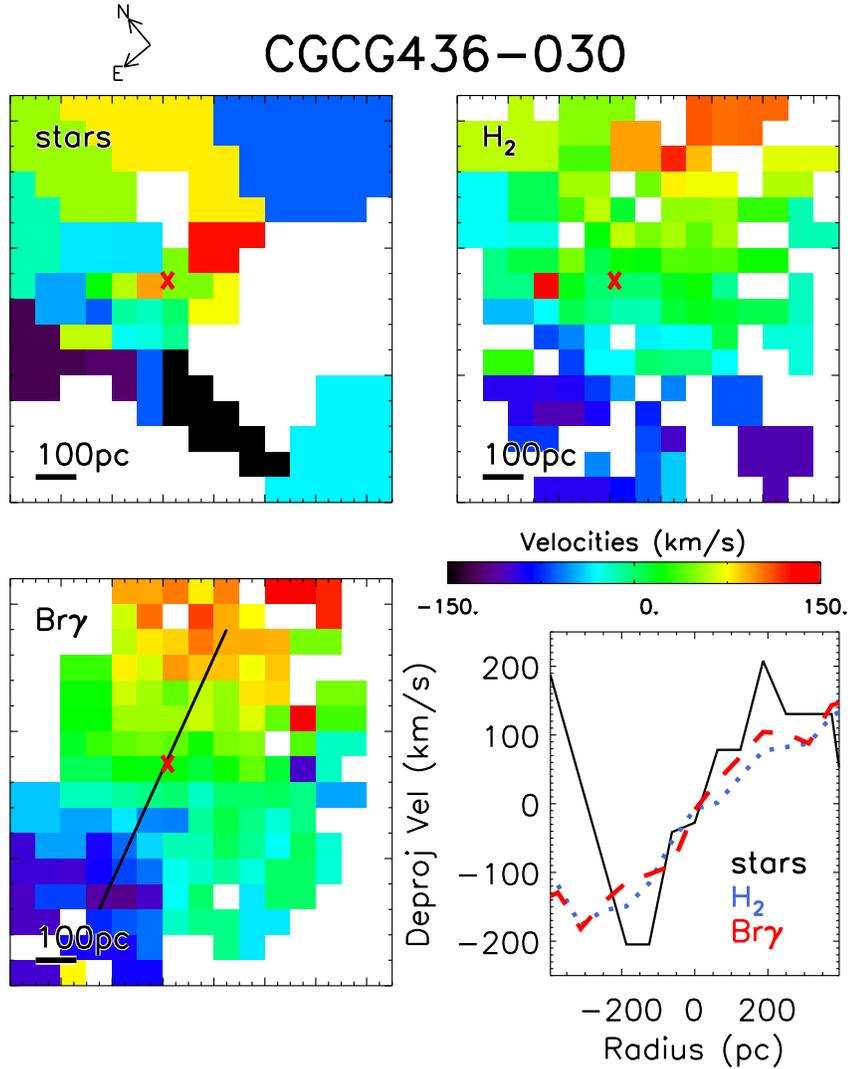}}
\caption{Observed velocity maps of stars (top left), molecular gas in \molhy~(top right), and ionized gas in \brg~(bottom left) for CGCG436-030.  All velocity maps use the same color bar; white pixels do not have sufficient signal to measure velocities accurately.  Bottom right panel shows the deprojected velocity profile cut through the major axis for each of the 3 tracers: stars (solid black), \molhy~(dotted blue),  \brg~(dashed red).  The major axis cut is indicated in black on the bottom left map for clarity.  Though the stellar kinematics are noisy, they appear to be consistent with the motions of the gas.}
\label{vels}
\end{figure}
\twocolumn

An interesting parameter of the disk to measure is $v/\sigma$, the ratio of peak velocity to velocity dispersion, which indicates how dynamically cold the orbits are.  For a thin disk, $v/\sigma >> 1$.  In turbulent or thick disks, this ratio can be closer to 1.  However, measuring the intrinsic $\sigma$ is difficult, because the measured width of a line includes peculiar velocity motions \citep{Kang13}.  In an unresolved disk spectrum, the width of the line is a light-weighted combination of velocity and dispersion at each point: $v_{rms} = \sqrt{v^2 + \sigma^2}$.  As one resolves the disk better, this width will shrink as the velocity motions are split into separate pixels.  Though our disks are resolved, our velocity dispersion measurements are still affected by our resolution on a low level.  Following a similar technique to that used by \citet{Green13}, we estimate the velocity shear contribution in each spaxel by the difference in measured velocity of its neighbor along the steepest gradient.  The velocity shear correction is approximately 5\% for most of our systems, though a few show steep velocity gradients that require a correction up to 20\%.  The corrected velocity dispersion $\sigma_{corr}$ is:
\begin{displaymath}
\sigma_{corr}^2 = (\sigma_{meas} *(1-\frac{v_{shear}}{\sigma_{meas}}))^2 - \sigma_{instr}^2
\end{displaymath}
where $\sigma_{meas}$ is the mean velocity dispersion measured within one effective radius, $v_{shear}$ is the mean velocity jump between spaxels within one effective radius, and $\sigma_{instr}$ is the resolution limit of OSIRIS (42 km s$^{-1}$ for 100 mas pixel$^{-1}$ data, 33.5 km s$^{-1}$ for 35 mas pixel$^{-1}$ data).  This $\sigma_{corr}$ is then reported in Table~\ref{tbl:vsigmatable}, representing an average value of velocity dispersion over the disk.  

The average $v/\sigma$ of these disks is $1.8\pm0.2$ for stars, $2.0\pm0.1$ for \molhy, and $2.7\pm0.1$ for atomic hydrogen emission though each tracer shows substantial variability in the range 0.8-4, with two \brg~disks falling between 4-6.  

Using the measured $v/\sigma$ and $R_{eff}$ values, we are able to estimate the scale heights $h$ for these disks.  The scale heights are calculated using the formula $\frac{h}{R_{eff}} = (\frac{\sigma}{v})^2 $ \citep{BinneyMerrifield}.  These are also included in Table~\ref{tbl:vsigmatable}.  Disk scale heights range from very thin ($\sim7$ pc) to quite puffy ($\sim 400$ pc).  

 \begin{deluxetable}{lcccccc}
    \centering
    \tabletypesize{\scriptsize}
    \tablewidth{0pt}
    \tablecolumns{6}
    \tablecaption{Disk Kinematics}
    \tablehead{   
      \colhead{Galaxy Name} &
      \colhead{Tracer} &
      \colhead{Peak Velocity}&
      \colhead{Velocity Dispersion\tablenotemark{a}} &
      \colhead{v/$\sigma$} & 
      \colhead{Scale Height} & \\
      \colhead{} &
      \colhead{} &
      \colhead{(Deprojected; km/s)}&
      \colhead{($\sigma_{corr}$; km/s)} &
      \colhead{} &
      \colhead{(pc)} &
      	}
    \startdata
    CGCG436-030 & stars & $165 \pm 25$ & $126 \pm 29$ & $ 1.3\pm0.4$ & $169\pm61$\\
       & \molhy & $165 \pm 9$ & $81\pm3$ & $2.0 \pm 0.1$&$188\pm51$ \\
       & \brg & $ 189 \pm 12 $ & $92 \pm 5$ & $ 2.1\pm0.2$ & $106\pm20$\\
    IRAS F01364-1042 & \molhy & $111\pm20$&$111\pm 3$&$1.0\pm0.2$&$173\pm38$\\
       & \paa & $123\pm10$&$137\pm2$&$0.9\pm0.1$ & $224\pm25$\\
    IIIZw035 & stars &$241\pm25$&$110\pm3$&$2.2\pm0.2$ & $40 \pm 9$ \\
       & \molhy & $127\pm7$&$112\pm2$&$1.1\pm0.1$&$103\pm7$\\
       & \brg & $190\pm15$&$94\pm6$&$2.0\pm0.2$ & $25\pm5$\\
    IRAS F03359+1523 & \molhy & $152\pm35$&$91\pm4$&$1.7\pm0.4$ & $258\pm101$\\
       & \brg & $161\pm20$&$106\pm8$&$1.5\pm0.2$ & $112\pm27$\\
    MCG+08-11-002 & stars & $236\pm20$&$72\pm3$&$3.3\pm0.3$ & $ 16\pm5$ \\
       & \molhy & $244\pm10$&$78\pm1$&$3.1\pm0.1$&$73\pm13$\\
       & \brg & $280\pm10$&$75\pm6$&$3.7\pm0.2$ & $7.4\pm2.5$\\
    NGC~2623 & stars & $188\pm35$ & $121\pm2 $& $1.6\pm0.3$ & $59\pm17$\\
       & \molhy & $195 \pm 14$ & $102 \pm 1$ & $1.9 \pm 0.1$ & $47\pm8$ \\
       & \brg & $164 \pm 11$ & $106 \pm 1$ & $1.6 \pm 0.1$ & $46\pm6$\\
    UGC5101 & stars & $272\pm25$&$155\pm7$&$1.75\pm0.2$ & $169\pm30$\\
       & \molhy & $367\pm15$&$100\pm2$&$3.7\pm0.2$ & $38\pm7$\\
       & \brg & $370\pm20$&$67\pm7$&$5.6\pm0.7$ & $15\pm10$\\
   Mrk231  & \molhy & $229\pm10$&$82\pm16$&$2.8\pm0.6$ & $15\pm9$\\
    Mrk273 N& \molhy & $247\pm5$ & $147\pm1$ & $1.7\pm0.1$ & $73\pm3$ \\
       & \brg & $306\pm6$&$84\pm2$&$3.7\pm0.1$&$17\pm2$\\
    Mrk273 SW& \molhy & ...\tablenotemark{b}&...&...& ...\\
       & \brg & ...&...&...&...\\
    VV340a & stars & $76\pm20$ & $92\pm41$ & $0.8\pm0.4$ & $349\pm154$\\
       & \molhy & $242\pm50$ & $158\pm28$ & $1.5\pm0.4$ & $90\pm39$ \\
       & \brg & $246\pm65$ & $46\pm11$& $5.3\pm1.9$ & $38\pm72$\\
    IRAS F15250+3608 NW & \molhy &...&...&... &...\\
       & \brg & ...&...&...&...\\
    NGC6090 & stars & $163\pm40$ & $163\pm11$ & $1.0\pm0.3$&$783\pm205$\\
    NGC6240N & stars & $230\pm20$&$210\pm4$&$1.1\pm0.1$&$296\pm119$\\
    NGC6240S & stars & $345\pm20$&$237\pm2$&$1.5\pm0.1$&$24\pm2$\\
    IRAS F17207-0014 E & \molhy & $282\pm15$&$104\pm21$&$2.7\pm0.6$ & $31\pm18$\\
       & \brg & $309\pm7$&$70\pm3$&$4.4\pm0.2$ & $25\pm6$\\
    IRAS F17207-0014 W & \molhy & $232\pm15$&$125\pm7$&$1.9\pm0.2$ & $97\pm17$\\
       & \brg & $223\pm15$&$93\pm4$&$2.4\pm0.2$ & $17\pm4$\\
    IRAS 20351+2521 & stars & $214\pm60$ & $75\pm21$ & $2.9\pm 1.1$ &$36\pm42$\\
       & \molhy & $133\pm20$&$133\pm17$&$1.0\pm0.2$ & $297\pm145$ \\
       & \brg & $182\pm15$&$108\pm18$&$1.7\pm0.3$ & $127\pm68$\\
    IRAS F22491-1808 E & \molhy &$211\pm20$& $131\pm6$&$1.6\pm0.2$ & $38\pm8$ \\
       & \pab & $325\pm25$&$97\pm3$&$3.3\pm0.3$ & $64\pm36$\\
    IRAS F22491-1808 W 
       & \pab & $64\pm20$&$64\pm4$&$1.0\pm0.3$& $400\pm175$\\
    \enddata
    \tablenotetext{a}{Velocity dispersion and $v/\sigma$ are reported here after correcting for the contributions of unresolved velocity shear and the instrumental profile; see text for details.}
    \tablenotetext{b}{Some nuclei show disky morphology but do not exhibit rotation in the velocity maps; or show rotation in some lines but not others.  These cases are indicated by `...'.}

    \label{tbl:vsigmatable}
  \end{deluxetable}


\section{Discussion}
\label{discussion}

\subsection{Nuclear Disks Commonly Occur in Gas-Rich Mergers}

Nuclear disks are common in our sample of late-stage gas-rich mergers.  As shown in \S\ref{results}, gas disks are observed kinematically in 
16 of 19 nuclei, and stellar disks in 11 out of 11.  Additionally, the SW nucleus of Mrk273 has a small stellar disk unresolved by our observations but revealed in the imaging of \citet{mrk273}; this brings our count of stellar disks to 12 of 12.  It may indeed have a gas disk on similar scales that we do not detect, suggesting that our fractions of disks are in fact lower limits to the true fraction.
These effective radii of these disks have a mean of $344\pm9$ pc and a mode of $211\pm20$ pc.  They are generally smaller than galactic-scale disks, with only two having effective radii larger than one kiloparsec.  

\subsubsection{How and When did These Disks Form?}
\label{formation}

\emph{How did the stellar disks form?}  A stellar disk begins as a ``cold" structure, with low velocity dispersion, because dynamical interactions over time can only heat a dynamical system, increasing its velocity dispersion \citep[see Chapter 7.5 of][]{BinneyTremaine}.  Therefore a stellar disk likely forms from a disk of gas, and may or may not slowly evolve dynamically into a thicker disk.  The stellar disks presented here are likely to have formed in one of two ways: 1) they may have formed from gas disks in the progenitor galaxies long ago, and have perhaps been partially-stripped during the current merger, or 2) they may have formed recently, during the merger, from the gas disks that are also currently observed.  Two galaxies in our sample, IRAS F03359+1523 and NGC~6090, have disks that are quite large, with continuum radii of 1765 and 780 parsecs, respectively.  These two galaxies also appear to be at relatively earlier stages of merging \citep[3 and 4 in the 0-6 scale of][respectively]{Haan11}; we suggest that their disks are more likely to be remnants of galactic-scale disks.  A third early-stage merging galaxy, VV340a \citep[merger class 1][]{Haan11}, appears to have both a large-scale disk of continuum radius 1782 parsecs and a small-scale disk of continuum radius 238 parsecs.  The remainder of our sample contains stellar disks which are small (of radius only a few hundred parsecs), and similar in both size and kinematics to the currently-observed gas disks.  This suggests that the majority of stellar disks were formed from these gas disks.

\emph{When did the stellar disks form?}  To test the hypothesis that these are young stars forming \textit{in situ} during the course of the merger, we measure the equivalent width $W_{CO}$ of the 2.292 $\mu$m absorption feature for the 10 nuclei whose observations cover that waveband.  Because these features are deepest in giants and supergiants, which evolve more quickly than their lower-mass siblings, $W_{CO}$ can be correlated with the age of a stellar population.  We compare our measurements to a fiducial model of an instantaneous starburst of metallicity $Z=0.02$ and $\alpha=2.35$ from Figure 101b of \cite{Leitherer99} to determine the corresponding starburst age.  It is important to note that an AGN or hot dust in the nuclear regions can produce nonstellar continuum that will dilute the $W_{CO}$ measurements.  As such, our $W_{CO}$ measurements are lower limits to the actual $W_{CO}$ of the stars in the nucleus.  This then corresponds only to an upper limit to the age of the stars.  Our measurements of $W_{CO}$ and the corresponding ages are listed in Table~\ref{tbl:coeqwidth}.  We find that our CO bandheads are generally deep, but two galaxies have markedly smaller $W_{CO}$ measurements: UGC5101 is known to have a central AGN and likely has continuum flux filling in the absorption band, artificially decreasing the $W_{CO}$ measurement; NGC~6090's older stellar populations suggest that it may indeed be a remnant galactic-scale disk.  Excluding UGC5101 and NGC~6090, our nuclear disks show equivalent widths ranging from 12.52 - 18.65 \AA~(corresponding to stellar ages of $<10-30$ Myr), with a mean of 15.7 \AA~and a median value of 16.2 \AA~(both indicating stellar ages $\sim 13$ Myr).  As galaxy mergers occur on much longer timescales \citep[$10^{8}-10^{9}$ years; e.g.][]{BarnesHernquist92,BarnesHernquist96}, these stars were formed comparatively recently in the disks, probably during the most recent peripassage.  Stars in a galactic-scale disk remnant would likely be considerably older.  

 \begin{deluxetable}{lccc}
    \centering
    \tabletypesize{\scriptsize}
    \tablewidth{0pt}
    \tablecolumns{3}
    \tablecaption{CO 2.292 $\mu$m Equivalent Widths}
    \tablehead{   
      \colhead{Galaxy Name} &
      \colhead{$W_{CO}$ (\AA)\tablenotemark{a}} &
      \colhead{Starburst Age (yr)\tablenotemark{a,b}} &
      	}
    \startdata
    CGCG436-030 & 18.65 & $<1\times10^{7}$ \\
    IIIZw035 & 15.30 & $1.5\times10^{7}$\\
    MCG+08-11-002 & 18.47 & $<1\times10^{7}$ \\
    NGC~2623 & 12.52 & $3\times10^{7}$\\
    UGC5101 & $>5.75$\tablenotemark{c} & $<6\times10^{8}$ \\
    UGC8387 & 18.39 & $<1\times10^{7}$ \\
    VV340a & 16.23 & $1\times10^{7}$\\
    NGC~6090 & 7.54 & $2\times10^{8}$\\
    NGC~6240N & 13.53 & $<2.5\times10^{7}$\\
    NGC~6240S & 12.80 & $<3\times10^{7}$\\
    \enddata
    \tablenotetext{a}{The presence of nonstellar continuum may confound some or all of these measurements; for that reason, these should be considered a lower limit to the $W_{CO}$ values and an upper limit to the stellar population ages.}
    \tablenotetext{b}{Starburst ages were estimated using Figure 101b from \cite{Leitherer99}.}
    \tablenotetext{c}{To minimize contribution from the nonstellar continuum, the central 3x3 pixel region was excluded for this calculation.}
    \label{tbl:coeqwidth}
  \end{deluxetable}
  
\begin{figure}[ht]
\centering
\includegraphics[scale=0.53]{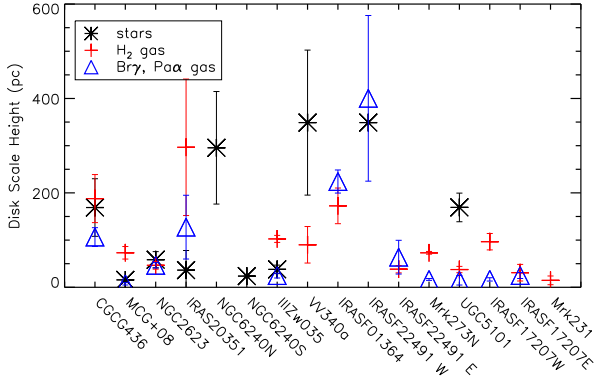}
\caption{Scale heights of nuclear disks, sorted by galaxy infrared luminosity.  Merging systems with two separate galaxy components have been split in $L_{IR}$ according to the components' flux ratio from MIPS 70 or 24 $\mu$m.  Systems with only one component in which we resolve two disks have been split in half evenly, to attribute half the infrared luminosity to each disk.  Points represent the scale heights measured from the stars (black asterisks), the \molhy~emission (red crosses), and the \brg~or \paa~emission (blue triangles).  If stars are recently formed from the gas disks, each galaxy ought to show similar scale heights for gas and stars.  The scale heights of atomic hydrogen disks are consistent with those of stellar disks in all cases except UGC5101.  \molhy~disks may differ because the \molhy~is shocked and may partially trace outflows rather than only disks.}
\label{scaleheights}
\end{figure}

If the stellar nuclear disks did indeed form from these gas disks, we would expect them to have similar disk scale heights.  That is, a consistent scale height suggests that the gas and stars are spatially mixed, which would likely be the case if the stars are recently or currently forming from the gas disks.  To investigate this, we plot the scale heights for each galaxy for each tracer in Figure~\ref{scaleheights}.  The galaxies are sorted by infrared luminosity for convenience.  We find that the scale heights of atomic hydrogen disks are consistent with those of stellar disks in all cases except UGC5101, lending support to the hypothesis that the stars are recently formed from the gas disks.  \molhy~disks often differ, showing larger or smaller scale heights.  This may be because the \molhy~is shocked and may partially trace outflows rather than only disks.

\emph{How and when did the gas disks form?} The gas disks observed here by OSIRIS may have formed in the progenitor galaxies and survived the merger (perhaps evolving dynamically) or may be new dynamical structures formed during the current interaction.  
We find gas disks both around individual nuclei before coalescence (e.g. IR17207-0014) and in the final coalesced nuclei (e.g. NGC~2623).  Though early stages of mergers may show larger scale disks, the disks observed at late stages of mergers do not appear to vary in structure before versus after coalescence.
High resolution hydrodynamic simulations of galaxy mergers (discussed more in \S\ref{sims}) show small gas disks being disrupted during coalescence and reforming on timescales of $\sim10^7$ years.  While our data are consistent with this scenario, we have no galaxies in our sample during the coalescence phase; to determine whether or not disks are disrupted and for how long would require a larger sample of merging galaxies with a more precise measurement of merger stage, or a method of estimating the age of a gas disk.

\subsubsection{Do Nuclear Disk Properties Correlate with $L_{IR}$?}

Though the disks we have measured in our OSIRIS data are traced by stars, warm molecular gas and ionized gas, there is no consistent trend in sizes between different disk tracers.  For an individual galaxy, differences in disk sizes measured by \molhy~versus atomic hydrogen can either be indicative of different conditions allowing gas to cool and form \molhy~or of different ranges of ionizing radiation versus shock-excitation.  
Because stars form from gas, a stellar disk larger than a gas disk indicates that the gas has been depleted, evaporated, or blown out.  A smaller stellar disk may indicate either that new gas is continually being added to the outside of the disk or that conditions for star formation are only met at inner radii. 

\begin{figure}[ht]
\centering
\includegraphics[scale=0.53]{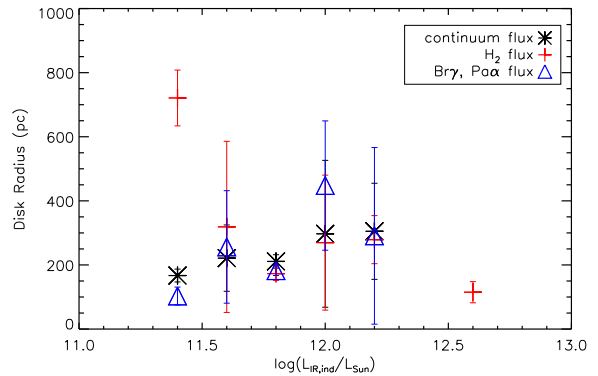}
\caption[Nuclear Disk Radius vs. Infrared Luminosity]{Radii of nuclear disks, as measured from GALFIT parameters.  Systems are binned by 0.2 dex in the infrared luminosity of the individual galaxy component.  Merging systems with two separate galaxy components have been split in $L_{IR}$ according to the components' flux ratio from MIPS 70 or 24 $\mu$m.  Systems with only one component in which we resolve two disks have been split in half evenly, to attribute half the infrared luminosity to each disk.  Points represent the disk sizes measured from the continuum flux map (black asterisks), the \molhy~flux map (red crosses), and the \brg~or \paa~flux map (blue triangles).  The error bars show the standard deviation of the disk sizes within each bin.}
\label{disksize_binnedLIR}
\end{figure}

Infrared luminosity has been shown to correlate with molecular gas content, AGN fraction, merger fraction, and merger stage \citep{Sanders88,Melnick90,Veilleux95, Clements96,Veilleux02,IshidaPhD,Ellison13, Koss13}.  It may be expected therefore that nuclear disk size might correlate as well.  However, the sizes of nuclear gas disks do not appear to be correlated with infrared luminosity, though stellar disks appear to grow with increasing $L_{IR}$ (see Figure~\ref{disksize_binnedLIR}).  These results do not change when plotting against the 13- or 24-$\mu$m luminosities from Spitzer IRS spectra (T. D\'iaz-Santos, private communication; \citealp[][]{Stierwalt13}).
It is not clear why the size of gas disks vary so much from stellar disks, but an increase of stellar disk size logically follows from the idea that a nuclear disk grows over the course of a merger.  However, since the correlation of merger stage and $L_{IR}$ is statistical, a larger sample or a precise measurement of merger stage is needed to confirm the underlying cause of the trend.  The current sample contains a substantial amount of variation, as demonstrated by the large standard deviations present in most bins.  Disk sizes are, however, not correlated with infrared color \citep[either $f_{25\mu m}/f_{60\mu m}$ or $f_{60\mu m}/f_{100\mu m}$ from][]{Sanders03}.  We found no trend with the morphological merger stage classifications of \citet{Haan11} and \citet{Kim13}, but this also may be due to the small number of systems measured or misclassifications from morphological degeneracies.  Additionally, if gas disks are disrupted by AGN or stellar feedback during the course of the merger, their properties may vary on timescales smaller than those we can probe, effectively washing out any trends. 

\begin{figure}[ht]
\centering
\includegraphics[scale=0.53]{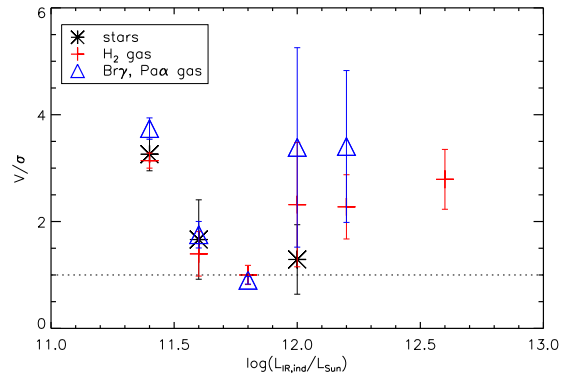}
\caption[$v/\sigma$ vs. Infrared Luminosity]{$v/\sigma$ for nuclear disks, plotted in bins of 0.2 dex in infrared luminosity of the host galaxy.  Merging systems with two separate galaxy components have been split in $L_{IR}$ according to the components' flux ratio from MIPS 70 or 24 $\mu$m.  Systems with only one component in which we resolve two disks have been split in half evenly, to attribute half the infrared luminosity to each disk.  Points represent the kinematics measured from the stars (black asterisks), the \molhy~emission (red crosses), and the \brg~or \paa~emission (blue triangles).  The error bars indicate the standard deviation of the $v/\sigma$ for each bin.  A horizontal dotted line has been drawn for $v/\sigma=1$ to show the value at which pressure support and rotational support are equal.}
\label{voversigma_binnedLIR}
\end{figure}

Because $L_{IR}$ is correlated with merger stage and gas fraction \citep[e.g.][]{IshidaPhD}, we might expect it also to correlate with $v/\sigma$ of the disks; more gas could mean diskier kinematics, and merger stage could indicate likely disruptions to the disk kinematics.  However, the $v/\sigma$ in our sample does not appear to correlate with the infrared luminosity of the host galaxy (see Figure~\ref{voversigma_binnedLIR}).  This could be because the $L_{IR}$ correlations are statistical in nature and our sample shows a lot of variability; in that case, comparing $v/\sigma$ to a precise measure of merger stage would reveal the underlying trends.  This quantity also does not correlate with the luminosities measured by Spitzer IRS in the 13- and 24-$\mu$m filters (T. D\'iaz-Santos, private communication; \citealp[][]{Stierwalt13}) or with the infrared color \citep[either $f_{25\mu m}/f_{60\mu m}$ or $f_{60\mu m}/f_{100\mu m}$ from][]{Sanders03}.
As with disk size, no trend was evident between $v/\sigma$ and the morphological merger stage classifications of \citet{Haan11} and \citet{Kim13}; this may be due to variations in our sample or to the difficulties of classifying mergers via morphology.  It could also be that disruptions to the disks' kinematics happen on rapid timescales which therefore wash out large-scale trends.  It is likely, however, that such disruptions would have a more profound affect on the gas kinematics than the stellar kinematics; future studies may be able to quantify this effect.

\subsubsection{Masses of the Disks}
\label{masssection}

To estimate the molecular gas mass in these disks, we use the average molecular gas surface mass density for (U)LIRGs empirically determined by \cite{BryantScoville99}, $<\Sigma_{g}> = 1.5\pm0.3 \times10^{4} M_{\sun} pc^{-2}$ (recalculated to include only (U)LIRGs), and multiply it by the deprojected area of the each disk.  Though the surface density was calculated for cold gas using interferometric observations, we consider this a lower limit to the molecular gas mass because it may cover a larger area than is traced by the \molhy~line. These derived molecular gas masses are listed in Table~\ref{tbl:diskmasses} and range from $9.2\times10^{8} - 5.8\times10^{10} M_{\sun}$.  It should be noted that because these numbers are calculated using an average surface mass density, they should be regarded only in a statistical sense: variations in the true density from the average will lead to an over- or under-estimate of the disk mass with this method.  \citet{Solomon92} measure high-density \molhy~gas in seven local (U)LIRGs using single-dish millimeter-wave measurements of the HCN(1-0) line.  They find dense \molhy~structures of masses $2\times10^9 - 3\times10^{10} M_{\sun}$ in the cores of these galaxies.  This is consistent with the range of \molhy~disk masses we find.  

We also make use of high spatial resolution kinematic modeling and report resulting dynamical mass profiles in Table~\ref{tbl:diskmasses}.  This modeling will be presented fully in a future paper studying black hole masses for a subset of these galaxies where data quality allows.  Briefly, the kinematics were modeled as $M(r) = \rho_0 r^{\gamma} + M_{BH}$.  Generally, the inclination and $\gamma$ were constrained using the light profile, assuming that the disk mass profile is smooth and follows the $K$-band light profile form; the remaining parameters, the disk mass density $\rho_0$ and the black hole mass $M_{BH}$, were fit from the kinematics.  These disk mass profiles encompass not just the molecular gas mass but all components of the disk.  The dynamical estimate is a 1-2 times higher than the molecular gas mass in two out of the three galaxies with both estimates, consistent with the presence of stars in the disk.  In the other case, the smaller dynamical mass may indicate that the systems' molecular gas surface densities are lower than average.

 \begin{deluxetable}{lccccc}
    \centering
    \tabletypesize{\scriptsize}
    \tablewidth{0pt}
    \tablecolumns{4}
    \tablecaption{Nuclear Disk Mass Estimates}
    \tablehead{   
      \colhead{Galaxy Name} &
      \colhead{Tracer} &
      \colhead{$M_{H_{2},disk}$\tablenotemark{a}} & 
      \colhead{$M_{dyn,disk}(r)$\tablenotemark{b}} & 
      \colhead{$M_{dyn,disk}$\tablenotemark{b}} \\
[0.5ex]      \colhead{} & 
      \colhead{} & 
      \colhead{($10^{9} M_{\sun}$)} & 
      \colhead{($10^{9} M_{\sun}$)} & 
      \colhead{($10^{9} M_{\sun}$)} & 
      	}
    \startdata 
    CGCG436-030 & \molhy & $58\pm30$ & & \\
    \hline
    IRAS F01364-1042 & \molhy & $2.8 \pm 0.9 $ & &\\
    \hline
    IIIZw035 & stars\tablenotemark{c} &  & $0.2~(\frac{r}{100 pc})^2 $&$1.5 \pm 0.3$ \\
       & \molhy & $1.6 \pm0.4 $ & $0.0001~(\frac{r}{100 pc})^2 $& $0.0003\pm0.0001$ \\ 
       & \brg      & & $ 0.19~(\frac{r}{100 pc})^2$ & $0.4 \pm 0.05$ \\
     \hline
    MCG+08-11-002 & \molhy &  $49 \pm 16$ & & \\
    \hline
    NGC~2623  & \molhy & $2.8 \pm 0.8 $ & &  \\
     \hline
    UGC5101 & stars &  & $0.95~(\frac{r}{100 pc})^2$ & $51 \pm 2$ \\
       & \molhy &  $24\pm5$\\
     \hline
    Mrk231 
       & \molhy & $1.2\pm0.8$ & & \\
    \hline
    Mrk273 N  & \molhy &$4.0\pm0.9$ & & \\
       & \feii      & & $1.1~(\frac{r}{100 pc}) $ & $6.6 \pm$0.2$$ \\
    \hline
    VV340a & \molhy & $4.2\pm1.2$& & \\
    \hline
    IRAS F15250+3608 NW\tablenotemark{d} & \molhy & $4.5\pm1.0$ & & \\
    \hline
    NGC~6240N & stars & & $0.1~(\frac{r}{100 pc})^{2.2} $ & $3.2 \pm2.8$ \\ 
    \hline
    NGC~6240S & stars & & $0.29~(\frac{r}{100 pc})^{1.5} $ & $ 0.2 \pm0.01$ \\ 
    \hline
    IRAS F17207-0014 E& \molhy & $4.8\pm1.1$ & &  \\
    \hline
    IRAS F17207-0014 W& \molhy & $10\pm3$ & & \\
    \hline
    IRAS 20351+2521 & \molhy &$8.3\pm7.6$& & \\ 
    \hline
    IRAS F22491-1808 E & \molhy  & $0.92\pm0.29$ & & \\
    \enddata
    \tablenotetext{a}{$M_{H_{2},disk} = 2\pi R_{eff}^2 <\Sigma_{g}>$, with $<\Sigma_{g}> = 1.5\pm0.3 \times10^{4} M_{\sun} pc^{-2}$ from \cite{BryantScoville99}.  Though we use the cold gas surface density, this is still likely a lower limit to the cold gas mass because we use the effective radius calculated from warm H$_{2}$, which may not trace the entire cold H$_{2}$ disk.}
   \tablenotetext{b}{Dynamical disk masses were calculated using kinematic fits that will be fully presented in a later paper for galaxies with high spatial resolution data, and are presented here for comparison.  These masses do not include black hole or other central masses.  In column 5, $M_{dyn,disk} = 2M(r_{eff})$ using the equation in column 4 and the appropriate $r_{eff}$ from GALFIT results.}
   \tablenotetext{c}{Tracers marked `stars' use the continuum radius for estimates of $M_{H_{2},disk}$ in column 3 and stellar kinematics for the $M_{dyn,disk}$ estimates in columns 4 and 5.}
    \tablenotetext{d}{Note that kinematic analysis of this galaxy does not show evidence of a disk (see \S\ref{kinematics}), but GALFIT parameters were included here for completeness.}
    \label{tbl:diskmasses}
  \end{deluxetable}

\subsection{Comparing Observed Disks to Simulations}
\label{sims}

Here we compare our observational results to two sets of hydrodynamic simulations which find nuclear disks.  The first set is published \citep{Mayer07}, but contains a limited treatment of star formation; to determine what effect a multiphase interstellar medium (ISM) has on disks, we also look at preliminary results from new simulations, to be published in Ro\v{s}kar et al. (in prep).

Previous work on modeling formation of nuclear disks in major merger
remnants employed simulations with a fixed equation of state and
largely isothermal gas due to the use of a $10^4$~K floor in the
cooling function \citep{Mayer07}. The runs were resampled by a
particle-splitting technique in order to increase the effective
spatial resolution to 2~parsecs in the central region and follow the decay
of the SMBH binary to parsec scales. Using this method, \citet{Mayer07}
 showed that dense nuclear disks that form from the rapid
accretion of gas into the central region after the completion of a
major merger can cause rapid decay of the binary on Myr
timescales. More recent simulations (Ro\v{s}kar et al. in prep) relax
the assumption of a single-phase ISM by implementing a cooling
function appropriate for the conditions expected in a merger remnant
\citep{Spaans00}. In addition, the simulations include the effects
of star formation and supernova feedback, which drastically affect the
ISM conditions following the starburst associated with the merger. The
merger is otherwise very similar to that used by \cite{Mayer07} and
a similar particle-splitting technique is used resulting in a final
resolution of 1~parsec.

\begin{figure}[ht]
\includegraphics[scale=.75]{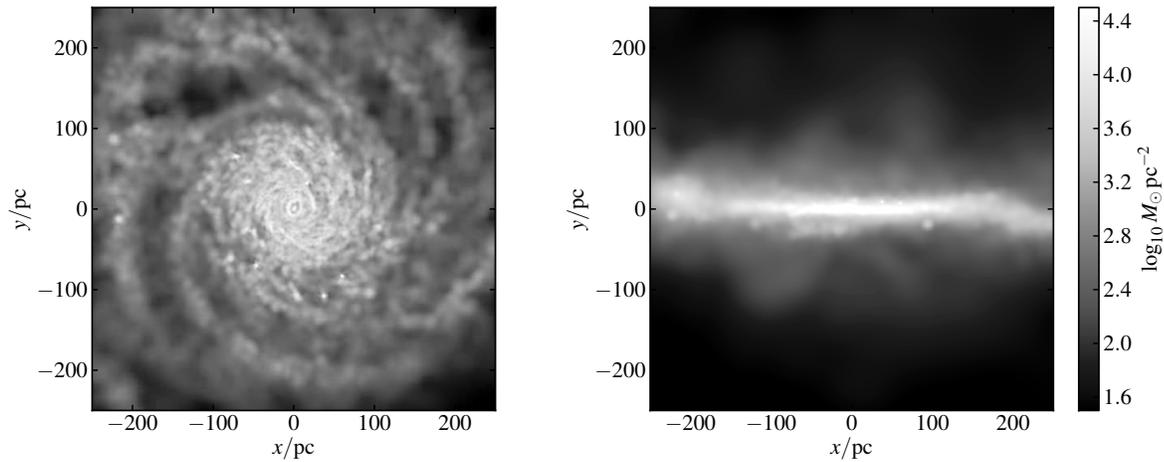}
\caption{Face-on and edge-on gas surface density maps of the central region approximately 10 Myr after the nuclei merge. During the final phases of the merger, the central parts are completely cleared of any coherent structure due to a vigorous starburst, but the rotationally-supported disk rebuilds in a few Myrs reaching a mass of $\sim10^8 \mathrm{~M_{\odot}}$.
}
\label{simdisk}
\end{figure}

The multiphase ISM results in a clumpy and disordered medium
immediately following the merging of the two nuclei compared to the system modeled by
Mayer et al.~\citep[see also][]{Fiacconi13}. Nevertheless, the
nuclear disk rebuilds quickly owing to strong gravitational torques
that drive material to the center. However, due to the fact that in
the new models some of the gas is consumed by star formation and a
large fraction of it is expelled due to stellar supernova feedback,
the mass of the nuclear disk is significantly lower than found
previously, making up only $\sim10^8 M_{\odot}$ 10 Myr after
the two nuclei merge. Immediately after the nuclei merge, the central region is
completely devoid of any coherent structure, but by 10 Myr after the
merger, it measures $\sim200$~pc across (see Figure \ref{simdisk}). In comparison,
the disks in \cite{Mayer07} reached maximum radii of $\sim80$~parsecs
and masses $\sim2\times 10^9 M_{\odot}$. Owing to
vigorous star formation, however, the stellar background is much more
important in the new models; within a sphere of 200~parsecs, the stars make
up $3\times10^9 M_{\odot}$. The nuclear disk has a line-of-sight
$v/\sigma$ of 1.5-2 depending on the chosen inclination.  Without the feedback
associated with star formation, the nuclear disk from \citet{Mayer07} has $v/\sigma$
a factor of a few higher, peaking at 6.

\onecolumn
\twocolumn

The nuclear gas disks observed here with OSIRIS have effective radii ranging from $\sim 100-800$ parsecs, with a mean size of approximately 300 parsecs and a median size of 220 parsecs.  These are considerably larger than the disk in \cite{Mayer07}, but the median size is consistent with the disk in the new multiphase ISM simulations.  The gas disks observed in (U)LIRGs have a mean $v/\sigma \sim 1.75$, similar to the disk seen in the multiphase ISM simulations, and a factor of few lower than the disk in \cite{Mayer07}.  Our mass measurements of disks show a large range, from $10^8 - 5\times10^{10} M_{\sun}$ , encompassing the disk masses from both sets of simulations.

\subsection{How Do Nuclear Disks Affect their Host Galaxies?}

The large disk mass measurements from \S\ref{masssection} reinforce the idea that these nuclear disks are substantial components.  Here we discuss what the presence of such structures may mean in the context of galaxy evolution.

\subsubsection{The Effects of Nuclear Disks on Merging Black Holes}
The presence of nuclear disks during galaxy mergers is likely to assist in the final coalescence of binary black holes.  Initially, the gas will provide substantial dynamical friction, enabling the black holes to lose angular momentum and spiral inwards at a more rapid rate.  In a dense circumnuclear disk such as those studied here, the sinking timescales of massive black holes down to the sub-parsec separations of a tight binary can be shorter than $10^{7}$ years \citep{Mayer07,Chapon13}, which is about two orders of magnitude shorter than in stellar spheroids.

Later, when dynamical friction becomes inefficient below parsec-scale separations, gravitational torques can enable further angular momentum to be removed from the binary \citep{Escala05}, perhaps analogously to the fast type-I migration of planets in protoplanetary disks \citep{Mayer13}.  Additionally, if the gas accretes on to one of the black holes, the binary will tighten \citep{Begelman80}.

Although our OSIRIS observations of the disks at the centers of (U)LIRGs don't have the spatial resolution to detect 
the presence of gas on scales of less than about 10 parsecs, we do find compelling evidence for gaseous disks on scales of hundreds of parsecs.  Models show that gas disks at tens or hundreds of parsecs funnel gas effectively to below parsec scales when sufficient turbulence exists \citep[$v/\sigma\lesssim1$;][]{Wada02, Schartmann09}, which is the case for many of our systems.  

Our observations reveal that star formation is occurring in these disks as well; it may be that star formation in these disks can replenish the region of stars ejected by the binary black holes \citep[the so-called ``loss cone''; see][and references therein]{Begelman80} to allow the binary to continue losing orbital energy through 3-body encounters and ultimately reach the regime dominated by gravitational wave emission and final coalescence \citep{Khan12}.

Simulations of gas disks with binary black holes have shown that the quantity of gas and/or new stars formed in the dense thin disk matters for the timescale of binary black hole coalescence \citep{Escala05,Dotti07,Mayer07}.  For example, by increasing the mass of the gas disk from 1\% to 10\% of the black hole mass, the binary separation can decrease by 75\% in half the time \citep{Escala05}.  Stochastic effects due to scattering with molecular clouds and spiral waves in the disks can broaden the sinking timescales but the trend with disk mass remains \citep{Fiacconi13}.  Our disk mass measurements in \S\ref{masssection} reveal substantial quantities of gas, potentially 10\% or higher of the black hole masses.  In a future paper, we will present black hole mass measurements for many of these galaxies from high spatial resolution kinematic maps.  Comparing those black hole masses to the disk masses presented here will determine a range of plausible merging timescales for binary black holes in (U)LIRGs.

\subsubsection{Are Nuclear Disks the Progenitors of Kinematically Decoupled Cores?}

Kinematically decoupled cores (KDCs) have been found in a number of isolated early-type galaxies as part of the SAURON survey \citep{deZeeuw02,McDermid04,Emsellem04,McDermid06NAR,Krajnovic08}.  \citet{Kormendy84} and \citet{Bender88} have suggested that galaxy mergers or interactions are likely to have formed these kinematic structures.  Indeed, simulations have supported their hypothesis: for at least some scenarios, KDCs can be formed by a merger of two disk galaxies \citep{Hernquist91}, a retrograde merger of two elliptical galaxies with different luminosities \citep{Balcells90}, or a retrograde fly-by encounter of two elliptical galaxies \citep{Hau94}.

The near-ubiquitousness of small-scale gaseous and stellar disks in the gas-rich major mergers observed here lends support to the hypothesis that KDCs can be formed in such mergers.  Though the systems as a whole have not yet dynamically relaxed, the central disks appear similar in size to the KDCs seen in other systems.  These observations only show that disks exist during mergers; future observations of merger remnants will have to be studied to confirm that the disks persist.  The disks seen here do not appear to be significantly decoupled from their host galaxies in position angle on the sky, though the direction of the final angular momentum vector is unknown.  Studying the kinematics on larger scales throughout the merger stage is necessary to determine if and when these disks decouple from the larger system.  In a future paper, we will expand this observational sample to include galaxies at a wider range of merger stage to trace the evolution of these nuclear disks and compare them to the kinematics of their host galaxies.  


\section{Conclusions}
\label{conclusions}

The formation of nuclear gas disks has been predicted by hydrodynamical simulations of gas-rich galaxy mergers.  In simulations, gas present in the progenitor galaxies is funneled towards the nuclei and forms a disk on scales of a few to a few hundred pc.  Stellar disks on these scales have been seen in a number of isolated galaxies. 
We present the first high spatial resolution integral field spectroscopy survey of a large sample of 17 nearby gas-rich major mergers, providing observational evidence of these nuclear disks in both gas and stars.  Because the stellar and gas kinematics match well in most cases, we suggest that the stars formed \textit{in situ}, from cold gas present in the disks.  This hypothesis is supported by the measured equivalent widths of the CO 2.292 $\mu$m absorption features, which indicate that the stellar populations found in the disks are young ($<30$ Myr for eight out of ten measured nuclei, a few dynamical timescales).

The formation of these disks seems to be a common occurrence in these gas-rich galaxy mergers.  We kinematically confirm gas disks in at least 16 of 19 nuclei, and stellar disks in 11 out of 11 nuclei.  Disks formed during the merger have radii ranging from 50 to 800 pc, and have $v/\sigma$ of generally between 1 and 5.   
This is in qualitative agreement with nuclear disks seen in simulations.  More detailed comparisons with simulations will be presented in a future paper.

  
\acknowledgements
We enthusiastically thank the staff of the W. M. Keck Observatory and its AO team, for their dedication and hard work.  Data presented herein were obtained at the W. M. Keck Observatory,
which is operated as a scientific partnership among the California Institute of Technology, the University of California, and the National Aeronautics and Space Administration.  The Observatory and the Keck Laser Guide Star AO systems were both made possible by the generous financial support of the W. M. Keck Foundation.  The authors wish to extend special thanks to those of Hawaiian ancestry on whose sacred mountain we are privileged to be guests.  Without their generous hospitality, the observations presented herein would not have been possible.  We also thank Nate McCrady of the University of Montana for providing spectral templates in the $H$-band, Nicholas McConnell of the Institute for Astronomy and others for making their OSIRIS calibrations public on the wiki, Loreto Barco and Aaron Evans for making their radio data and preliminary analyses available for comparison, Andrew Green and Massimo Dotti for helpful conversations, and the anonymous referee for thoughtful feedback.  This work was supported in part by the National Science Foundation Science and Technology Center for Adaptive Optics, managed
by the University of California at Santa Cruz under cooperative agreement AST 98-76783.  This material is based in part upon work supported by the National Science Foundation under award number AST-0908796.
AM was supported in part by a Graduate Research Fellowship from the National Science Foundation and by a fellowship from the Achievement Rewards for College Scientists Foundation.  
VU acknowledges funding support from the NASA Harriet G. Jenkins Predoctoral Fellowship Project and the Smithsonian Astrophysical Observatory Predoctoral Fellowship.
JG acknowledges funding from the ETH Z\"urich Postdoctoral Fellowship and the Marie Curie Actions for People COFUND Program.  

{\it Facility:} \facility{Keck:I,II (Laser Guide Star Adaptive Optics, OSIRIS)}

\appendix
\section{Additional Figures and Notes on Individual Galaxies}
\label{extrafig}

For clarity, galaxy-specific figures were only included in the main text for the first galaxy, CGCG436-030; here we include the appropriate GALFIT modeling and velocity map figures for the remainder of the sample.  

\textbf{CGCG436-030:} This galaxy shows a star-forming clump to the northwest of the nucleus.  Though the stellar kinematics are noisy, they appear to be consistent with the motions of the gas.  Flux and velocity maps are shown in Figures~\ref{galfit} and \ref{vels}, respectively.

\textbf{IRAS F01364-1042:} This galaxy shows slight evidence of extended \molhy~emission to the south; this extension is between the major and minor axis of the disk.  Though a similar velocity gradient appears in \molhy~and \paa~on axis, the shape of the gradient varies somewhat.  This, plus the off-axis differences in velocity structure, may be related to outflowing structure traced by \molhy~and not by \paa.  This and other potential indicators of outflows will be analyzed in a future paper.  Flux and velocity maps are shown in Figure~\ref{IR01364}.

\textbf{IIIZw035:} This galaxy shows strong extensions in \molhy~flux along the minor axis of the galaxy.  The bend in the continuum emission may be evidence of a warped disk, which is not modeled here.  It should also be noted that the rotation of stars and \brg~match well over most of our field of view, but the \molhy~kinematics only match over $\lesssim200$ pc.  Because the GALFIT models do not indicate that the \molhy~disk is smaller in flux, we suggest that the velocity discrepancy may be due to additional \molhy~emission that is outflowing, or otherwise not part of the disk, along the line of sight.  This potential outflow will be studied in a future paper.  We see no sign of structure associated with the 22 pc OH maser ring offset in position angle by $40^{\circ}$ \citep{Pihlstrom01}.  Flux and velocity maps are shown in Figure~\ref{IIIZw035}.

\textbf{IRAS F03359+1523:} This galaxy has a large, nearly-edge on disk with a string of star clusters along it.  Taking this in conjunction with a previous estimate of the merger stage \citep[3, as compared to 4-6 for most of our sample;][]{Haan11}, we suggest that this disk is not a nuclear disk formed during the merger, but is more likely galactic-scale progenitor galaxy's disk that has not yet been disrupted.  As such, its parameters are left out of the average nuclear disk calculations.  Flux and velocity maps are shown in Figure~\ref{IR03359}.

\textbf{MCG+08-11-002:} The extended bright \brg~emission shows that star formation in the nucleus of this galaxy is considerably strong and clumpy.  The flux map \brg~also suggests the presence of spiral arms which are fainter in the continuum map and not present in the \molhy~map.  Flux and velocity maps are shown in Figure~\ref{MCG08}.

\textbf{NGC~2623:} This galaxy has smooth flux profiles and similar kinematics in all tracers.  Flux and velocity maps are shown in Figure~\ref{NGC2623}.

\textbf{UGC5101:} The residuals of the continuum fit in this galaxy show faint structures resembling spiral arms.  Additionally, this galaxy is the only one in our sample to require a central PSF in the \molhy~flux GALFIT model.  Flux and velocity maps are shown in Figure~\ref{UGC5101}.

\textbf{Mrk231:} The central quasar in this galaxy produced an excess of noise in the central regions, making GALFIT models of the continuum emission impossible.  Indeed, to obtain even the noisy \molhy~map, we binned the data by a factor of two, yielding 0\farcs07 pixels.  However, the results found here are consistent with the \molhy~disk measured by \cite{Davies04}.  Flux and velocity maps are shown in Figure~\ref{Mrk231}.

\textbf{UGC8387:} Unfortunately, observing conditions cut short our observations of this galaxy, preventing us from completing the mosaic and obtaining details on the center of this galaxy.  Because of the incomplete data, we refrain from analysis which requires knowledge of the central position and kinematics.  For completeness and to aid future observations of this object, maps of flux and velocity are included in Figure~\ref{UGC8387}.

\textbf{Mrk273:} A thorough analysis of this galaxy was presented in \cite{mrk273}, including GALFIT modeling of both nuclei.    Though rotation is seen in the northern nucleus, the southwestern nucleus shows none.  However, their high resolution NIRC2 images (0.01\arcsec pixel$^-1$) reveal a morphological disk on scales unresolvable by OSIRIS' kinematics.  Here we include the southwest nucleus as a stellar disk in Table~\ref{tbl:galfitresults} even though these OSIRIS data cannot provide a kinematic confirmation.  As there are no kinematic data for the SW disk, it is left out of Table~\ref{tbl:vsigmatable} and Figure~\ref{voversigma_binnedLIR}.  Velocity maps of the northern nucleus are shown in Figure~\ref{Mrk273}.

\textbf{VV340a:} Even in $K$-band a solid dust lane is clearly evident in the large-scale disk, which may affect GALFIT measurements.  As an early-stage merger \citep[merger class of 1 in][]{Haan11}, it has retained its large-scale disk.  However, it appears also to be building an additional distinct nuclear disk on a smaller scale, seen in continuum emission and \molhy~emission; \brg~emission does not trace this additional structure.  Due to modeling uncertainties from the GALFIT measurements, we constrain the axial ratios of the two disks to be the same.  Flux and velocity maps are shown in Figure~\ref{VV340a}.

\textbf{IRAS F15250+3608:} The continuum image reveals a second clump to the southeast of the main nucleus.  However, due to the lack of emission line detections, it is not clear whether or not this is a second nucleus or an off-nuclear clump.  Velocity maps focus on the main component.  This is the only galaxy in our sample for which we see no rotation in either stars or gas.  GALFIT model residuals show that multiple components may be present in the main nucleus, but because no kinematic angle has been measured, constraints for additional components are not present.  Since the velocity maps do not confirm rotation, even the morphological parameters are left out of the final nuclear disk averages.  However, these parameters are included in Table~\ref{tbl:galfitresults} for completeness.  Flux and velocity maps are shown in Figure~\ref{IR15250}.

\textbf{NGC~6090:} This galaxy's continuum map shows a low-inclination disk with large spiral arms and possibly a bar.  As with IRAS F013359+1523, the size of the disk and previous estimates of the merger stage \citep[4, on the low side of our mainly 4-6 sample, according to][]{Haan11,Kim13} suggest that this disk was not formed during the merger, but rather is a large galactic disk that has not yet been disturbed.  As such, its parameters are left out of the average nuclear disk calculations.  It is also interesting to note that, while the nucleus shows strong continuum, minimal line flux is seen at the nucleus of the disk.  \brg~flux is, however, strong in the spiral arms.  Because of this lack of flux, and the limited field of view of these OSIRIS data, obtaining velocity information for the gas across the major axis of the disk is difficult, and provides little insight.  Flux and velocity maps are shown in Figure~\ref{NGC6090}.

\textbf{NGC~6240:} Observations of this galaxy were taken in the Kn5 narrowband, which does not cover the spectral regions of \molhy~or \brg.  Masks shown in GALFIT fits were implemented due to the excess of dust present between the two nuclei \citep{Max07}.  Though the narrow-band OSIRIS data presented here do not include emission lines with which to measure gas kinematics, we note that the SINFONI data presented by \cite{Hauke} suggests that the gas kinematics are dominated by disrupted tidal streaming and shocks, not disks.  However, their stellar kinematics are consistent with those presented here.  Flux and velocity maps are shown in Figure~\ref{NGC6240}.

\textbf{IRAS F17207-0014:} This galaxy shows two overlapping nuclear disks.  They are likely overlapping in superposition rather than actually colliding; such collisions would presumably disrupt the disks.  Here as well the \molhy~flux map shows morphologies that do not match continuum emission; this may be signs either of outflows or, more likely, of collisional shocks related to the larger nuclear orbits.  Flux and velocity maps are shown in Figure~\ref{IR17207}.

\textbf{IRAS 20351+2521:} The clumpy nature of this nucleus is reminiscent of spiral arms; with a large clump of emission appearing in continuum and both line emission maps.  The \molhy~and \brg~velocity maps show similar kinematics.  Flux and velocity maps are shown in Figure~\ref{IR20351}.

\textbf{IRAS F22491-1808:} This galaxy also shows two nuclei; in \paa, the strongest emission line, both disks appear to show rotating kinematic signatures.  However, though \molhy~is detected in the E disk, it does not appear to share the kinematics of \paa.  A velocity gradient is seen, but because it isn't centered about the same velocity as \paa, which is likely due to a strong component of streaming or outflowing shocked gas.  The \molhy~emission is extended along the minor axis of the E disk.  Flux and velocity maps are shown in Figure~\ref{IR22491}.

\begin{figure}[ht]
\centering
\subfloat{\includegraphics[scale=0.85]{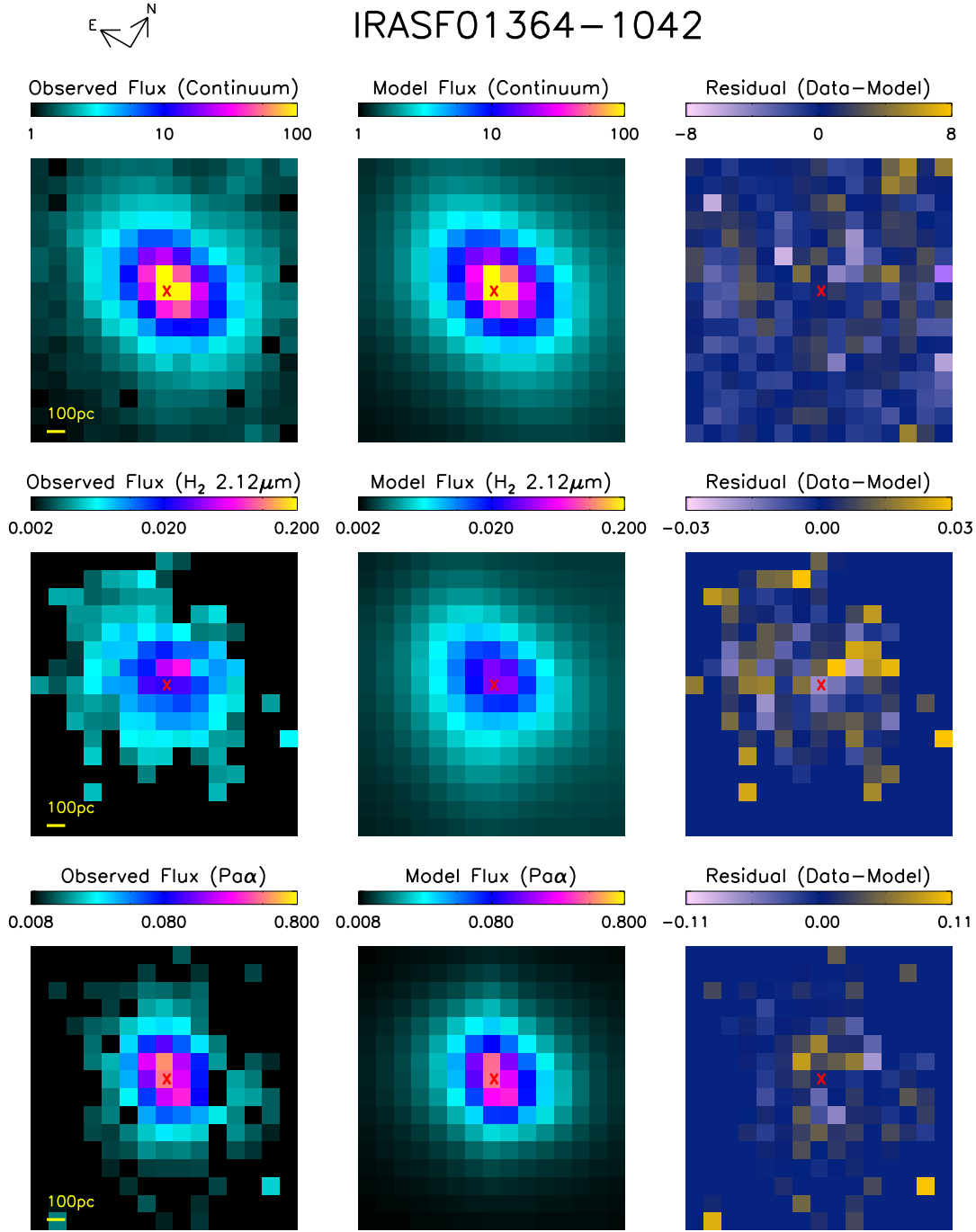}}
\caption{Flux map (left panel), GALFIT model (center panel) and residual (right panel) for IRAS F01364-1042.  The top row is the continuum flux, the middle row is \molhy~flux, and the bottom row is \paa~flux.  The flux maps (left and center panels) are shown on a log scale, while the residual map is shown on a linear scale, in units of counts per second.  This galaxy shows some evidence of extended \molhy~emission to the south, between the major and minor axis of the disk.}
\label{IR01364}
\end{figure}

\begin{figure}[ht]
\centering
\ContinuedFloat
\subfloat{\includegraphics[scale=1.7]{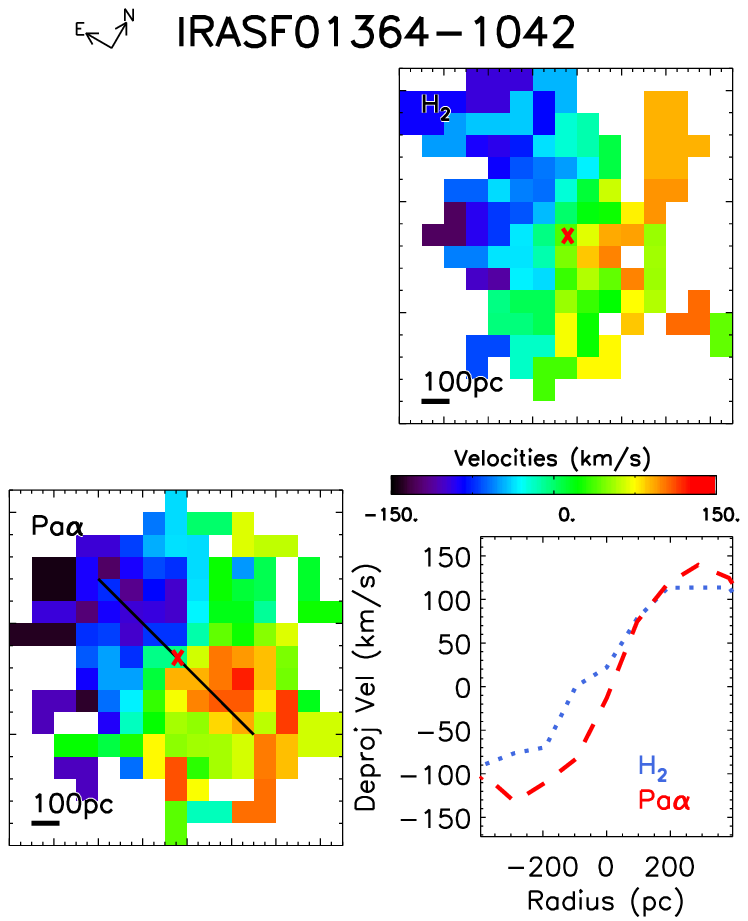}}
\caption{Observed velocity maps of molecular gas in \molhy~(top right) and ionized gas in \paa~(bottom left) for IRAS F01364-1042.  All velocity maps use the same color bar; white pixels do not have sufficient signal to measure velocities accurately.  Bottom right panel shows the deprojected velocity profile cut through the major axis for each of the two tracers: \molhy~(dotted blue) and \paa~(dashed red).  The major axis cut is indicated in black on the bottom left map for clarity.  Though a similar velocity gradient appears in \molhy~and \paa~here, the differences may be related to outflowing structure traced by \molhy~and not by \paa.}
\end{figure}

\begin{figure}[ht]
\centering
\subfloat{\includegraphics[scale=0.85]{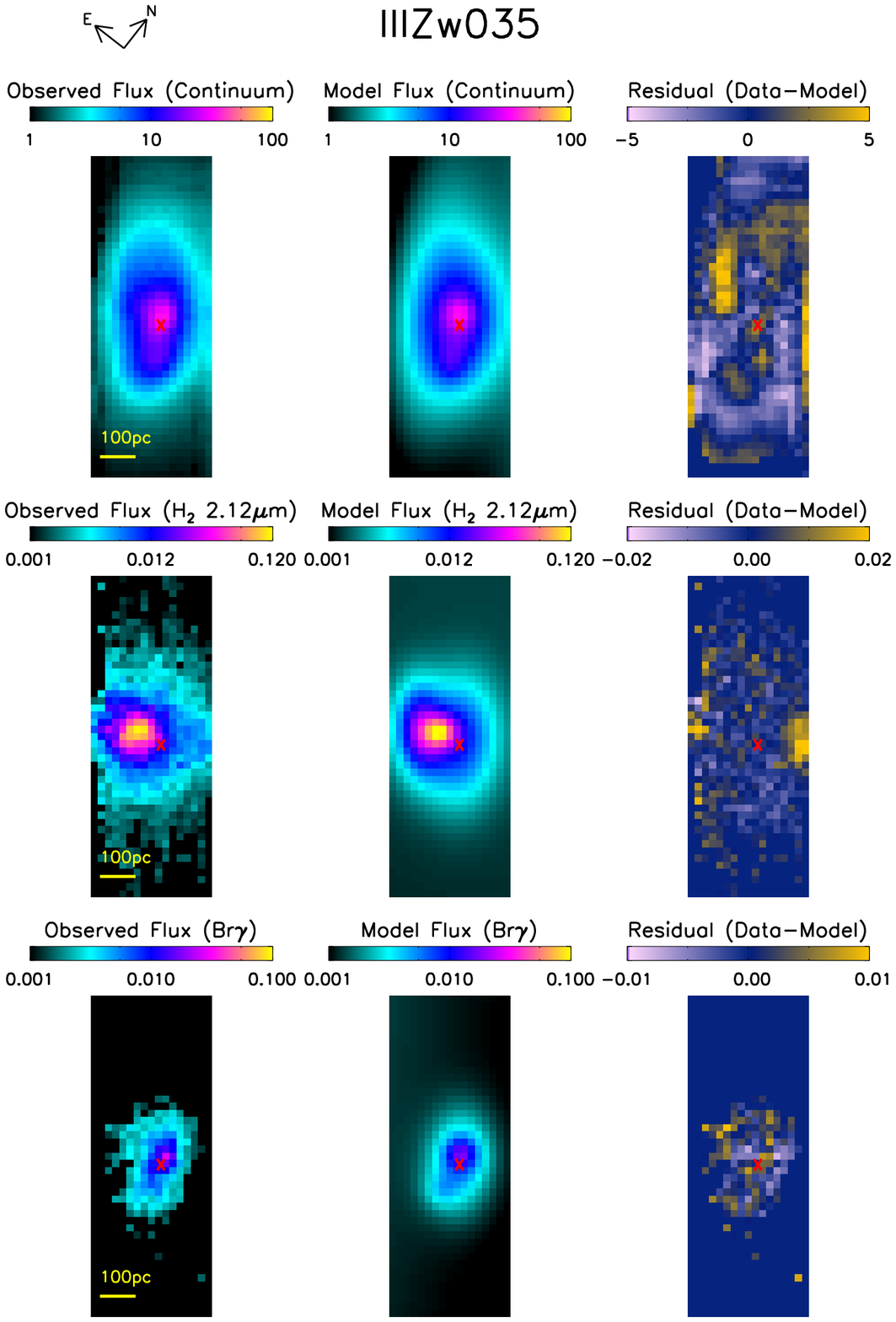}}
\caption{Flux map (left panel), GALFIT model (center panel) and residual (right panel) for IIIZw035.  The top row is the continuum flux, the middle row is \molhy~flux, and the bottom row is \brg~flux.  The flux maps (left and center panels) are shown on a log scale, while the residual map is shown on a linear scale, in units of counts per second.  This galaxy shows strong extensions in \molhy~flux along the minor axis of the galaxy.  The bend in the continuum emission may be evidence of a warped disk, which is not modeled here.}
\label{IIIZw035}
\end{figure}

\begin{figure}[ht]
\centering
\ContinuedFloat
\subfloat{\includegraphics[scale=1.1]{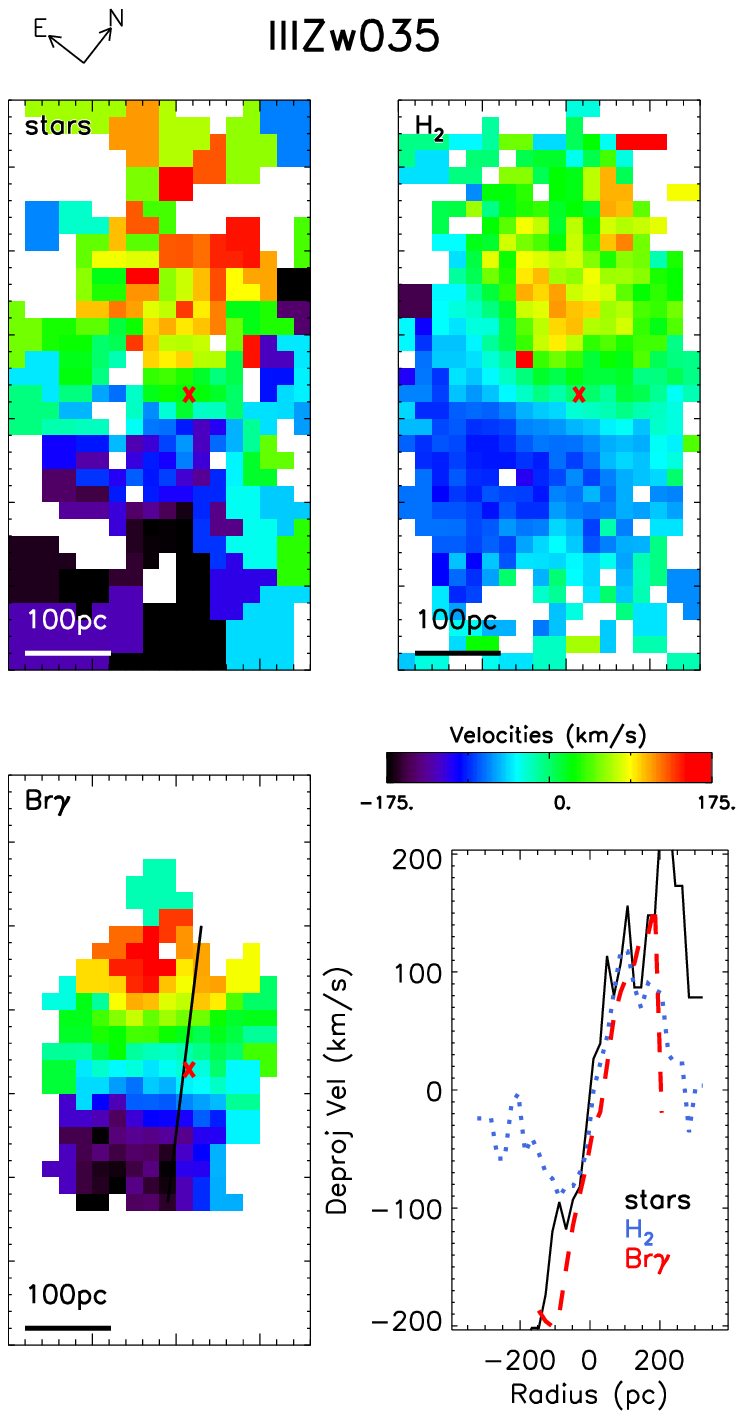}}
\caption{Observed velocity maps of molecular gas in stars (top left), \molhy~(top right), and ionized gas in \brg~(bottom left) for IIIZw035.  All velocity maps use the same color bar; white pixels do not have sufficient signal to measure velocities accurately.  Bottom right panel shows the deprojected velocity profile cut through the major axis for each of the 3 tracers: stars (solid black), \molhy~(dotted blue),  \brg~(dashed red).  The major axis cut is indicated in black on the bottom left map for clarity.  The on-axis velocity profiles of \molhy~only match other tracers over $\lesssim200$ pc; \molhy~also shows warped off-axis velocities.  Because the GALFIT models do not indicate that the \molhy~disk is smaller in flux, we suggest that the velocity discrepancy may be due to additional \molhy~emission that is outflowing, or otherwise not part of the disk, along the line of sight. }
\end{figure}

\begin{figure}[ht]
\centering
\subfloat{\includegraphics[scale=0.7]{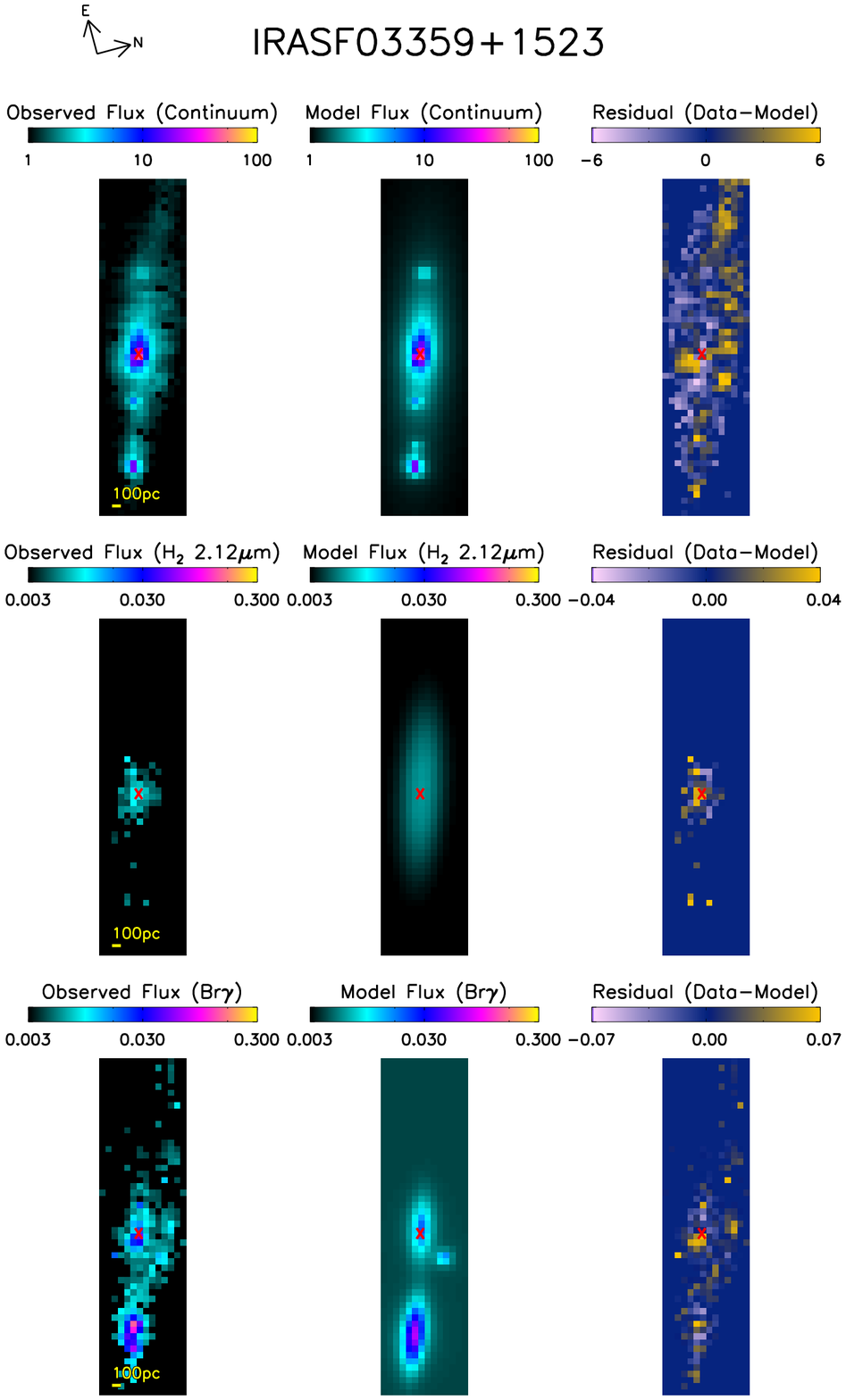}}
\caption{Flux map (left panel), GALFIT model (center panel) and residual (right panel) for IRAS F03359+1523.  The top row is the continuum flux, the middle row is \molhy~flux, and the bottom row is \brg~flux.  The flux maps (left and center panels) are shown on a log scale, while the residual map is shown on a linear scale, in units of counts per second.  We suggest that this disk is not a nuclear disk formed during the merger, but is more likely galactic-scale progenitor galaxy's disk that has not yet been disrupted.  As such, its parameters are left out of the average nuclear disk calculations.  (See text for details.)}
\label{IR03359}
\end{figure}

\begin{figure}[ht]
\centering
\ContinuedFloat
\subfloat{\includegraphics[scale=1.25]{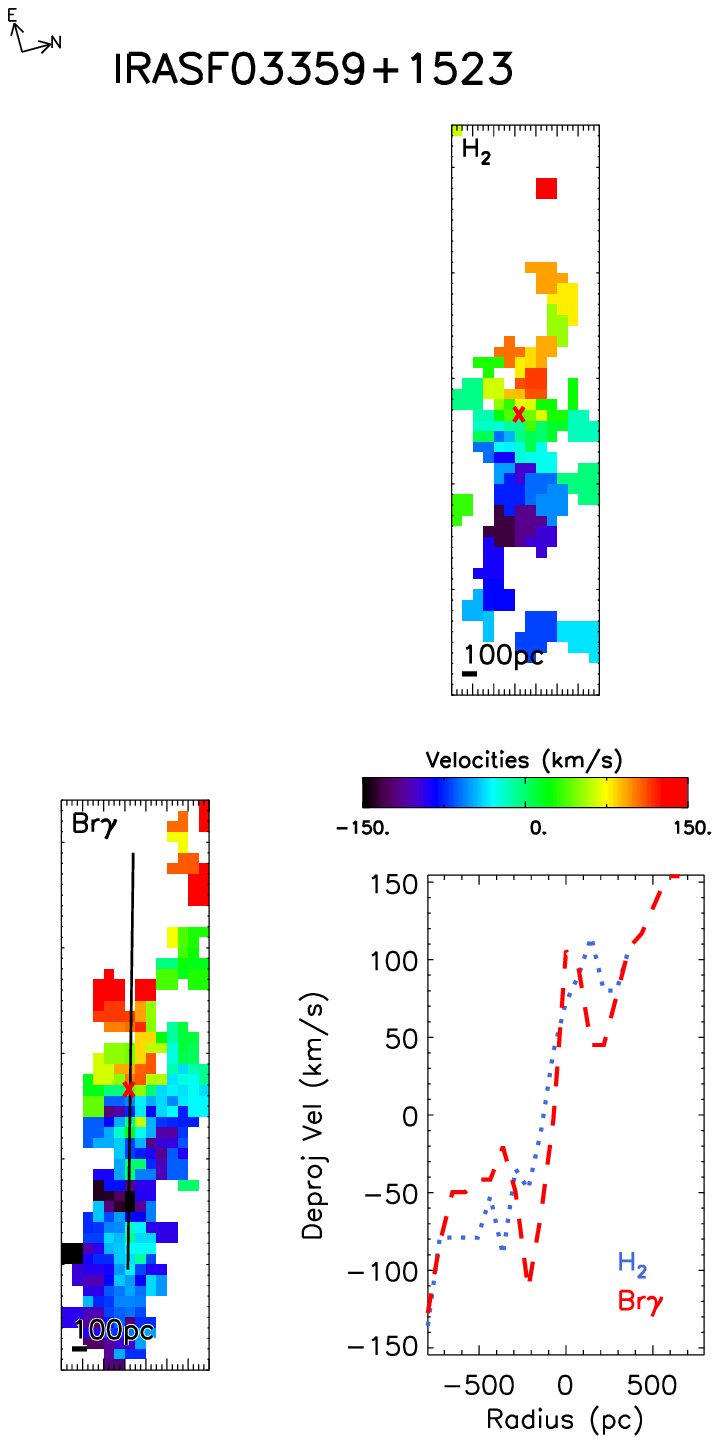}}
\caption{Observed velocity maps of molecular gas in \molhy~(top right) and ionized gas in \brg~(bottom left) for IRAS F03359+1523.  All velocity maps use the same color bar; white pixels do not have sufficient signal to measure velocities accurately.  Bottom right panel shows the deprojected velocity profile cut through the major axis for each of the two tracers: \molhy~(dotted blue) and \brg~(dashed red).  The major axis cut is indicated in black on the bottom left map for clarity.}
\end{figure}

\begin{figure}[ht]
\centering
\subfloat{\includegraphics[scale=0.8]{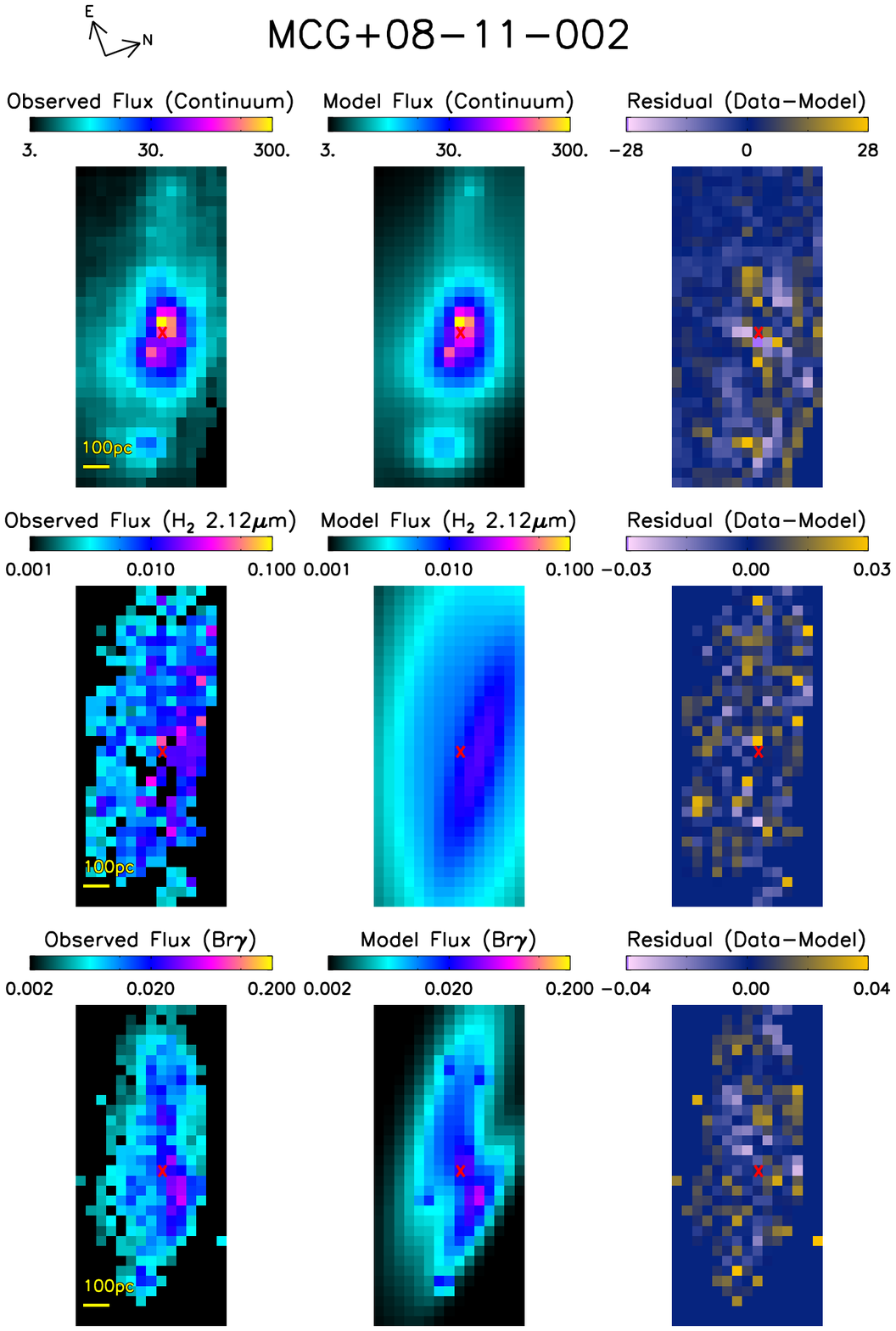}}
\caption{Flux map (left panel), GALFIT model (center panel) and residual (right panel) for MCG+08-11-002.  The top row is the continuum flux, the middle row is \molhy~flux, and the bottom row is \brg~flux.  The flux maps (left and center panels) are shown on a log scale, while the residual map is shown on a linear scale, in units of counts per second.  The extended bright \brg~emission shows that star formation in the nucleus of this galaxy is considerably strong and clumpy.  The flux map \brg~also suggests the presence of spiral arms which are fainter in the continuum map and not present in the \molhy~map.  }
\label{MCG08}
\end{figure}

\begin{figure}[ht]
\centering
\ContinuedFloat
\subfloat{\includegraphics[scale=1.2]{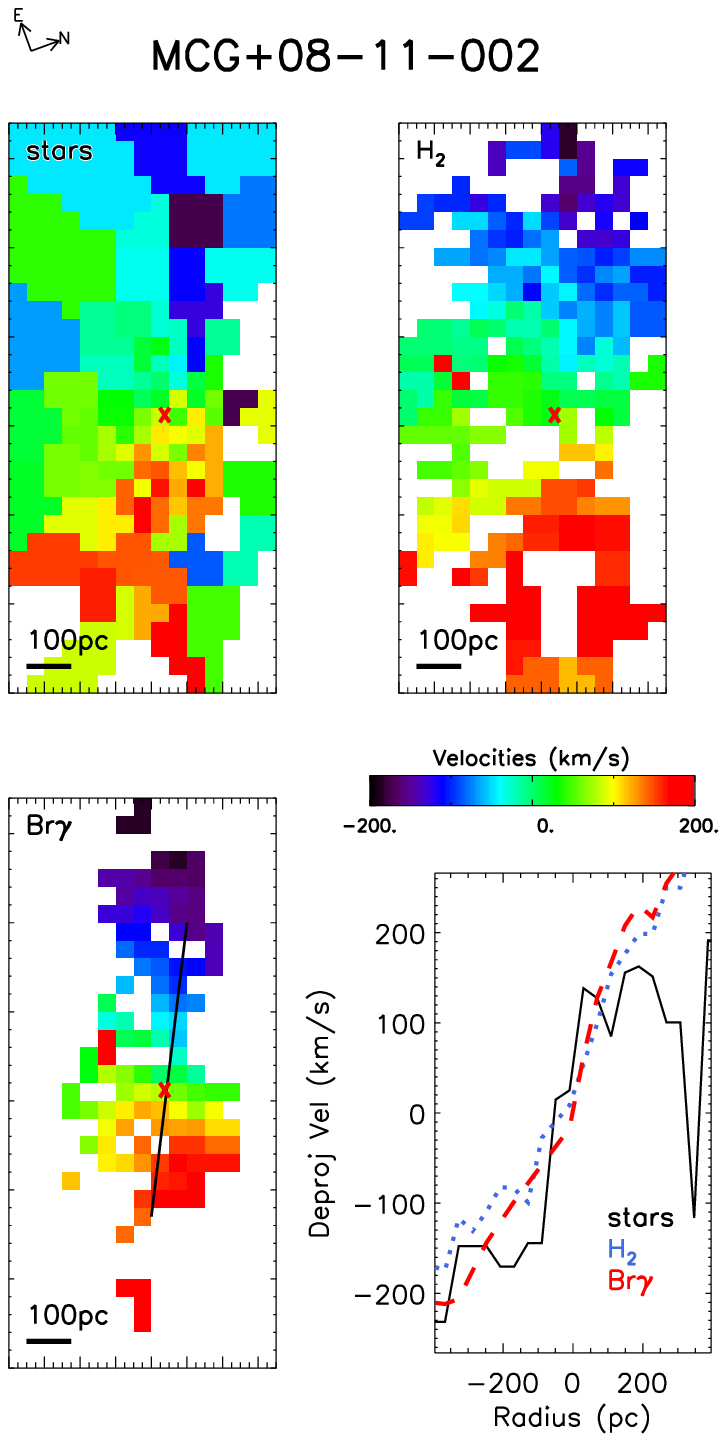}}
\caption{Observed velocity maps of molecular gas in stars (top left), \molhy~(top right), and ionized gas in \brg~(bottom left) for MCG+08-11-002.  All velocity maps use the same color bar; white pixels do not have sufficient signal to measure velocities accurately.  Bottom right panel shows the deprojected velocity profile cut through the major axis for each of the 3 tracers: stars (solid black), \molhy~(dotted blue),  \brg~(dashed red).  The major axis cut is indicated in black on the bottom left map for clarity.}
\end{figure}

\begin{figure}[ht]
\centering
\subfloat{\includegraphics[scale=0.85]{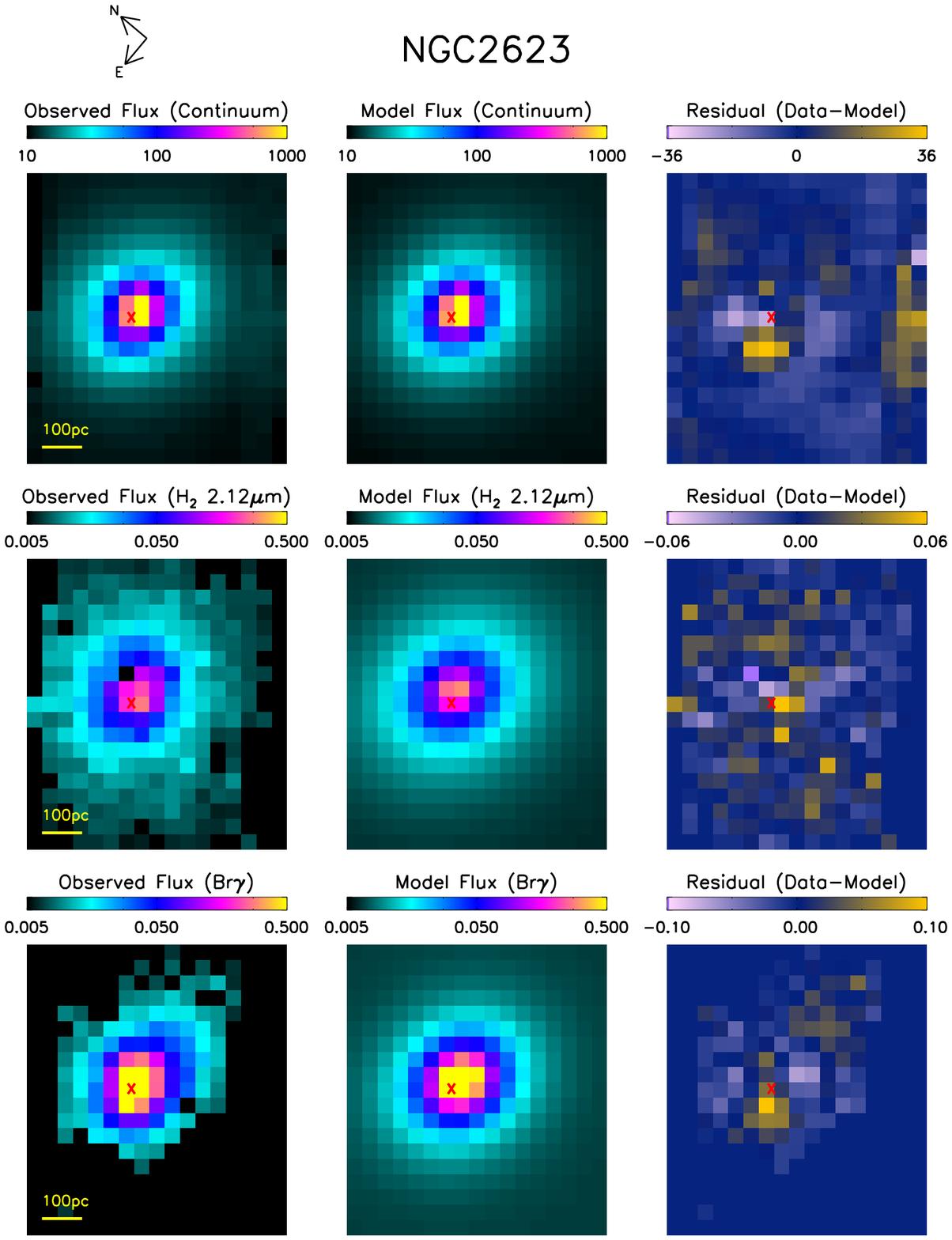}}
\caption{Flux map (left panel), GALFIT model (center panel) and residual (right panel) for NGC~2623.  The top row is the continuum flux, the middle row is \molhy~flux, and the bottom row is \brg~flux.  The flux maps (left and center panels) are shown on a log scale, while the residual map is shown on a linear scale, in units of counts per second.  }
\label{NGC2623}
\end{figure}

\begin{figure}[ht]
\centering
\ContinuedFloat
\subfloat{\includegraphics[scale=1.7]{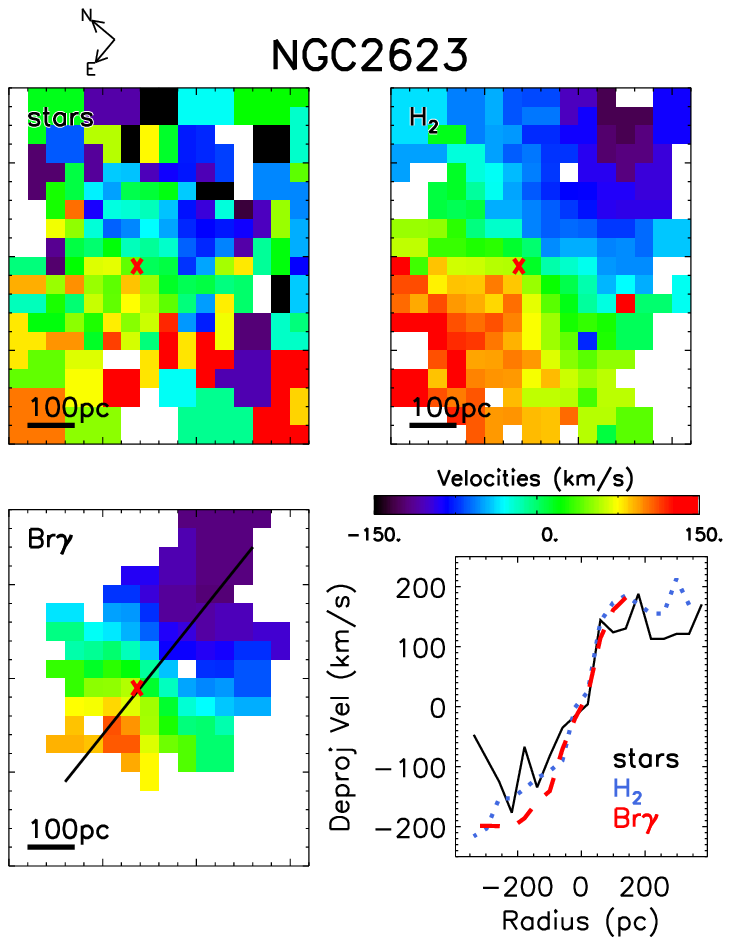}}
\caption{Observed velocity maps of molecular gas in stars (top left), \molhy~(top right), and ionized gas in \brg~(bottom left) for NGC2623.  All velocity maps use the same color bar; white pixels do not have sufficient signal to measure velocities accurately.  Bottom right panel shows the deprojected velocity profile cut through the major axis for each of the 3 tracers: stars (solid black), \molhy~(dotted blue),  \brg~(dashed red).  The major axis cut is indicated in black on the bottom left map for clarity.}
\end{figure}

\begin{figure}[ht]
\centering
\subfloat{\includegraphics[scale=0.8]{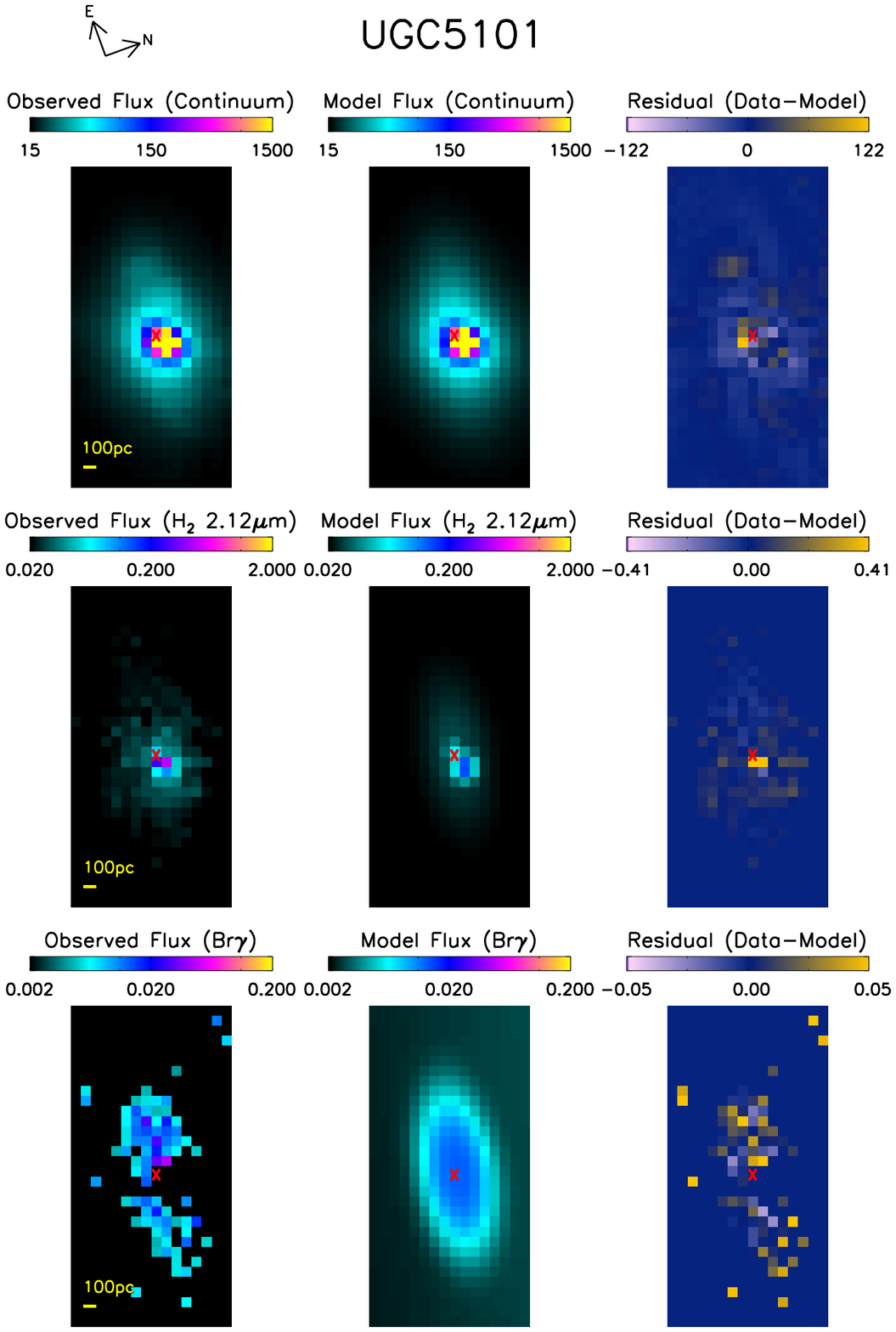}}
\caption{Flux map (left panel), GALFIT model (center panel) and residual (right panel) for UGC5101.  The top row is the continuum flux, the middle row is \molhy~flux, and the bottom row is \brg~flux.  The flux maps (left and center panels) are shown on a log scale, while the residual map is shown on a linear scale, in units of counts per second.  The residuals of the continuum fit in this galaxy show faint structures resembling spiral arms.  Additionally, this galaxy is the only one in our sample to require a central PSF in the \molhy~flux GALFIT model.}
\label{UGC5101}
\end{figure}

\begin{figure}[ht]
\centering
\ContinuedFloat
\subfloat{\includegraphics[scale=1.2]{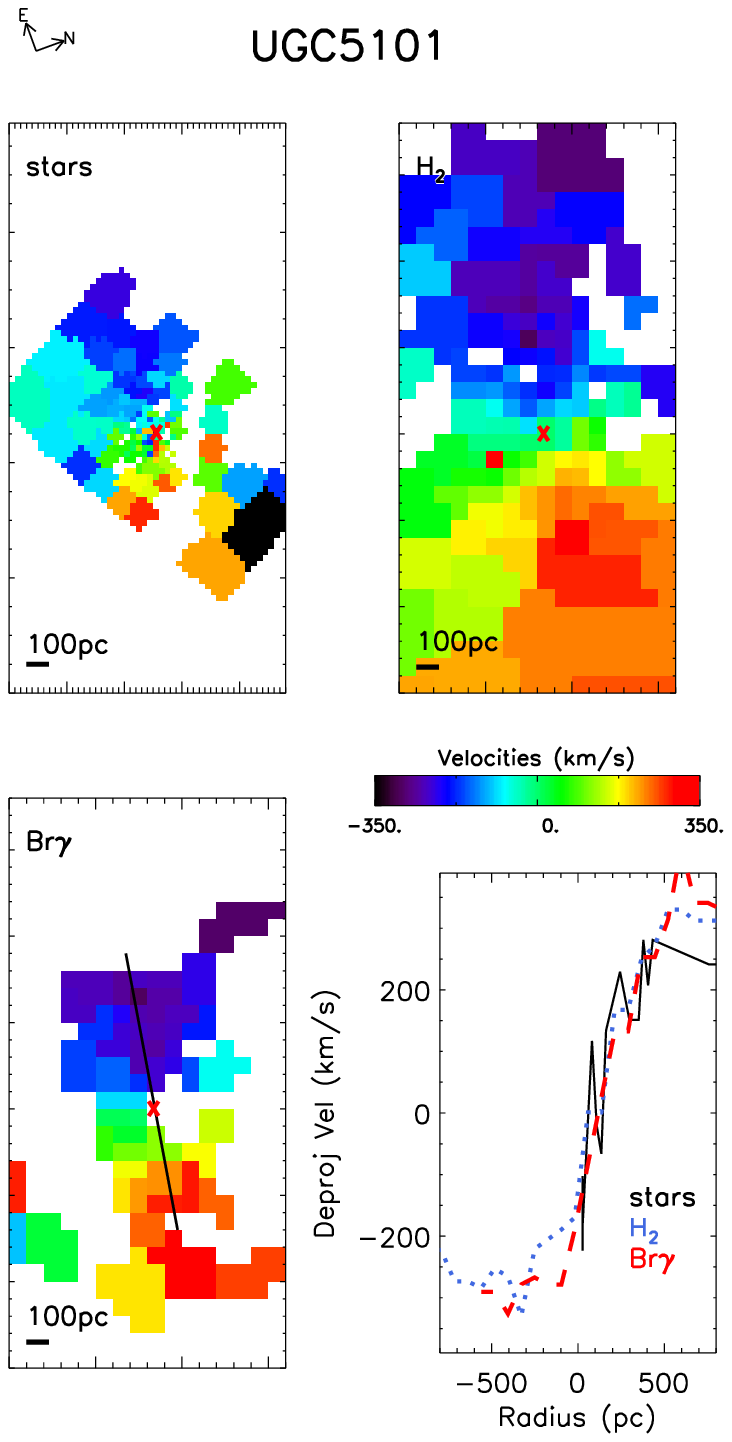}}
\caption{Observed velocity maps of molecular gas in stars(top left), \molhy~(top right), and ionized gas in \brg~(bottom left) for UGC5101.  All velocity maps use the same color bar; white pixels do not have sufficient signal to measure velocities accurately.  Bottom right panel shows the deprojected velocity profile cut through the major axis for each of the 3 tracers: stars (solid black), \molhy~(dotted blue),  \brg~(dashed red).  The major axis cut is indicated in black on the bottom left map for clarity.}
\end{figure}

\begin{figure}[ht]
\centering
\subfloat{\includegraphics[scale=0.9]{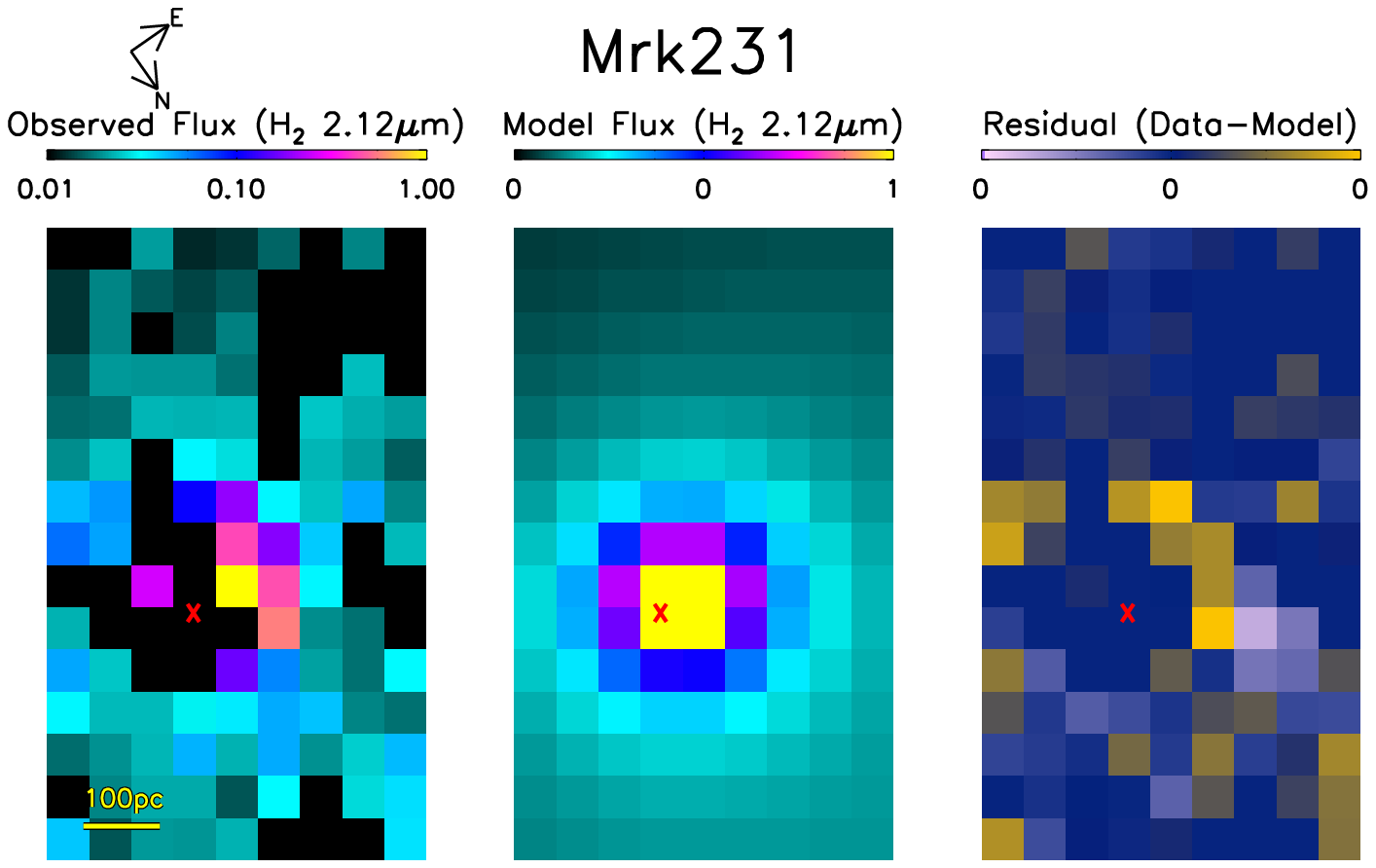}}
\caption{Flux map (left panel), GALFIT model (center panel) and residual (right panel) for \molhy~in Mrk231.  The flux maps (left and center panels) are shown on a log scale, while the residual map is shown on a linear scale, in units of counts per second.  The central quasar in this galaxy produced an excess of noise in the central regions, making GALFIT models of the continuum emission impossible.  To obtain this \molhy~map, we binned the data by a factor of two, yielding 0\farcs07 pixels.  The results found here are consistent with the \molhy~disk measured by \cite{Davies04}.}
\label{Mrk231}
\end{figure}

\begin{figure}[ht]
\centering
\ContinuedFloat
\subfloat{\includegraphics[scale=1.2]{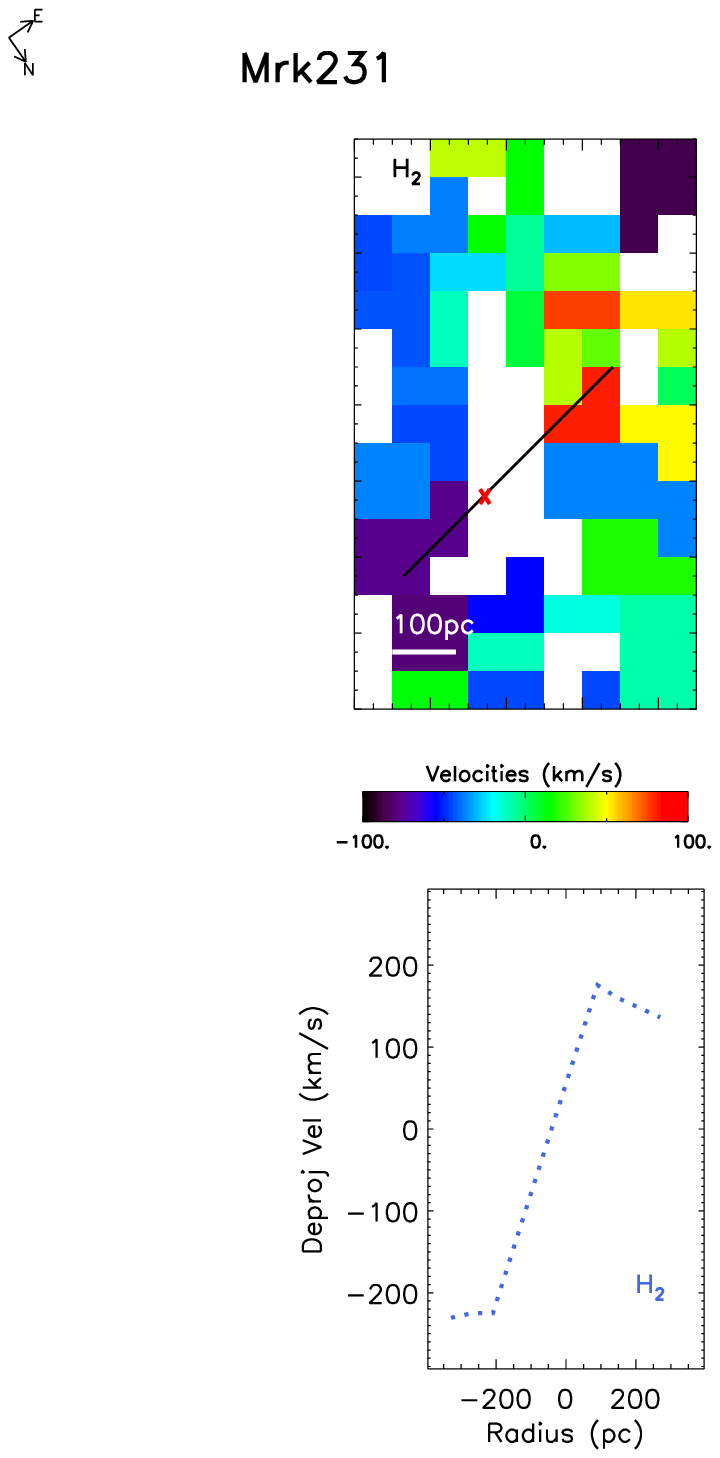}}
\caption{Observed velocity maps of molecular gas in \molhy~(top right) for Mrk231, with data binned by a factor of two, yielding 0\farcs07 pixels.  White pixels do not have sufficient signal to measure velocities accurately.  Bottom right panel shows the deprojected velocity profile cut through the major axis.  The major axis cut is indicated in black in the velocity map for clarity.}
\end{figure}

\begin{figure}[ht]
\centering
\subfloat{\includegraphics[scale=0.85]{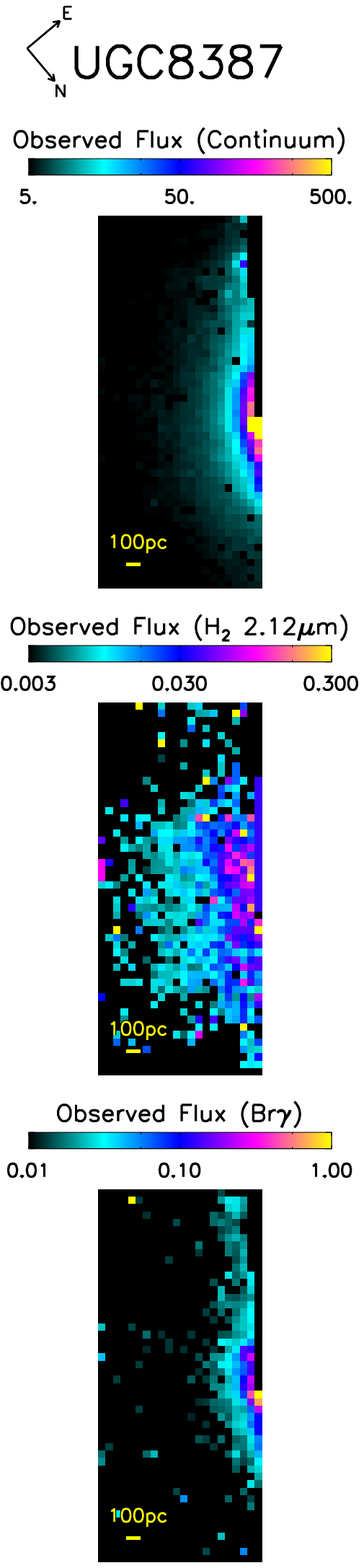}}
\caption{Flux maps for UGC8387.  The top row is the continuum flux, the middle row is \molhy~flux, and the bottom row is \paa~flux.  The flux maps are shown on a log scale, in units of counts per second.  Because observations of this galaxy were off-center, GALFIT analysis of the data was not possible.}
\label{UGC8387}
\end{figure}

\begin{figure}[ht]
\centering
\ContinuedFloat
\subfloat{\includegraphics[scale=1.2]{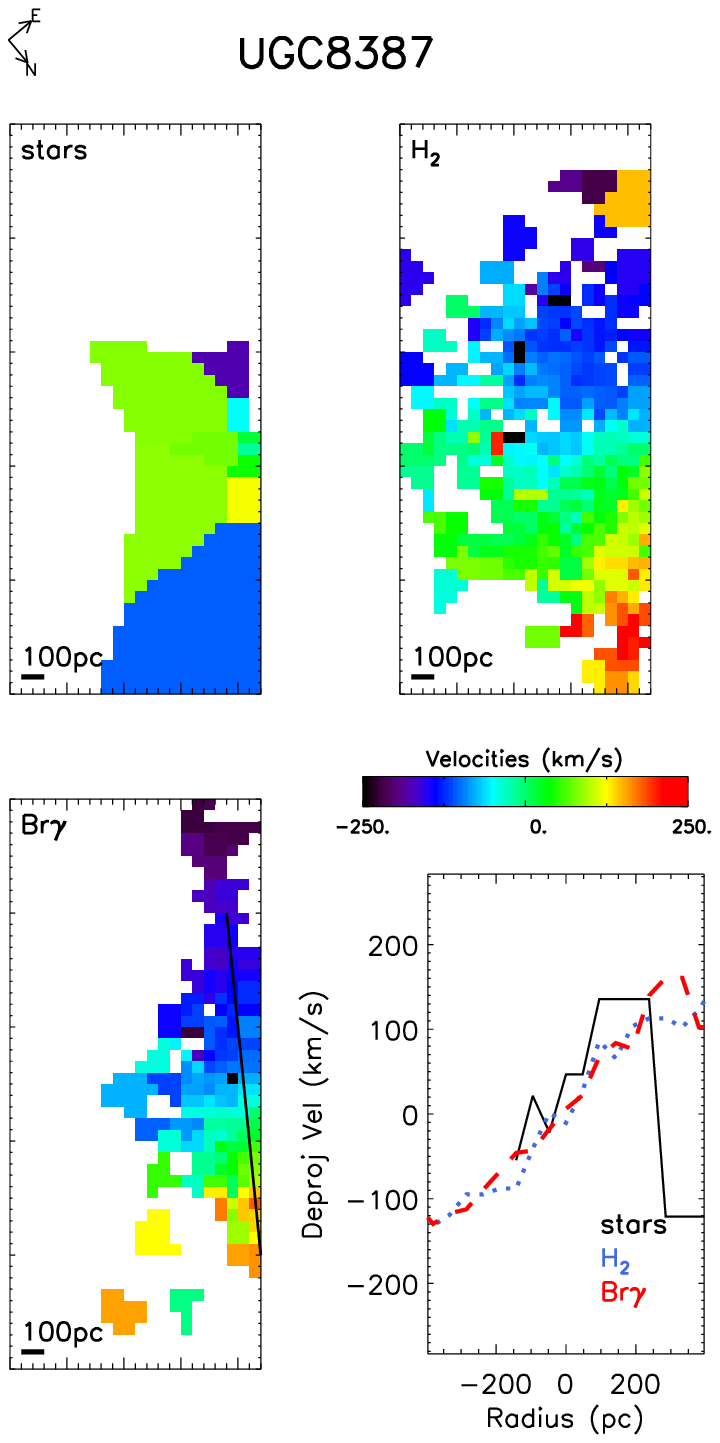}}
\caption{Observed velocity maps of stars (top left), molecular gas in \molhy~(top right), and ionized gas in \brg~(bottom left) for UGC8387.  All velocity maps use the same color bar; white pixels do not have sufficient signal to measure velocities accurately.  Bottom right panel shows the deprojected velocity profile cut through a possible major axis for each of the 3 tracers: stars (solid black), \molhy~(dotted blue),  \brg~(dashed red); the deprojected velocity profile was calculated using the axial ratio (0.47) found from the GALFIT modeling of \citet{Kim13}.  The major axis cut adopted is indicated in black on the bottom left map for clarity. }
\end{figure}

\begin{figure}[ht]
\centering
\subfloat{\includegraphics[scale=1.5]{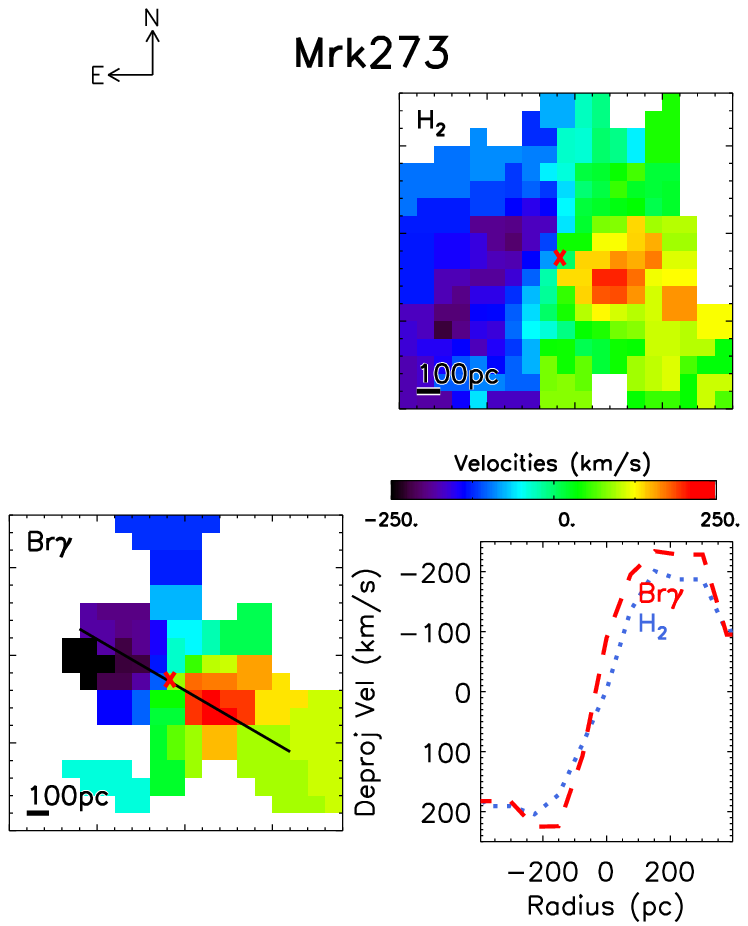}}
\caption{Observed velocity maps of molecular gas in \molhy~(top right) and ionized gas in \brg~(bottom left) for Mrk273.  All velocity maps use the same color bar; white pixels do not have sufficient signal to measure velocities accurately.  Bottom right panel shows the deprojected velocity profile cut through the major axis for each of the two tracers: \molhy~(dotted blue) and \brg~(dashed red).  The major axis cut is indicated in black on the bottom left map for clarity.  Though rotation is seen in the northern nucleus, the southwestern nucleus shows none observable on these scales, and is not shown.}
\label{Mrk273}
\end{figure}

\clearpage

\begin{figure}[ht]
\centering
\subfloat{\includegraphics[scale=0.75]{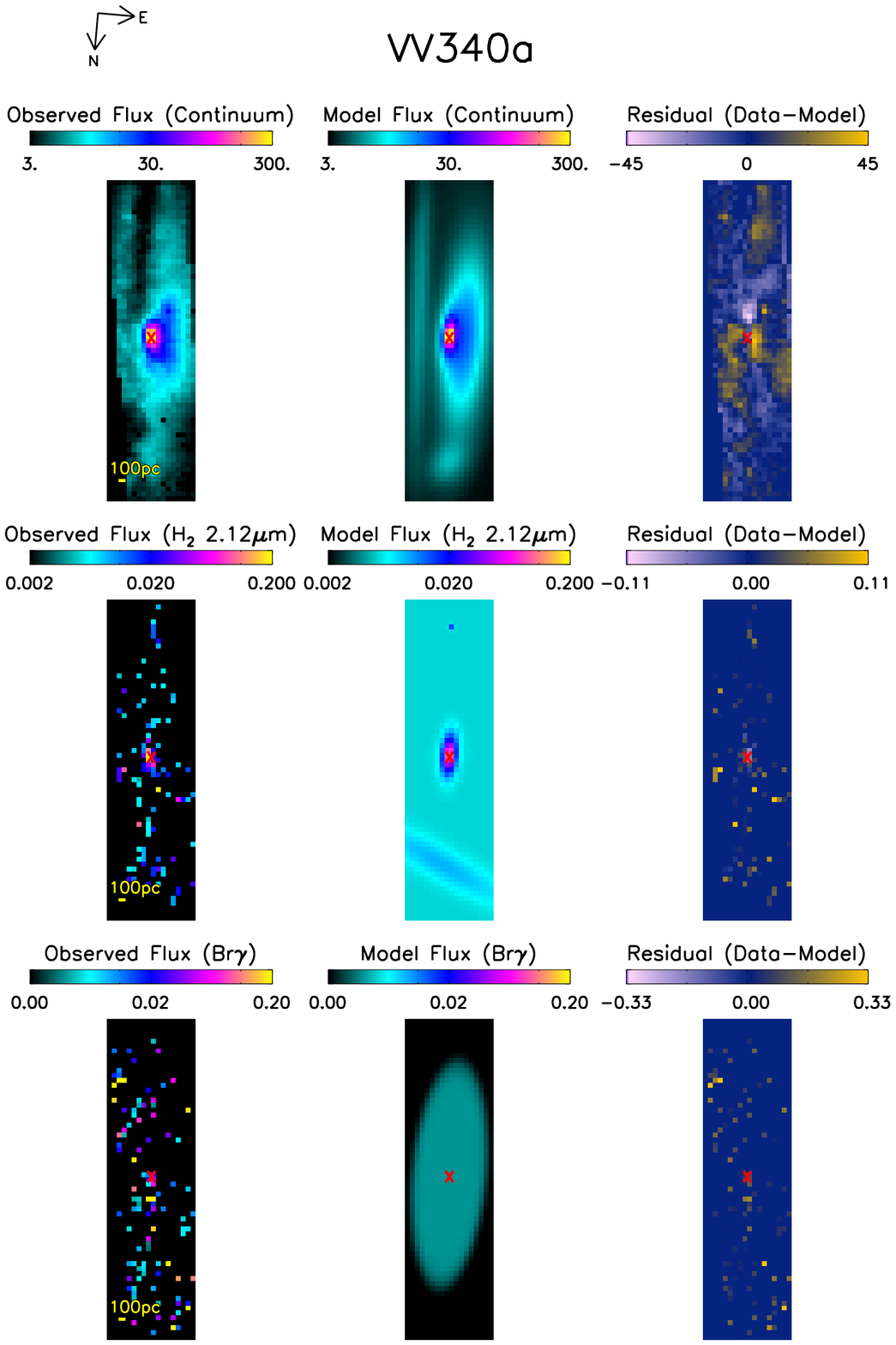}}
\caption{Flux map (left panel), GALFIT model (center panel) and residual (right panel) for VV340a.  The top row is the continuum flux, the middle row is \molhy~flux, and the bottom row is \brg~flux.  The flux maps (left and center panels) are shown on a log scale, while the residual map is shown on a linear scale, in units of counts per second.  Even in $K$-band a solid dust lane is clearly evident in the large-scale disk, which may affect GALFIT measurements.  As an early-stage merger \citep[merger class of 1 in][]{Haan11}, it has retained its large-scale disk.  However, it appears also to be building an additional distinct nuclear disk on a smaller scale, seen in continuum emission and \molhy~emission; \brg~emission does not trace this additional structure.  Due to modeling uncertainties from the GALFIT measurements, we constrain the axial ratios of the two disks to be the same.}
\label{VV340a}
\end{figure}

\begin{figure}[ht]
\centering
\ContinuedFloat
\subfloat{\includegraphics[scale=1.15]{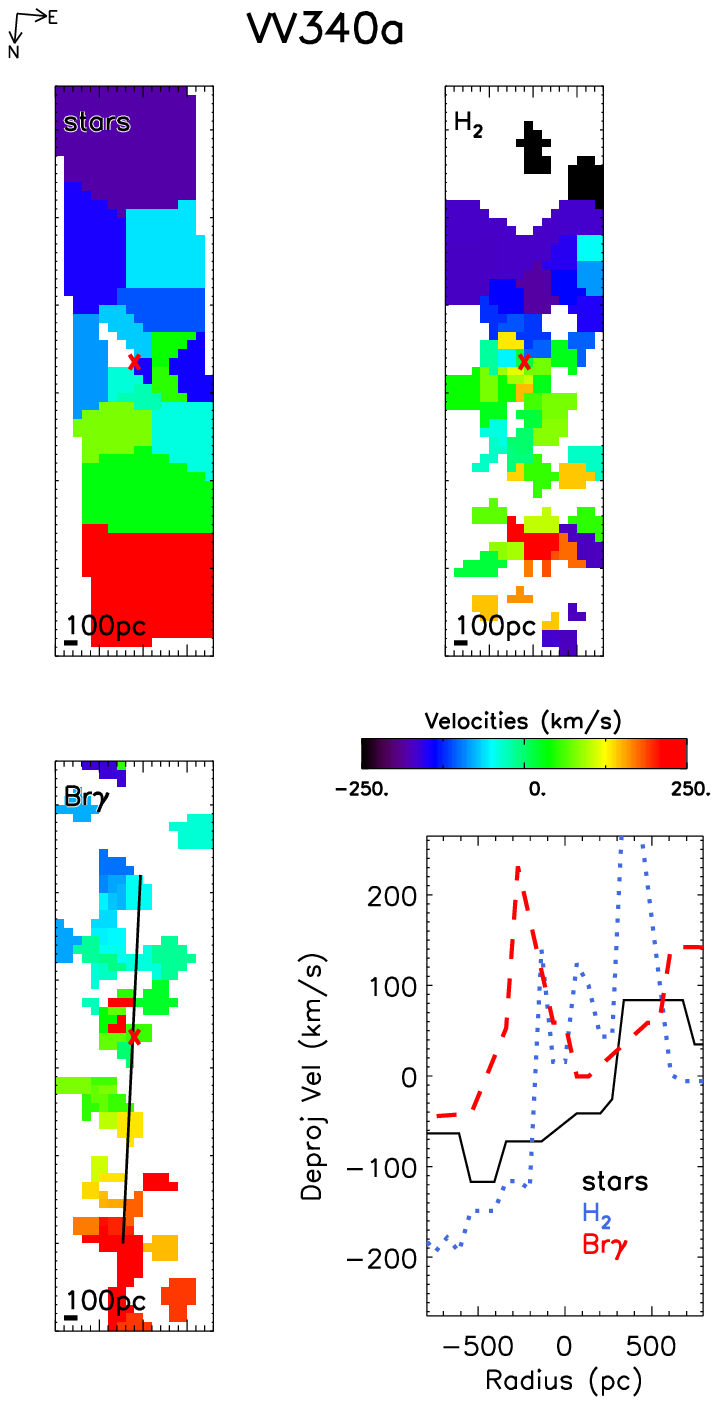}}
\caption{Observed velocity maps of stars (top left), molecular gas in \molhy~(top right), and ionized gas in \brg~(bottom left) for VV340a.  All velocity maps use the same color bar; white pixels do not have sufficient signal to measure velocities accurately.  Bottom right panel shows the deprojected velocity profile cut through the major axis for each of the 3 tracers: stars (solid black), \molhy~(dotted blue),  \brg~(dashed red).  The major axis cut is indicated in black on the bottom left map for clarity.}
\end{figure}

\begin{figure}[ht]
\centering
\subfloat{\includegraphics[scale=0.75]{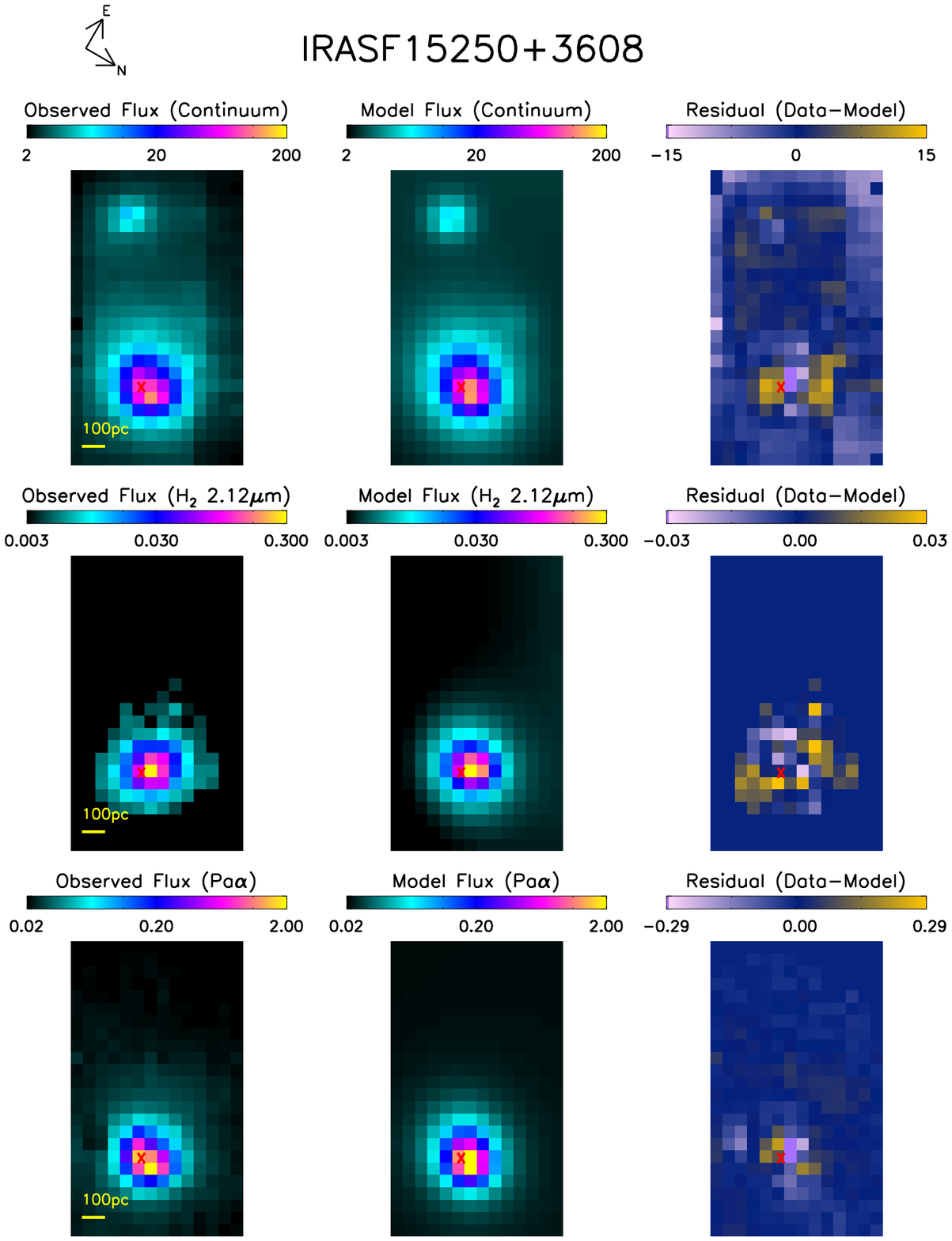}}
\caption{Flux map (left panel), GALFIT model (center panel) and residual (right panel) for IRAS F15250+3608.  The top row is the continuum flux, the middle row is \molhy~flux, and the bottom row is \paa~flux.  The flux maps (left and center panels) are shown on a log scale, while the residual map is shown on a linear scale, in units of counts per second.  The continuum image reveals a second clump to the southeast of the main nucleus; it is not clear whether or not this is a second nucleus or an off-nuclear clump.  GALFIT model residuals show that multiple components may be present in the main nucleus, but because no kinematic angle has been measured, constraints for additional components are not present.  }
\label{IR15250}
\end{figure}

\begin{figure}[ht]
\centering
\ContinuedFloat
\subfloat{\includegraphics[scale=1.6]{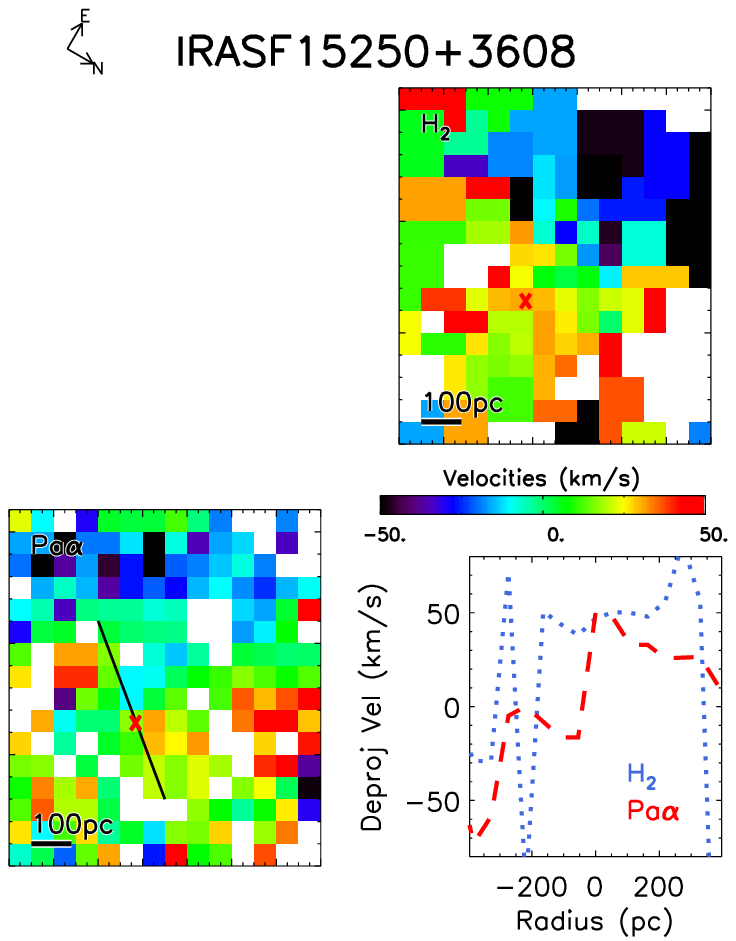}}
\caption{Observed velocity maps of molecular gas in \molhy~(top right) and ionized gas in \paa~(bottom left) for IRAS F15250+3608.  All velocity maps use the same color bar; white pixels do not have sufficient signal to measure velocities accurately.  Bottom right panel shows the deprojected velocity profile cut through the major axis for each of the two tracers: \molhy~(dotted blue) and \paa~(dashed red).  The major axis cut is indicated in black on the bottom left map for clarity.  Since the second component from the continuum GALFIT models had no counterpart in emission lines, these velocity maps focus on the main component only.  This is the only galaxy in our sample for which we see no rotation in either stars or gas.}
\end{figure}

\begin{figure}[ht]
\centering
\subfloat{\includegraphics[scale=0.75]{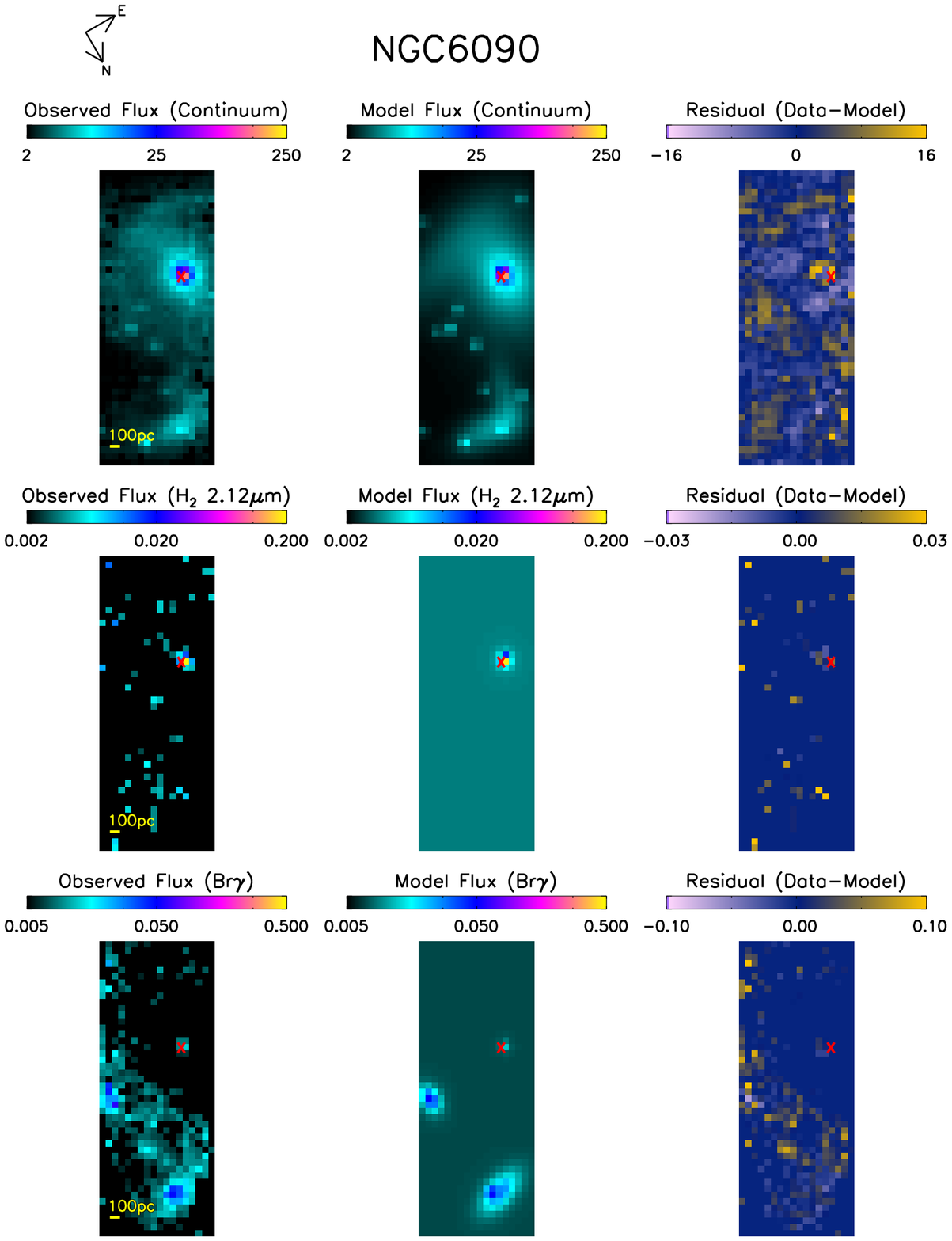}}
\caption{Flux map (left panel), GALFIT model (center panel) and residual (right panel) for NGC~6090.  The top row is the continuum flux, the middle row is \molhy~flux, and the bottom row is \brg~flux.  The flux maps (left and center panels) are shown on a log scale, while the residual map is shown on a linear scale, in units of counts per second.  This galaxy's continuum map shows a low-inclination disk with large spiral arms and possibly a bar.  The size of the disk and previous estimates of the merger stage suggest that this disk was not formed during the merger, but rather is a large galactic disk that has not yet been disturbed.  Note that, while the nucleus shows strong continuum, minimal line flux is seen at the nucleus of the disk.  \brg~flux is, however, strong in the spiral arms.  }
\label{NGC6090}
\end{figure}

\begin{figure}[ht]
\centering
\ContinuedFloat
\subfloat{\includegraphics[scale=1.3]{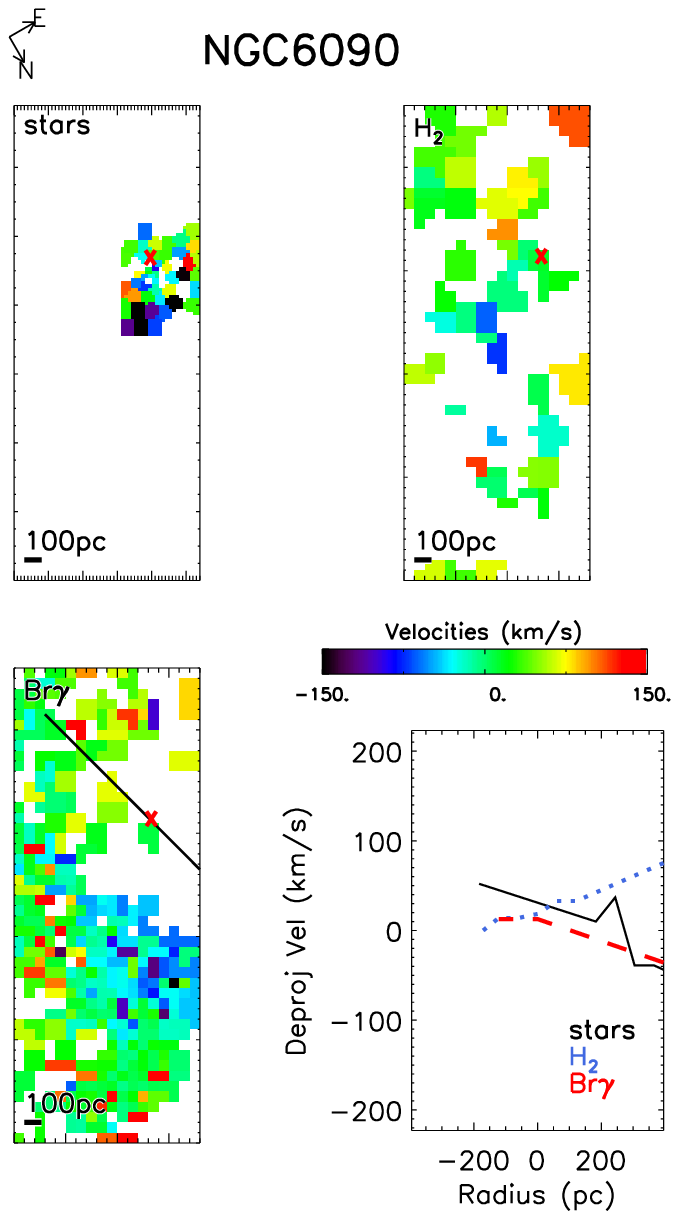}}
\caption{Observed velocity maps of stars (top left), molecular gas in \molhy~(top right), and ionized gas in \brg~(bottom left) for NGC~6090.  All velocity maps use the same color bar; white pixels do not have sufficient signal to measure velocities accurately.  Bottom right panel shows the deprojected velocity profile cut through the major axis for each of the 3 tracers: stars (solid black), \molhy~(dotted blue),  \paa~(dashed red).  The major axis cut is indicated in black on the bottom left map for clarity.  Because of the lack of flux in emission lines, and the limited field of view of these OSIRIS data, obtaining velocity information for the gas across the major axis of the disk is difficult, and provides little insight.  We conclude that we see no gaseous nuclear disk, although a stellar disk is measured and included in this analysis.}
\end{figure}

\begin{figure}[ht]
\centering
\subfloat{\includegraphics[scale=0.9]{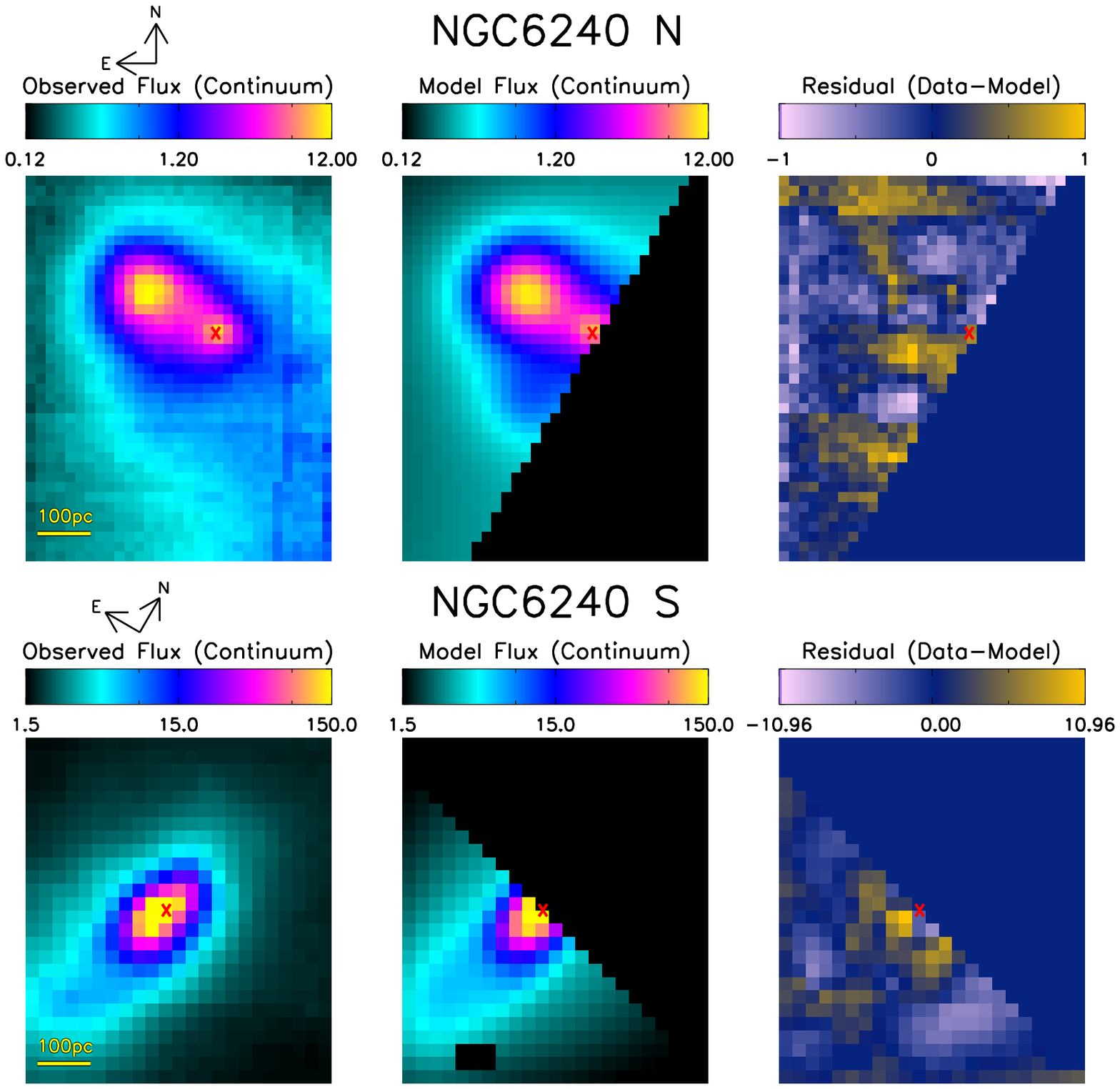}}
\caption{Flux map (left panel), GALFIT model (center panel) and residual (right panel) for the continuum levels of the two nuclei of NGC~6240.  The flux maps (left and center panels) are shown on a log scale, while the residual map is shown on a linear scale, in units of counts per second.  Masks shown in GALFIT fits were implemented due to the excess of dust present between the two nuclei \citep{Max07}.  }
\label{NGC6240}
\end{figure}

\begin{figure}[ht]
\centering
\ContinuedFloat
\subfloat{\includegraphics[scale=1.5]{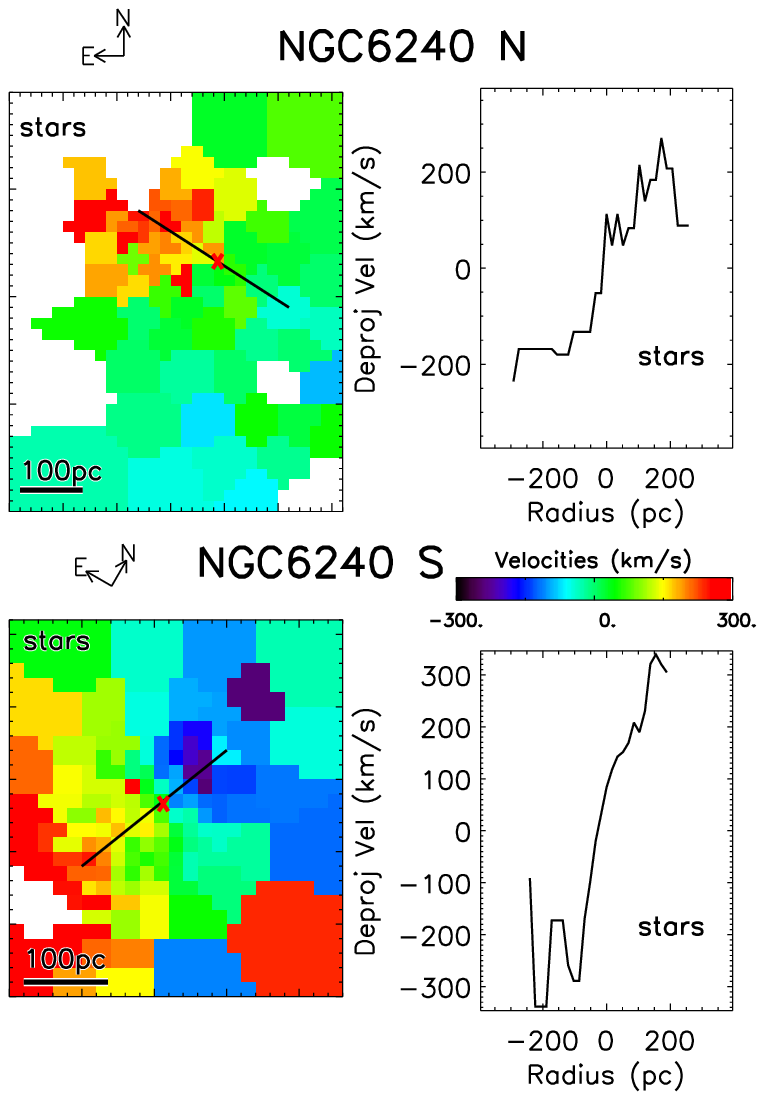}}
\caption{Observed velocity maps of stars (left panels) for the two nuclei of NGC~6240.  Both velocity maps use the same color bar; white pixels do not have sufficient signal to measure velocities accurately.  The two right panels show the deprojected velocity profile cut through the major axis for each nucleus.  The major axis cut is indicated in black on the left maps for clarity.}
\end{figure}

\begin{figure}[ht]
\centering
\subfloat{\includegraphics[scale=0.8]{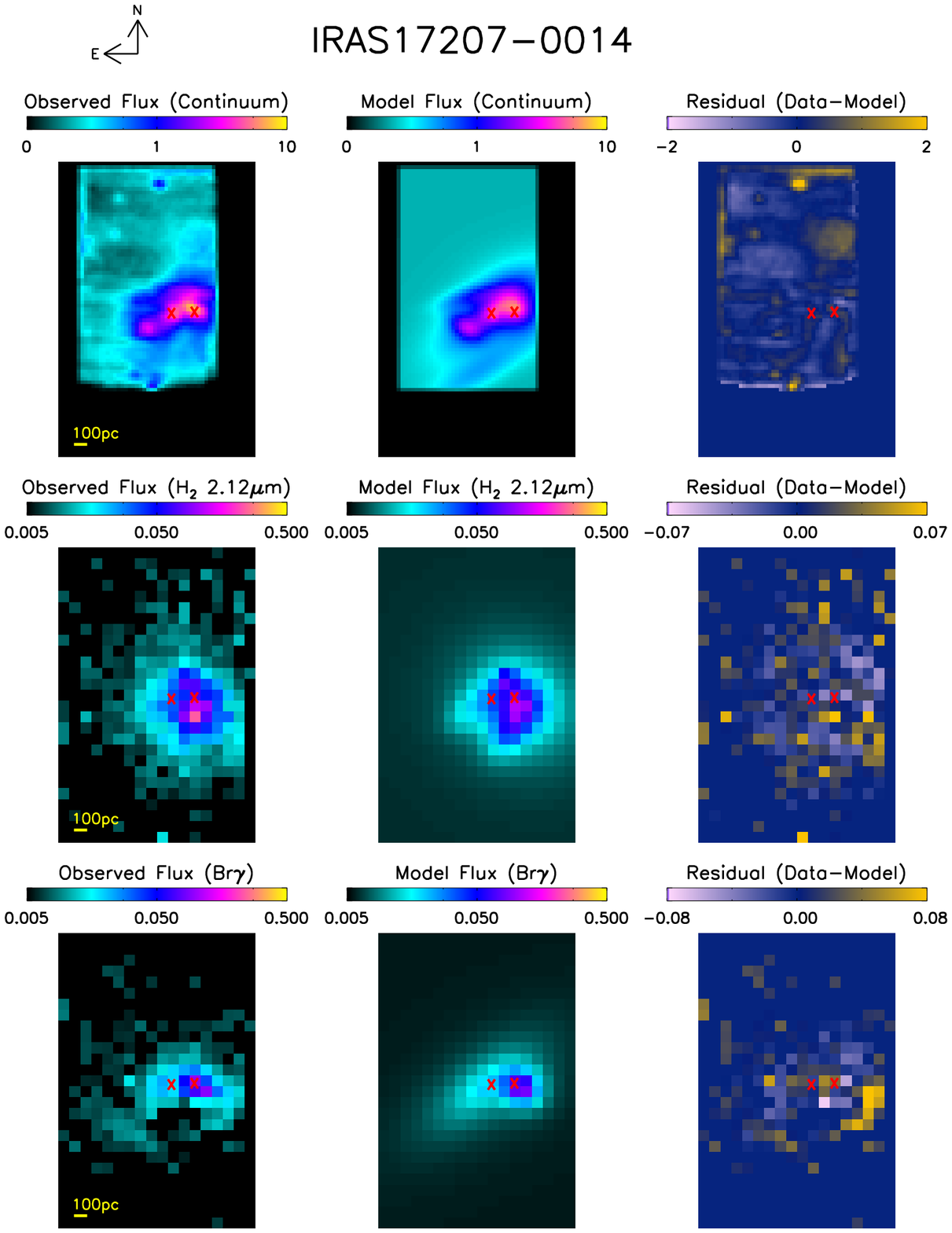}}
\caption{Flux map (left panel), GALFIT model (center panel) and residual (right panel) for IRAS F17207-0014.  The top row is the continuum flux, the middle row is \molhy~flux, and the bottom row is \brg~flux.  The flux maps (left and center panels) are shown on a log scale, while the residual map is shown on a linear scale, in units of counts per second.  This galaxy shows two overlapping disks.  Here as well the \molhy~flux map shows morphologies that do not match continuum emission; this may be signs either of outflows or of collisional shocks related to the larger nuclear orbits.
}
\label{IR17207}
\end{figure}

\begin{figure}[ht]
\centering
\ContinuedFloat
\subfloat{\includegraphics[scale=1.6]{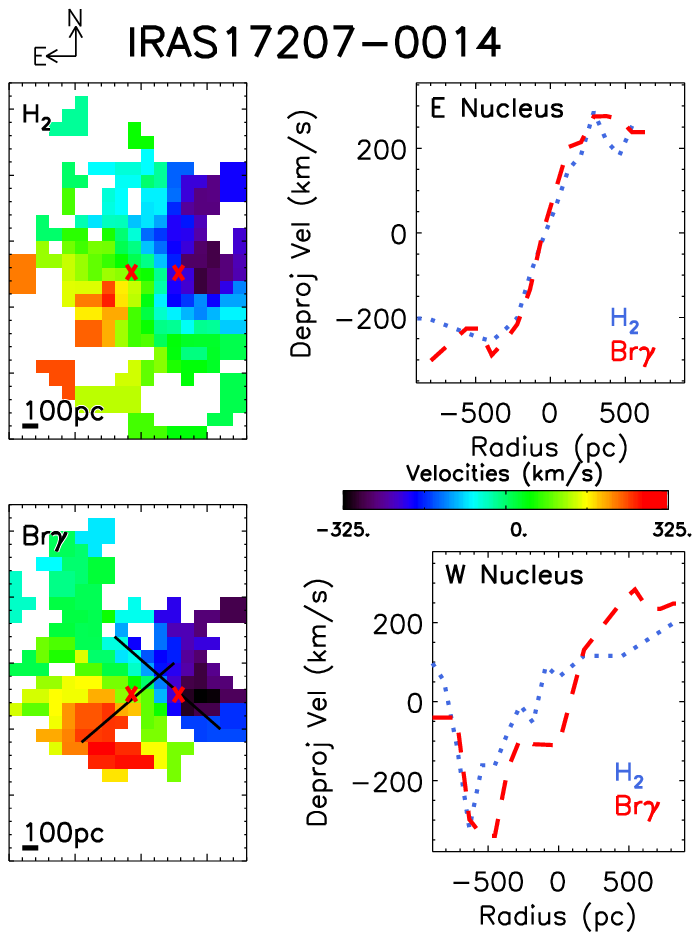}}
\caption{Observed velocity maps of molecular gas in \molhy~(top left) and ionized gas in \brg~(bottom left) for IRAS F17207-0014.  Both velocity maps use the same color bar; white pixels do not have sufficient signal to measure velocities accurately.  The two right panels shows the deprojected velocity profile cut through the major axis of each nucleus for each of the two tracers: \molhy~(dotted blue) and \brg~(dashed red).  The major axis cuts are indicated in black on the bottom left map for clarity.}
\end{figure}

\begin{figure}[ht]
\centering
\subfloat{\includegraphics[scale=0.75]{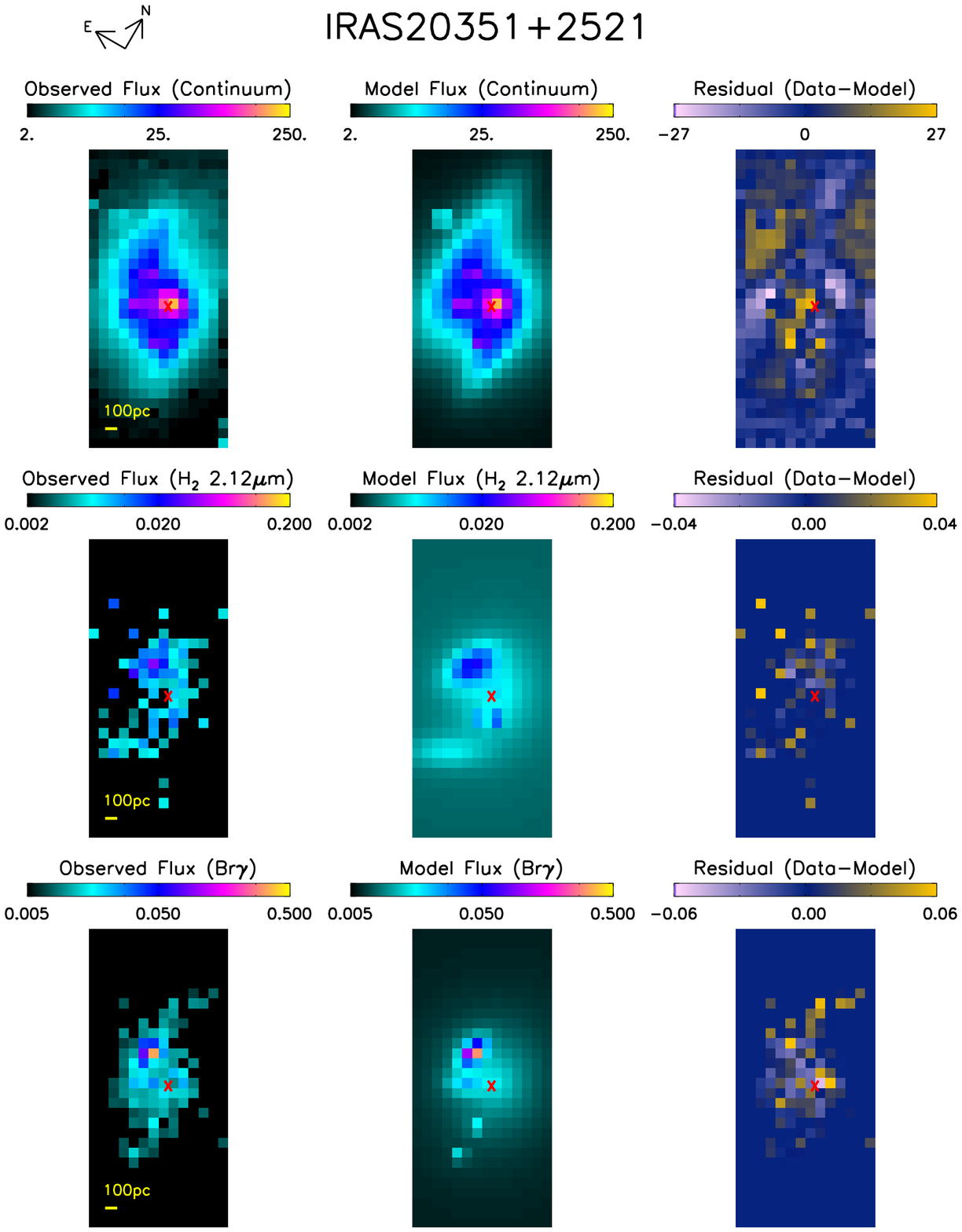}}
\caption{Flux map (left panel), GALFIT model (center panel) and residual (right panel) for IRAS 20351+2521.  The top row is the continuum flux, the middle row is \molhy~flux, and the bottom row is \brg~flux.  The flux maps (left and center panels) are shown on a log scale, while the residual map is shown on a linear scale, in units of counts per second.  The clumpy nature of this nucleus is reminiscent of spiral arms; with a large clump of emission appearing northeast of center in continuum and both line emission maps.  }
\label{IR20351}
\end{figure}

\begin{figure}[ht]
\centering
\ContinuedFloat
\subfloat{\includegraphics[scale=1.6]{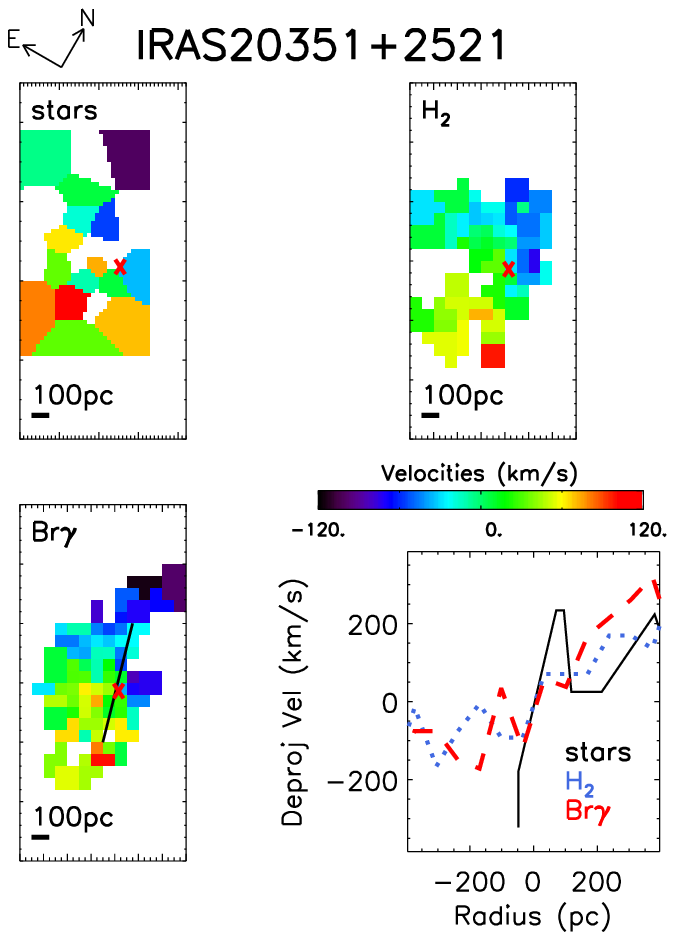}}
\caption{Observed velocity maps of molecular gas in stars (top left), \molhy~(top right), and ionized gas in \brg~(bottom left) for IRAS 20351+2521.  All velocity maps use the same color bar; white pixels do not have sufficient signal to measure velocities accurately.  Bottom right panel shows the deprojected velocity profile cut through the major axis for each of the 3 tracers: stars (solid black), \molhy~(dotted blue),  \brg~(dashed red).  The major axis cut is indicated in black on the bottom left map for clarity.}
\end{figure}

\begin{figure}[ht]
\centering
\subfloat{\includegraphics[scale=0.8]{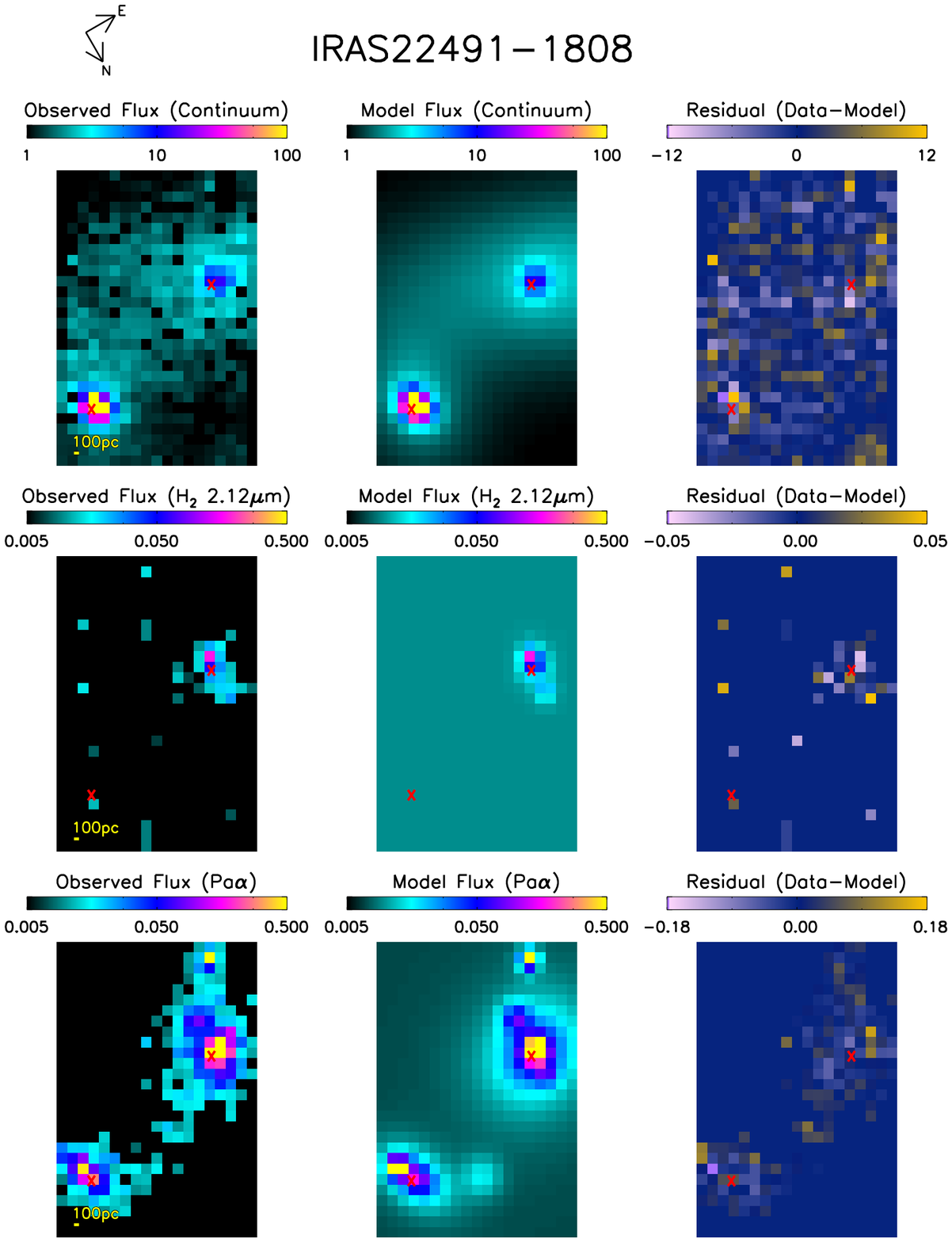}}
\caption{Flux map (left panel), GALFIT model (center panel) and residual (right panel) for IRAS22491-1808.  The top row is the continuum flux, the middle row is \molhy~flux, and the bottom row is \paa~flux.  The flux maps (left and center panels) are shown on a log scale, while the residual map is shown on a linear scale, in units of counts per second.  This galaxy also shows two nuclei.  The eastern nucleus shows extended \molhy~emission along the minor axis.}
\label{IR22491}
\end{figure}

\begin{figure}[ht]
\centering
\ContinuedFloat
\subfloat{\includegraphics[scale=1.5]{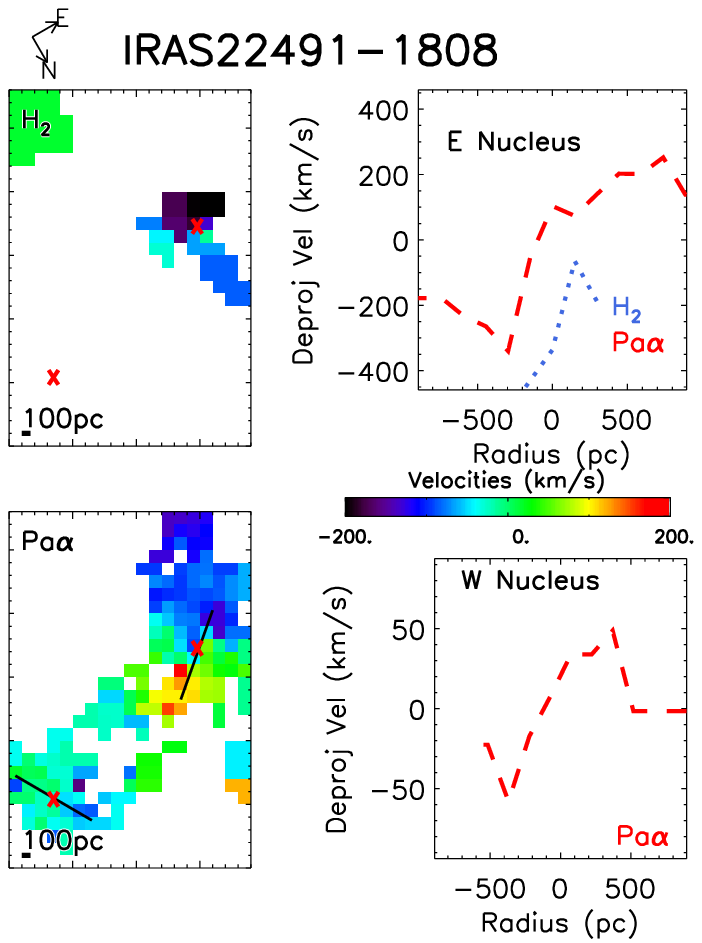}}
\caption{Observed velocity maps of molecular gas in \molhy~(top left) and ionized gas in \paa~(bottom left) for IRAS22491-1808.  Both velocity maps use the same color bar; white pixels do not have sufficient signal to measure velocities accurately.  The two right panels show the deprojected velocity profile cut through the major axis of each nucleus for each of the two tracers: \molhy~(dotted blue) and \paa~(dashed red).  The major axis cut is indicated in black on the bottom left map for clarity.  in \paa, the strongest emission line, both disks appear to show rotating kinematic signatures.  However, though \molhy~is detected in the E disk, it does not appear to share the kinematics of \paa.  A velocity gradient is seen, but because it isn't centered about the same velocity as \paa, it likely has a strong component of streaming or outflowing shocked gas.}
\end{figure}

\clearpage

\section{Details of GALFIT Fits}
\label{galfitdetails}
We include here the details of the GALFIT fits presented in \S\ref{galfitsection}.  In order to obtain a robust measurement for the nuclear disk component, we included additional components in the fits.  These components take the place of components such as star clusters, spiral arms, and tidal streams in order to avoid biasing the fitting of the nuclear disks.  Table~\ref{tbl:giantgalfittable} lists the detailed parameters of the fitted components in each galaxy.  The nuclear disks themselves have been presented in Table~\ref{tbl:galfitresults} but their parameters are also included in this table for reference.  For each galaxy and tracer, the nuclear disk component is listed first, with additional components listed afterwards with brightest first.


Though GALFIT and similar modeling schemes are the most robust methods to measure sizes and other parameters of morphological features in these data, it is important to consider the limitations of such fits.  Because GALFIT fits the image $\chi^{2}$ without regard to physical meaning, it is easily distracted by components that may not be of interest to the science question at hand.  Consider a modeler looking to measure the size of the galactic disk for two galaxies: one which is only a disk, and one which has both a disk and a bulge.  In the first case, a GALFIT model of a single disk will probably fit the diskÕs size well.  In the second case, GALFIT may be confused by the bulge component, and if its luminosity is high, the Òfitted diskÓ may be actually closer in size and shape to the bulge itself.  Now fit both galaxies with a two-component model.  The first case will fit the same disk as one component, with the second component either very faint or with high errors in parameters, indicating that the initial component was a complete model.  In the case of the second galaxy, the two component fit will now accurately measure the bulge and the disk, reporting appropriate measures for both components.  Allowing GALFIT to add a third component will not significantly alter the parameters of the bulge or the disk, unless either of them were poorly fit to begin with (if, for example, a point source exists at the center).  Following this example, we have added components to the GALFIT models until the disk parameters stabilized.  Though we have not vetted each of these additional components in the same way, we include them for completeness, as they may aid future researchers in setting up their own GALFIT models to study additional components if they so desire.  In other words, we include these components in our fits because the GALFIT models may find fitting their morphological features more important than fitting the disk.  Removing them from the fit can skew the disk model parameters significantly; on the other hand, adding additional parameters will not.

\begin{landscape}
 \begin{deluxetable}{lcccccccc}
    \centering
    \tabletypesize{\scriptsize}
    \tablewidth{0pt}
    \tablecolumns{8}
    \tablecaption{Parameters of Additional GALFIT Components}
    \tablehead{   
      \colhead{Galaxy Name} &
      \colhead{Tracer} &
      \colhead{Component Type} &
      \colhead{Position x,y} &
      \colhead{$R_{eff}$ or FWHM} &
      \colhead{S\'ersic index} &
      \colhead{Axis Ratio} & 
      \colhead{Position Angle} & \\
      \colhead{} &
      \colhead{} &
      \colhead{} &
      \colhead{(pc, relative to disk center)} &
      \colhead{(pc)} &
      \colhead{$n$} &
      \colhead{(b/a) } & 
      \colhead{($^\circ$ E of N)} &
      	}
    \startdata 
    CGCG436-030 & continuum & S\'ersic & 0,0 & $293\pm6$ & $0.6\pm0.05$&$0.76\pm0.01$ & -71 \\
           & & Gaussian & -15,-120 & $65\pm2$ & - &  $0.89\pm0.04$ & -37\\
           & & PSF & -150,5 & - & - & - & - \\
           & & PSF & 125,-105 & - & - & - & - \\
           & & PSF & 50,230 & - & - & - & - \\
       & \molhy & S\'ersic & 0,0 & $775\pm200$&$0.42\pm0.17$&$0.76$ & -71\\
	  & & Gaussian & -5,-50 & $280\pm10$ & - & $0.62\pm0.03$ & 45 \\
	  & & PSF & -30,-160 & - & - & - & - \\
	  & & PSF & -150,-275 & - & - & - & - \\	  
       & \brg      & S\'ersic & 0,0 & $450\pm30$ & $0.87\pm0.1$ & $0.76$ & -71 \\
           & & Gaussian & -70,-45 & $210\pm10$ & - & $0.55\pm0.03$ & 17 \\
           & & PSF & -25,-95 & - & - & - & - \\
    IRAS F01364-1042 & continuum  & S\'ersic & 0,0 & $211\pm21$ & $0.4\pm0.2$ & $0.49\pm0.03$ & 75 \\
	  & & Gaussian & -90,5 & $1060\pm85$ & - & $0.36\pm0.02$ & 55 \\
	  & & PSF & 0,0 & - & - & - & - \\
	  & & PSF & -10,-155 & - & - & - & - \\
       & \molhy & S\'ersic & 0,0 & $173\pm20$&$1.2\pm0.5$&$0.49$ & 75 \\
	  & & Gaussian & -175,-60 & $800\pm110$ & - & $0.37\pm0.05$ & 39\\
       & \paa     & S\'ersic & 0,0 & $182\pm15$ & $0.75\pm0.25$&$0.49$ & 75 \\
	  & & Gaussian & -70,30 & $645\pm215$ & - & $0.1\pm6.6$ & 29 \\
	  & & PSF & -5,-165 & - & - & - & - \\
	  & & PSF & -85,15 & - & - & - & - \\
    IIIZw035 & continuum & S\'ersic & 0,0 & $182\pm6$& $0.67\pm0.04$&$0.51\pm0.01$ & 35\\
    	  & & Gaussian & -125,25 & $975\pm85$ & - & $0.25\pm0.01$ & 32\\
	  & & Gaussian & -30,-120 & $50\pm390$ & - & $0.06\pm6.4$ & 18 \\
	  & & PSF & -5,25 & - & - & - & - \\  
       & \molhy & S\'ersic & 0,0 & $132\pm5$&$0.45\pm0.04$&$0.51$ & 35\\
	  & & Gaussian & -115,30 & $190\pm10$ & - & $0.36\pm0.03$ & 40\\
	  & & PSF & -55,25 & - & - & - & - \\
       & \brg      & S\'ersic & 0,0 & $103\pm5$&$0.5\pm0.1$&$0.51$ & 35 \\ 
	  & & Gaussian & -55,-80 & $40\pm600$ & - & $0.05\pm0.8$ & 55\\
	  & & PSF & 0,20& - & - & - & - \\ 
    IRAS F03359+1523 & continuum & S\'ersic & 0,0 & $1765\pm2$&$2.9\pm0.5$&$0.24\pm0.02$ & 72\\
	  & & Gaussian & -75,-1360 & $97\pm2$ & - & $0.51\pm0.03$ & 107\\
	  & & PSF & -35,-115 & - & - & - & - \\
	  & & Gaussian & 35,895 & $79\pm2$ & - & 0.20 & -3\\
	  & & PSF & -60,-610 & - & - & - & - \\
	  & & PSF & -55,-1130 & - & - & - & - \\
	  & & PSF & -115,85 & - & - & - & - \\
       & \molhy & S\'ersic & 0,0 & $725\pm25$&$0.4\pm0.1$&$0.24$ & 72\\
       & \brg & S\'ersic & 0,0 & $259\pm24$&$0.7\pm0.2$&$0.24$ & 72\\
	  & & Gaussian & -85,-1235 & $568\pm22$& - & $0.22\pm0.01$ 69\\
	  & & PSF & 260,-5 & - & - & - & - \\
    MCG+08-11-002 & continuum & S\'ersic & 0,0 & $167\pm20$&$0.9\pm0.1$&$0.66\pm0.02$ & 65\\
	  & & Gaussian & -345,-165 & $3760\pm10,000$ & - & $0.17\pm0.46$ &24\\
	  & & Gaussian & -35,-170 & $318\pm15$ & - & $0.75\pm0.03$& 158\\
	  & & Gaussian & -85,-475 & $196\pm5$ & - & $0.81\pm0.02$ & -18\\
	  & & Gaussian & -10,500 & $331\pm46$ & - & $0.43\pm0.06$ 60\\
	  & & PSF & -15,30 & - & - & - & - \\
	  & & PSF & -75,-115 & - & - & - & - \\
	  & & PSF & 30,545 & - & - & - & - \\
       & \molhy & S\'ersic & 0,0 & $720\pm85$&$0.4\pm0.1$&$0.4$ & 65\\
	 & & Gaussian & 60,-20 & $810\pm180$ & - & $0.17\pm0.04$ & 53\\
       & \brg      & S\'ersic & 0,0 & $105\pm30$&$0.8\pm0.3$&$0.4$ & 65\\
	 & & Gaussian & -50,190 & $860\pm30$ & - & $0.26\pm0.01$ & 54\\
	 & & Gaussian & 40,-175 & $462\pm24$ & - & $0.23\pm0.01$ & 42\\
	 & & PSF & -85,-470 & - & - & - & - \\
	 & & PSF & 70,350 & - & - & - & - \\
	 & & PSF & -75,390 & - & - & - & - \\
	 & & PSF & -160,-130 & - & - & - & - \\
	 & & PSF & -85,-470 & - & - & - & - \\
    NGC~2623 & continuum & S\'ersic & 0,0 & $142\pm3 $& $2.0\pm0.1$ & $0.81\pm0.01$ & -110 \\
	 & & PSF & 5,35 & - & - & - & - \\ 
       & \molhy & S\'ersic & 0,0 & $172\pm18 $& $1.25\pm0.15$ & $0.81\pm0.01$ -110\\
       & \brg      & S\'ersic & 0,0 & $112\pm7 $& $0.94\pm0.10$ & $0.81\pm0.01$ -110\\
    UGC5101 & continuum & S\'ersic & 0,0 & $518\pm5$& $0.6\pm0.01$&$0.44\pm0.01$ & 79\\
	 & & S\'ersic & 40,-90 & $5600\pm3600$ & $9.3\pm1.4$ & $0.75\pm0.01$ & 145\\
	 & & Gaussian & 70,-55 & $125\pm2$ & - & $0.34\pm0.01$ & 136 \\
       & \molhy & S\'ersic & 0,0 &  $505 \pm 16 $&$0.6\pm0.1$&$0.44$ & 79\\
	 & & Gaussian & 60,-110, & 124 & - & $1.0\pm0.1$ & 132\\
	 & & PSF & -35,30 & - & - & - & - \\
       & \brg & S\'ersic & 0,0 & $459 \pm 15$ &$0.3\pm0.1$&$0.44$ & 79\\
    Mrk231 
       & \molhy & S\'ersic & 0,0 & $115\pm33$&$4.2\pm1.8$&$1.0\pm0.1$ & 100\\
    VV340a & continuum & S\'ersic & 0,0 & $238\pm23$ & $1.1\pm0.1$ & $0.33\pm0.02$ & 180 \\
	 & & Gaussian & 215,-40 & $1898\pm50$ & & $0.25\pm0.01$ & 180  \\
	 & & S\'ersic & 0,0 & $1782\pm595$ & $0.05\pm0.04$ & $0.33\pm0.02$ & 180 \\
	 & & Gaussian & -410,410 & $3298\pm416$ & & $0.06\pm0.01$ & 183\\
	 & & Gaussian & -25,-1745 & $485\pm44$ & & $0.76\pm0.05$ & 150\\
       & \molhy & S\'ersic & 0,0 & $211\pm20$ & $0.6\pm0.2$ & $0.33$ & 180\\
	 & & Gaussian & -100,-1490 & $1800\pm1572$ & & $0.19\pm0.17$ & 245\\
	 & & Gaussian & 35,1765 & $7\pm700$ & & $0.08\pm1.0$ & 206\\
       & \brg & S\'ersic & 0,0 & $1078\pm19$ & $0.1\pm0.1$ & $0.33$ & 180  \\
    IRAS F15250+3608 & continuum & S\'ersic & 0,0 & $246\pm41$&$1.9\pm0.3$&$0.78\pm0.01$ & 136\\
	 & & PSF & 30,-75 & - & - & - & - \\
	 & & PSF & -70,720 & - & - & - & - \\
	 & & PSF & 210,260 & - & - & - & - \\
       & \molhy & S\'ersic & 0,0 & $220\pm10$&$2.0\pm0.9$&$0.78$ & 136\\
	 & & PSF & 55,-15 & - & - & - & - \\
       & \paa     & S\'ersic & 0,0& $135\pm30$&$2.2\pm0.7$&$0.78$ & 136\\
	 & & PSF & 20,-65 & - & - & - & - \\
    NGC~6090 & continuum & S\'ersic & 0,0  & $780\pm50$&$1.5\pm0.1$&$0.74\pm0.01$ & 174\\
          & & Gaussian & -600,135 & $1600\pm350$ & - & $0.26\pm0.06$ &106\\
          & & Gaussian & -125,-1450 & $535\pm25$ & - & $0.43\pm0.02$ & 85\\
          & & Gaussian & 10,-1170 & $550\pm50$ & - & $0.32\pm0.03$ & 173\\
	 & & PSF & -15,15 & - & - & - & - \\
	 & & PSF & -515,-505 & - & - & - & - \\
	 & & PSF & -380,-1585 & - & - & - & - \\
	 & & PSF & -30,-1455 & - & - & - & - \\
	 & & PSF & 160,-1130 & - & - & - & - \\
	 & & PSF & -275,-655 & - & - & - & - \\
	 & & PSF & -450,-320 & - & - & - & - \\
	 & & PSF & -195,-1085 & - & - & - & - \\
	 & & PSF & -640,-530 & - & - & - & - \\
	 & & PSF & 150,730 & - & - & - & - \\
	 & & PSF & -775,-1470 & - & - & - & - \\
	 & & PSF & -770,510 & - & - & - & - \\
       & \brg & Gaussian & -45,-1385 & $349\pm24$ & - & $0.44\pm0.03$ & 109\\
          & & Gaussian & -705,-510 & $164\pm16$ & - & $0.53\pm0.10$ & 179\\
          & & PSF & -145,-1385 & - & - & - & - \\
    NGC6240N & continuum & S\'ersic & 0,0 &$350\pm140$&$1.9\pm0.4$&$0.61\pm0.02$ & 61\\
          & & Gaussian & -125,70 & $110\pm2$ & - & 1.0 & -12\\
          & & Gaussian & -120,-120 & 5.0 & - & 1.0 & -12\\
          & & PSF & 0,0 & - & - & -  & - \\ 
    NGC6240S & continuum & S\'ersic & 0,0 & $50\pm1$&$0.4\pm0.1$&$0.50\pm0.01$ & -15\\
           & & Gaussian & -135,-125 & $177\pm11$ & - & $0.67\pm0.04$ & 2\\
           & & Gaussian & -15,-75 & $106\pm21$ & - & $0.3$ & 30 \\
    IRAS F17207-0014 & continuum & S\'ersic & 0,0 (E) & $410\pm15$&$0.8\pm0.05$&$0.4\pm0.01$ & -52\\
     & & S\'ersic & 200,15 (W) &$200\pm15$&$0.9\pm0.05$&$0.85\pm0.05$ & 40\\
           & & Gaussian & 170,-500 & $770\pm130$ & - & $0.26\pm0.04$ & -59\\
           & & Gaussian & -180,120 & $410\pm40$ & - & $0.28\pm0.02$ & -75\\
           & & Gaussian & -235,-125 & $240\pm20$ & - & $0.55\pm0.04$ & 55\\
           & & PSF & 5,-45 & - & - & - & - \\
       & \molhy & S\'ersic & 0,0 (E) & $225\pm10$&$0.2\pm0.1$&$0.41$ & -52\\
       & & S\'ersic & 200,15 (W) &$330\pm30$&$1.2\pm0.2$&$0.85$ & 40\\
	 & & Gaussian & 220,-195 & $325\pm25$ & - & $0.51\pm0.04$& -27\\
	 & & PSF & 195,-85 & - & - & - & - \\
       & \brg & S\'ersic & 0,0 (E) $485\pm50$&$1.1\pm0.2$&$0.041$ & -52\\
       & & S\'ersic & 200,15 (W) & $96\pm7$&$0.75\pm0.24$&$0.85$& 40 \\
    IRAS 20351+2521 & continuum & S\'ersic & 0,0 & $296\pm40$&$1.4\pm0.2$&$0.81\pm0.1$ & 15\\
	 & & Gaussian & -170,340 & $840\pm30$ & - & $0.28$ & 0\\
	 & & Gaussian & -235,-35 & $510\pm60$ & - & $1.0$ & 76\\
	 & & Gaussian & -70,-370 & $400\pm40$ & - & $0.35$ & 51\\
 	 & & PSF & -240,-25 & - & - & - & - \\
	 & & PSF & -110,-250 & - & - & - & - \\
	 & & PSF & -370,580 & - & - & - & - \\
	 & & PSF & -165,220 & - & - & - & - \\
       & \molhy & S\'ersic & 0,0 & $295\pm130$&$0.85\pm0.4$&$0.81$ & 15\\
	 & & Gaussian & -170,185 & $ 267$ & - & 0.52 & -28\\
	 & & Gaussian & -340,-395 & $400\pm125$ & - & $0.35\pm0.12$ & -60\\
	 & & PSF & 0,-175 & - & - & - & - \\
	 & & PSF & -130,-175 & - & - & - & - \\
       & \brg & S\'ersic & 0,0 & $360\pm155$&$1.1\pm0.4$&$0.81$ & 15\\
	 & & Gaussian & -160,215 & $125\pm5$ & - & $0.42\pm0.03$ & -14\\
	 & & Gaussian & -220,5 & $170\pm25$ & - & $0.55\pm0.09$ & 118\\
	 & & PSF & -140,-295 & - & - & - & - \\
	 & & PSF & -255,-490 & - & - & - & - \\
    IRAS F22491-1808 & continuum & S\'ersic & 0,0 (E) & $213\pm137$&$1.9\pm1.6$&$0.9\pm0.1$ & 120\\
    & & S\'ersic & -1670,-1755 (W) & $138\pm10$&$1.0\pm0.2$&$0.6\pm0.1$ & 205\\
	 & & Gaussian & 690,85 & $4500\pm1000$ & - & $0.44\pm0.09$ & 65 \\
	 & & S\'ersic & -1680,-1650 & $1100\pm200$ & $0.8\pm0.2$ & $0.73\pm0.05$ & 167\\
       & \molhy  &  S\'ersic & 0,0 (E) &$100 \pm 10$&$1.0\pm0.9$&$0.85$ & 120 \\
	 & & PSF & -60,155 & - & - & - & -  \\
	 & & PSF & 150,-340 & - & - & - & - \\ 
       & \paa &  S\'ersic & 0,0 (E) &$710\pm350$&$1.2\pm0.07$&$0.85$ & 120\\
       &  & S\'ersic & -1670,-1755 (W) & $400\pm120$&$1.2\pm0.8$&$0.6$ & 205\\
	 & & PSF & 15,35 & - & - & - & - \\ 
	 & & PSF & -90,1300 & - & - & - & - \\
	 & & PSF & -1935,-1593 & - & - & - & - \\
	 & & Gaussian & -270,490 & $220\pm190$ & - &$1.0$ & 198\\
	 & & Gaussian & -740,-1730 & $380\pm230$ & - & 1.0 & 155\\
	 & & PSF & 160,-325 & - & - & - & - 
    \enddata
    \label{tbl:giantgalfittable}
  \end{deluxetable}
\end{landscape}


\bibliographystyle{apj}


\end{document}